\documentclass[pre,amsmath,amssymb,onecolumn, showpacs, superscriptaddress,10pt]{revtex4-2}

\usepackage[english]{babel}
\usepackage[utf8]{inputenc}
\usepackage[T1]{fontenc}
\usepackage{amsmath}
\usepackage[colorlinks=true,citecolor=blue]{hyperref}
\usepackage{graphicx}
\usepackage{amsfonts}
\usepackage{amsthm}

\newcommand{\be}{\begin{equation}}
\newcommand{\ee}{\end{equation}}
\newcommand{\bea}{\begin{eqnarray}}
\newcommand{\eea}{\end{eqnarray}}

\newcommand{\cP}{{\cal P}}

\usepackage{color}
\definecolor{Blue}{rgb}{0.00, 0.00, 1.00}
\definecolor{Red}{rgb}{1.00, 0.00, 0.00}
\definecolor{Green}{rgb}{0.00, 0.70, 0.00}

\begin{document}

\title{Burgers dynamics for Poisson point process initial conditions}

\author{Patrick Valageas}
\affiliation{Universit\'e Paris-Saclay, CEA, CNRS, Institut de Physique th\'eorique, 91191 Gif-sur-Yvette, France}

\begin{abstract}

We investigate the statistical properties of one-dimensional Burgers dynamics evolving from 
stochastic initial conditions defined by a Poisson point process for the velocity potential,
with a power-law intensity.
Thanks to the geometrical interpretation of the solution in the inviscid limit, in terms
of first-contact parabolas, we obtain explicit results for the multiplicity functions of shocks
and voids, and for velocity and density one- and two-point correlation functions and power
spectra.
These initial conditions gives rise to self-similar dynamics with probability distributions
that display power-law tails.
In the limit where the exponent $\alpha$ of the Poisson process that defines the initial conditions
goes to infinity, the power-law tails steepen to Gaussian falloffs and we recover the spatial 
distributions obtained in the classical study by Kida (1979) of Gaussian initial conditions with 
vanishing large-scale power.

\end{abstract}

\maketitle

\section{Introduction}

Nonlinear transport processes governed by advection and dissipation occupy a central place in 
statistical physics. A paradigmatic example is the Burgers equation 
\cite{Burgers1974,Hopf1950,Cole1951}, which offers a minimal setting where 
steepening nonlinearities and dissipation coexist. 
In this paper, we focus on the deterministic Burgers equation without external noise,
so that the randomness only comes from the stochasticity of the initial conditions.
In the inviscid limit, where the viscosity coefficient $\nu$ becomes infinitesimally small, 
velocity gradients intensify until shocks form \cite{Burgers1974,Gurbatov1981,Whitham1999}, 
with linear velocity ramps in-between.
This produces strongly intermittent fields with highly inhomogeneous density structures. 
This makes Burgers turbulence a valuable theoretical laboratory for nonlinear stochastic dynamics
in far-from-equilibrium systems \cite{Frisch2001,Bec2007}, including turbulence modeling 
\cite{Kraichnan1968,Kida1979} 
or irreversible aggregation processes \cite{Frachebourg2000,Valageas2009}.

The inviscid Burgers equation also appears in the context of the formation of cosmological
large-scale structures. 
The well-known Zel’dovich approximation \cite{Zeldovich1970,Shandarin1989}, 
which corresponds to a linear approximation in the Lagrangian picture, models collisionless 
gravitational dynamics by pure advection in comoving coordinates. 
However this leads to multistreaming and unphysical particle escape from gravitational potential wells. 
The adhesion model was then introduced to mimic gravitational trapping within overdensities 
\cite{Gurbatov1989,Gurbatov1991,Vergassola1994,Valageas2011} by adding an infinitesimal viscosity term, 
which prevents shell crossing.
This again leads to the Burgers equation in the inviscid limit and is then able to capture 
the emergence of the cosmic web \cite{Melott1994,Sahni1994}.

Much of this research focuses on Gaussian initial conditions with scale-free power spectra 
\cite{Gurbatov1981,She1992,Gurbatov1997}. 
This leads to a self-similar evolution characterized by scaling laws relating shock densities,
velocity increments, and a growing integral length scale.
Although the well-known Hopf-Cole transformation \cite{Hopf1950,Cole1951}
provides an explicit representation of the solution at any time $t$ in terms of the initial
conditions, the statistical properties of the dynamics can only be explicitly derived
in a few cases, such as for Brownian initial velocity \cite{Bertoin1998,Valageas2009a}
or white-noise initial velocity \cite{Frachebourg2000a,Valageas2009b}.
Such Gaussian initial conditions generically lead to tails of the various probability distributions
that decay as the exponential of a power-law, with an exponent that depends on the slope of the
initial power spectrum \cite{Avellaneda1995,Molchan1997,Valageas2009c}.

However, in many physical systems the dominant contribution comes from rare, large events rather than 
Gaussian fluctuations, with probability distributions that show heavy tails.
Such initial conditions have already been considered by 
\cite{Carraro1998,Bertoin1998,Bertoin2001} for Levy processes
and by \cite{Bernard1998,Gurbatov1999,Gueudre2014}
for Poisson point processes.
In the present work, following \cite{Molchanov1997,Gueudre2014}, 
we investigate the case where the initial velocity potential $\psi_0(q)$ is a Poisson point process
in the $(q,\psi_0)$ plane with a power-law intensity of exponent $\alpha > 3/2$,
(see also \cite{Albeverio1994} for a study of more general Poisson point processes).
This Poisson point process leads to a discrete set of points $\{(q,\psi_0)_i\}$ instead of a continuous
function, but we can imagine that from each point we draw two almost vertical lines of slopes
$\pm \gamma$ and next join together these triangles to form a continuous function.
In the limit $\gamma \to \infty$ only the upper summits $\{(q,\psi_0)_i\}$ will be relevant.
In practice, this class of initial conditions is analytically tractable because the inviscid 
Burgers solution can be expressed as a maximization problem over the initial potential, 
using the well-known Hopf–Cole transformation \cite{Hopf1950,Cole1951}.
In particular, it admits a geometrical interpretation in terms of the first-contact points of 
parabolas with the initial potential $\psi_0(q)$.
Then, the dynamics remain well defined despite the strong singularity of the initial field: shocks and 
voids naturally emerge as first-contact structures between upward-opening parabolas and the random 
point landscape.
Moreover, through this geometrical construction the Poisson point process provides the simplest
class of initial conditions that allows for explicit analytical expressions, as all statistical 
distributions can be derived from the probabilities that there are no initial points above
a set of parabolic arcs.

Whereas \cite{Gurbatov1999} considered the decay of the energy integral $E(t) = \langle v^2 \rangle$
and the one-point distribution of the velocity potential $\psi$,
\cite{Gueudre2014} also derived the one- and two-point distributions of the Lagrangian coordinate $q$
(i.e., the initial position of the particle observed at position $x$ at time $t$), as well as the
mass function of the shocks.
In this paper, we extend these works by providing explicit expressions and numerical computations
for the distribution of voids, the velocity correlation and power spectrum, 
the distribution of the mass within a given spatial domain, and the two-point distribution of the
particle displacements.
We also explicitly show that these results converge to the classical case of Gaussian initial conditions
with vanishing large-scale power studied by \cite{Kida1979}.

This paper is organized as follows. 
In Section~\ref{sec:Initial} we review the equations of motion and the geometrical interpretation 
of the system, we describe the class of initial conditions that we consider and we provide two numerical
realizations to illustrate the dependence on the exponent $\alpha$.
In Section~\ref{sec:one-point-Eulerian} we derive the one-point Eulerian probability distribution 
$P_0(v)$ of the velocity field. 
In Section~\ref{sec:two-point} we present the two-point Eulerian distributions of the velocity and 
density fields, as well as the distribution of voids and the energy power spectrum.
We briefly discuss higher-order distributions in Section~\ref{sec:higher-order}.
In Section~\ref{sec:Lagrangian} we turn to the Lagrangian distributions of the particle displacements
and we obtain the multiplicity function of shocks. 
In Section~\ref{sec:alpha-infty} we discuss the limit $\alpha \to \infty$ and its convergence toward the 
well-known Gaussian regime. 
We conclude in Section~\ref{sec:conclusion}.

\section{Equations of motion and Initial conditions}
\label{sec:Initial}

\subsection{Equations of motion}
\label{sec:Equations-of-motion}

We consider in this article the one-dimensional Burgers equation \cite{Burgers1974} for the velocity 
field $v(x,t)$ in the limit of vanishing viscosity,
\be
\frac{\partial v}{\partial t} + v \frac{\partial v}{\partial x} = \nu \frac{\partial^2 v}{\partial x^2}
\hspace{1cm} \mbox{with} \hspace{1cm} \nu \rightarrow 0^+ ,
\label{eq:Burg}
\ee
and the associated density field $\rho(x,t)$, which is governed by the continuity equation,
\be
\frac{\partial\rho}{\partial t} + \frac{\partial}{\partial x}(\rho v) = 0 , \;\;\; \mbox{with} \;\;\;
\rho(x,0) = \rho_0 ,
\label{eq:continuity}
\ee
with the uniform initial density $\rho_0$.
In the cosmological context, the three-dimensional version of 
Eqs.(\ref{eq:Burg})-(\ref{eq:continuity}) with $\nu=0$ (and $t$ stands for the linear growing
mode $D_+(t)$ and $\vec x$ is a comoving coordinate) is the well-known Zeldovich approximation 
\cite{Zeldovich1970,Valageas2007}, where particles always keep their initial velocity 
and merely follow straight trajectories. 
To prevent particles from escaping to infinity after crossing each other and to mimic 
the gravitational trapping within the potential wells formed by the overdensities,
one adds the diffusive term of Eq.(\ref{eq:Burg}). This gives the "adhesion model" 
\cite{Gurbatov1989,Vergassola1994}, which cannot describe the inner structure of collapsed
objects (e.g., galaxies) but provides a good description of the large-scale structure
of the cosmic web \cite{Melott1994}.
In this context \cite{Valageas2011}, one is actually more interested in the properties of the density
field $\rho(\vec x,t)$ than in the velocity field $\vec v(\vec x,t)$, whereas turbulence studies
only consider the velocity field.
This motivates our consideration of the density field in Section~\ref{sec:Density-field} below.

As is well known \cite[]{Hopf1950,Cole1951,Gurbatov1991}, introducing the velocity potential 
$\psi(x,t)$, with $v= - \partial\psi/\partial x$, and making the change of variable 
$\psi(x,t)=2\nu\ln\theta(x,t)$, the Burgers equation yields the linear heat equation for $\theta$. 
This gives the explicit solution 
\be
v(x,t) = - \frac{\partial\psi}{\partial x} \hspace{0.5cm} \mbox{with} \hspace{0.5cm} 
\psi(x,t)= 2\nu \ln \int_{-\infty}^{\infty} \frac{d q}{\sqrt{4\pi\nu t}} \;
\exp\left[-\frac{(x-q)^2}{4\nu t} + \frac{\psi_0(q)}{2\nu}\right] ,
\label{eq:psinu}
\ee
where we introduced the initial condition $\psi_0(q)=\psi(q,t=0)$. 
Then, in the limit $\nu \rightarrow 0^+$ the steepest-descent method
gives \cite{Gurbatov1991,Bec2007}
\be
\psi(x,t) = \max_q \left[ \psi_0(q) - \frac{(x-q)^2}{2t} \right]
\hspace{0.5cm} \mbox{and} \hspace{0.5cm} v(x,t) = \frac{x-q(x,t)}{t} ,
\label{eq:psinu0}
\ee
where we introduced the Lagrangian coordinate $q(x,t)$ defined by
\be
\psi_0(q) - \frac{(x-q)^2}{2t} \hspace{0.5cm} \mbox{is maximum at the point} 
\hspace{0.5cm} q = q(x,t) .
\label{eq:qmin}
\ee
The Eulerian locations $x$ where there are two solutions, $q_-<q_+$, to the
maximization problem (\ref{eq:psinu0}) correspond to shocks (and all the matter
initially between $q_-$ and $q_+$ is gathered at $x$). The application
$q \mapsto x(q,t)$ is usually called the Lagrangian map, and
$x \mapsto q(x,t)$ the inverse Lagrangian map (which is discontinuous at
shock locations) \cite{Bec2007,Vergassola1994}.

\subsection{Geometrical interpretation and Legendre transform}
\label{sec:Geometrical-interpretation}

As is well known \cite{Burgers1974,Gurbatov1991}, the minimization problem (\ref{eq:qmin})
has a nice geometrical solution. Indeed, let us consider the 
upward parabola $\cP_{x,c}(q)$ centered at $x$ and of minimum $c$, of equation
\be
\cP_{x,c}(q) = c+ \frac{(q-x)^2}{2 t} .
\label{eq:paraboladef}
\ee
Then, starting from above with a large positive value of $c$, such that the
parabola is everywhere well above $\psi_0(q)$, we decrease $c$ until the parabola touches
the initial potential $\psi_0(q)$.
Then, the abscissa of the point of first contact is the Lagrangian coordinate
$q(x,t)$ and the potential is given by $\psi(x,t)=c$.

The expression (\ref{eq:psinu0}) for the velocity potential can also be written
in terms of a Legendre-Fenchel transform \cite{She1992,Vergassola1994,Valageas2011}.
Thus, let us define the linear Lagrangian potential $\varphi_L(q,t)$ and Lagrangian map
$x_L(q,t)$ by
\be
\varphi_L(q,t) = \frac{q^2}{2} - t \psi_0(q), \;\;\;
x_L(q,t) = \frac{\partial\varphi_L}{\partial q} = q + t v_0(q) ,
\ee
which would describe the system in the absence of shocks.
Introducing the function
\be
H(x,t) = \frac{x^2}{2} + t \psi(x,t) ,
\ee
the maximum (\ref{eq:psinu0}) can be written as
\be
H(x,t) = \max_q \left[ x q - \frac{q^2}{2} + t \psi_0(q) \right] = {\cal L}_x[ \varphi_L(q,t) ] ,
\label{eq:H-def}
\ee
where ${\cal L}_x$ is the Legendre transform evaluated at point $x$.
In this manner, $\psi(x,t)$ is obtained from $\psi_0(q)$ through a Legendre transform.
This also provides the inverse Lagrangian map $q(x,t)$ and the velocity field $v(x,t)$.

\subsection{Initial condition}
\label{sec:Initial-condition}

In this paper, following \cite{Molchanov1997,Gueudre2014}, 
we consider stochastic initial conditions for $\psi_0(q)$ that are given by a Poisson
point process with intensity $\lambda(\psi_0)$, in the upper half-plane $(q,\psi_0>0)$.
Thus, we initially have a set of points $\{(q,\psi_0)_i\}$, with a probability 
$P(N_{\cal B}=n)$ to have $n$ points within any domain $\cal B$ given by the Poisson distribution
\be
P(N_{\cal B}=n) = \frac{\Lambda({\cal B})^n}{n!} e^{-\Lambda({\cal B})} , \;\;\;
\mbox{with} \;\;\; \Lambda({\cal B}) = \int_{\cal B} dq d\psi_0 \, \lambda(\psi_0) .
\ee
As in \cite{Gueudre2014}, we focus on the case of power-law intensity $\lambda(\psi_0)$,
\be
\psi_0 > 0 : \;\;\; \lambda(\psi_0) = a \, \psi_0^{-\alpha} , \;\;\; a > 0 , \;\;\; \alpha > 3/2 .
\label{eq:lambda-psi}
\ee
The condition $\alpha>3/2$ ensures that $\Lambda({\cal B}_{x,c})$ is finite when ${\cal B}_{x,c}$
is the domain above the parabola $\cP_{x,c}(q)$ defined in Eq.(\ref{eq:paraboladef}), with $c>0$,
\be
\Lambda( {\cal B}_{x,c} ) = \int_{-\infty}^{\infty} dq \int_{c+(q-x)^2/(2t)}^{\infty} d\psi \, a \psi^{-\alpha} =
 \Lambda_{\alpha} a \sqrt{t} \, c^{3/2-\alpha} ,
\;\;  \mbox{with} \;\; \Lambda_{\alpha} = \frac{\Gamma(\alpha-3/2)}{\Gamma(\alpha)} \sqrt{2\pi} .
\label{eq:Lambda-Parabola}
\ee
For $c \to 0^+$ we have $\Lambda({\cal B}_{x,c}) \to \infty$. This means that, because of the
accumulation of points near the horizontal axis $\psi_0=0$, the first contact of the parabola
$\cP_{x,c}(q)$ with the potential $\psi_0$ will almost surely occur at a point $(q,\psi_0)$
with $\psi_0>0$. Thus, only the upper half-plane $(q,\psi>0)$ is relevant.

This Poisson point process gives a discrete set of points $\{(q,\psi_0)_i\}$.
This initial condition is not a continuous function $\psi_0(q)$ for the velocity potential.
We can imagine that from each point $\{(q,\psi_0)_i\}$ we draw two almost vertical lines
of slopes $\pm \gamma$ that connect to the horizontal axis $\psi_0$, and next join these triangles
by running along the horizontal axis $\psi_0$.
For a finite set of initial points over a domain $[q_1,q_2]$, by restricting to points with 
$\psi_{0,i} \geq \psi_{\min}>0$, this would define a continuous potential $\psi_0(q)$. 
Then, in the limits $\gamma \to \infty$ and $\psi_{\min} \to 0$ the almost vertical lines would
become irrelevant as the parabolas (\ref{eq:paraboladef}) would only make first contacts with the upper 
summits $\{(q,\psi_0)_i\}$ of the narrow triangles.
However, thanks to the geometrical construction (\ref{eq:psinu0})-(\ref{eq:paraboladef}),
this step is not really necessary.
The maximum in Eq.(\ref{eq:psinu0}) can be directly defined as taken over the points
$\{(q,\psi_0)_i\}$,
\be
\psi(x,t) = \max_{i} \left[ \psi_{0,i} - \frac{(x-q_i)^2}{2t} \right]
\hspace{0.5cm} \mbox{and} \hspace{0.5cm} v(x,t) = \frac{x-q_{i\star}}{t} ,
\label{eq:psi-qi}
\ee
where $i_\star$ is the point where the maximum is reached.
This provides a well-defined function $\psi(x,t)$ at all times $t>0$.
In particular, the result at some early time $t_1$ could be considered as the initial condition,
which would thus be a continuous initial potential.

Thus, in this paper we study the velocity and density fields associated with 
Eq.(\ref{eq:psi-qi}), for the power-law intensity (\ref{eq:lambda-psi}).
Because the intensity $\lambda(\psi_0)$ does not depend on $q$, the system is statistically
homogeneous at all times.

\subsection{Self-similar dynamics}
\label{sec:self-similar}

Let us define the rescaled coordinates
\be
x = a^{1/[2(\alpha-3/2)]} t^{1/2+1/[4(\alpha-3/2)]} X , \;\; 
v= a^{1/[2(\alpha-3/2)]} t^{-1/2+1/[4(\alpha-3/2)]} V , \;\; 
\psi = a^{1/(\alpha-3/2)} t^{1/[2(\alpha-3/2)]} \Psi .
\label{eq:re-scaling}
\ee
We also rescale $\psi_0$ to $\Psi_0$ in the same manner, for any given time $t$.
Then, we obtain at any time $t$
\be
\Psi(X,t) = \max_{i} \left[ \Psi_{0,i} - \frac{(X-Q_i)^2}{2} \right]
\hspace{0.5cm} \mbox{and} \hspace{0.5cm} V(X,t) = X-Q_{i\star} ,
\label{eq:psi-Qi}
\ee
and the Poisson process intensity measure reads
\be
\Lambda({\cal B}) = \int_{\cal B} dQ d\Psi_0 \, \Psi_0^{-\alpha} .
\label{eq:Lambda-Psi0}
\ee
Thus, this rescaling has fully absorbed the time $t$. Therefore, the dynamics are statistically
self-similar for all times $t>0$.
In particular, the integral length scale $L(t)$, defined for instance as the transition between the
large-scale and small-scale asymptotic regimes of the system, grows with time as
\be
L(t) \propto t^{\gamma} \;\; \mbox{with} \;\; \gamma = \frac{1}{2} + \frac{1}{4(\alpha-3/2)} .
\label{eq:L-t}
\ee
We can see that $1/2 < \gamma < \infty$, $L(t)$ always grows faster than $\sqrt{t}$, which is reached
in the limit $\alpha \to \infty$.
We note that the normalization factor $a$ has also been fully absorbed by the rescaling
(\ref{eq:re-scaling}).
Thus, the properties of the dynamics only depend on the exponent $\alpha$.
In the following, we focus on equal-time statistics. Therefore, we work with the rescaled coordinates
(\ref{eq:re-scaling}) and to simplify the notations we use lower case letters instead of upper case letters.

\subsection{Limit $\alpha \to \infty$}
\label{sec:alpha-infty-rescale}

In the limit $\alpha \to \infty$, the Poisson intensity (\ref{eq:Lambda-Psi0}) implies that there are
very few points above $\Psi_0=1$ and many points below.
Therefore, the first-contact parabolas have $c \simeq 1$.
On the other hand, from Eq.(\ref{eq:q2-variance}) below, the variance $\langle q^2 \rangle$ 
of the Lagrangian coordinate $q$ found at the Eulerian position $x=0$ scales as $1/\alpha$
in the limit $\alpha \to \infty$.
Therefore, typical displacements scale with a factor $1/\sqrt{\alpha}$ and we define the rescaled
coordinate and probability distribution
\be
q = \tilde q / \sqrt{\alpha} , \;\;\; P_0(\tilde q) = P_0(q) / \sqrt{\alpha} .
\label{eq:tilde-def}
\ee
In the analytical computations we also make the change of variable
$c=1+u/\alpha$ for the height of the parabolas to obtain the asymptotic value
in the limit $\alpha \to \infty$ of the various probability distributions that we consider.
Although in the analytical expressions that we provide for finite values of $\alpha$ we keep the
coordinates (\ref{eq:re-scaling}), in the figures we rescale all results by appropriate powers of
$\sqrt{\alpha}$ as in (\ref{eq:tilde-def}), so that the convergence to the limit $\alpha \to \infty$
can be clearly seen in the tilde coordinates (\ref{eq:tilde-def}).

We present the results we obtain in this limit $\alpha \to \infty$ in 
Section~\ref{sec:alpha-infty} below.
It turns out that we recover the spatial distributions obtained at late times
for Gaussian initial conditions with vanishing large-scale power 
\cite{Kida1979,Gurbatov1981,Gurbatov1991},
$E_0(k) \propto k^n$ with $n>1$ \cite{Gurbatov1997},
or in the hyperbolic asymptotic scaling \cite{Molchanov1995}.
This is not surprising as in these regimes the nonlinear potential (\ref{eq:psi-qi})
is dominated by the rare peaks of the initial potential $\psi_0$, which asymptotically behave
as a Poisson point process for such initial conditions.
By considering the Poisson point process (\ref{eq:lambda-psi}) for the initial potential itself,
one generalizes this regime to the full class $\alpha > 3/2$, with power-law tails that vary
with $\alpha$ and steepen to the Gaussian case for $\alpha \to \infty$.
Moreover, the dynamics are now fully self-similar and there is no need to look for a late-time
asymptotic regime.
In particular, at finite $\alpha$ we do not have logarithmic corrections, such as 
$E(t) = \langle v^2(t) \rangle \propto t^{-1} \ln(t)^{-1/2}$ for Gaussian initial conditions
\cite{Kida1979,Gurbatov1997}.
Another difference is that while in the Gaussian case $E(k)$ goes to zero for $k \to 0$
faster than $k$ for the results of \cite{Kida1979} and Section~\ref{sec:alpha-infty}
to apply \cite{Gurbatov1997}, this is not the case for the full class of finite values
$\alpha > 3/2$ studied in this paper.
As seen in Fig.~\ref{fig:Ek} and Eq.(\ref{eq:Ek-small-k-power}) below, for $\alpha <3$ we have 
$E(0)=\infty$ and for $\alpha < 5/2$ the velocity correlation and the energy spectrum do not exist
as the velocity variance is infinite (because of heavy power-law tails).

\subsection{Numerical realizations}
\label{sec:realizations}

\begin{figure}
\centering
\includegraphics[height=5.cm,width=0.24\textwidth]{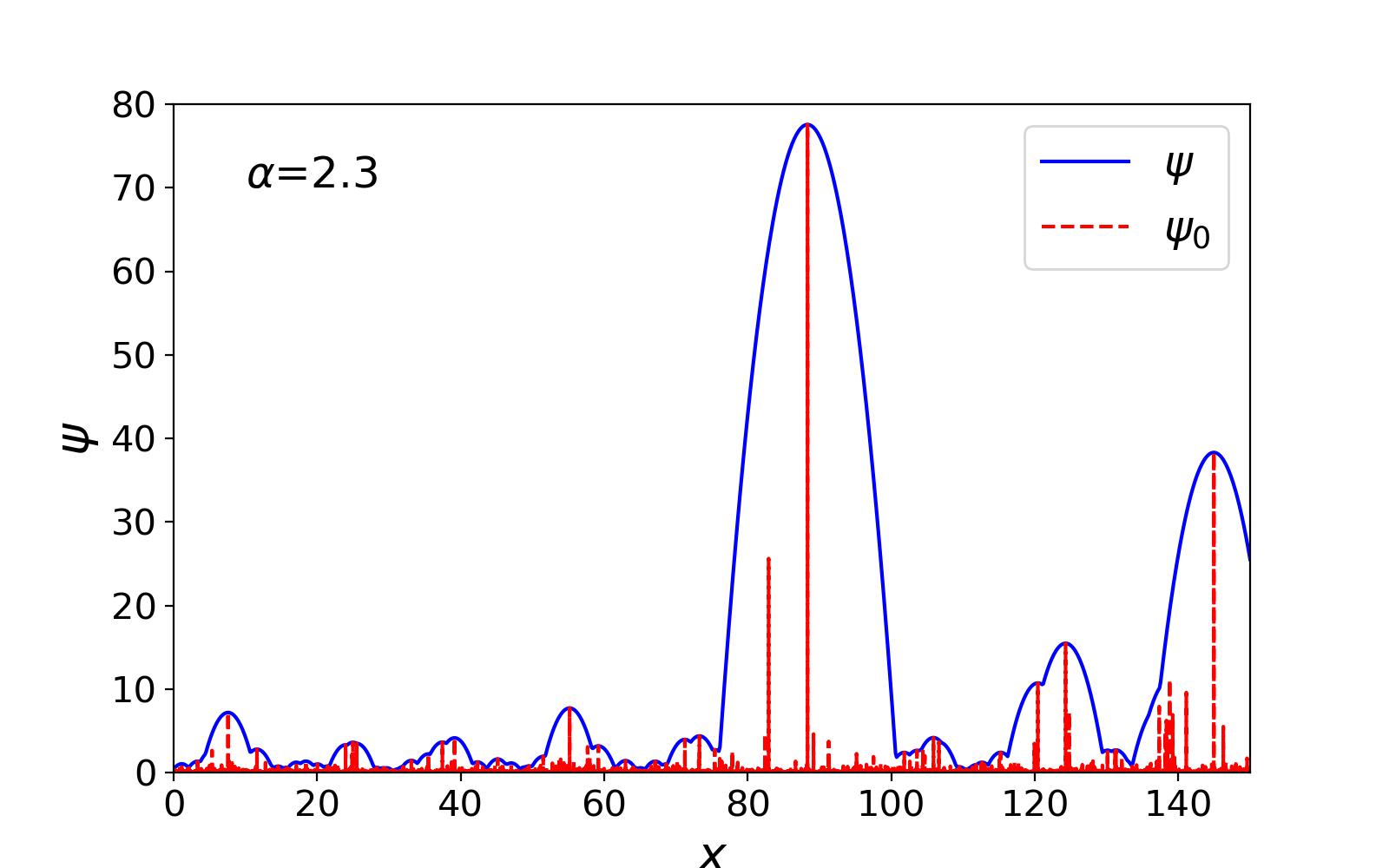}
\includegraphics[height=5.cm,width=0.24\textwidth]{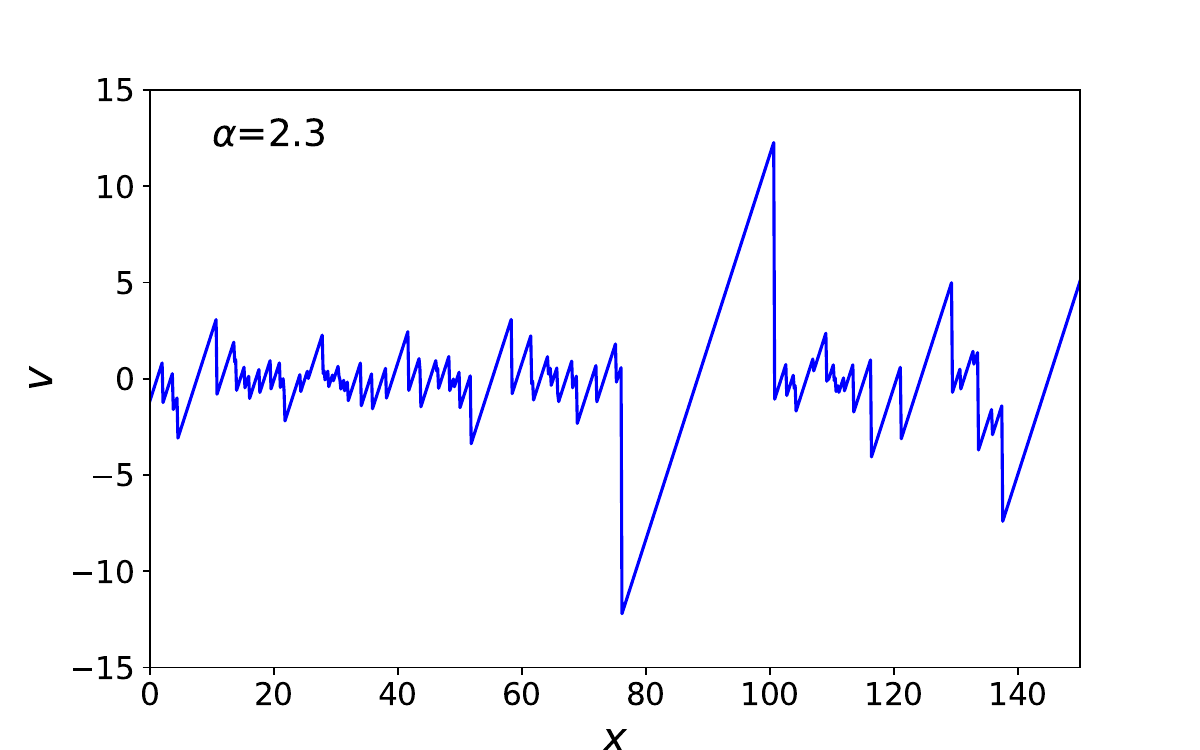}
\includegraphics[height=5.cm,width=0.24\textwidth]{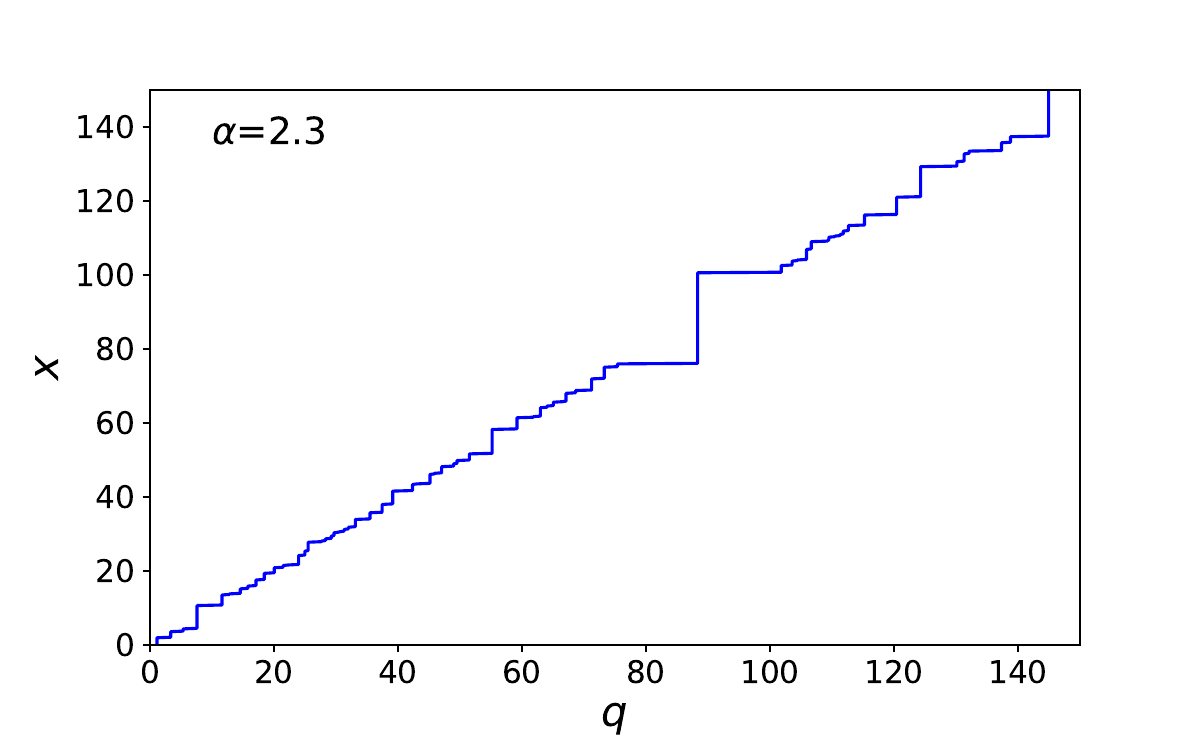}
\includegraphics[height=5.cm,width=0.24\textwidth]{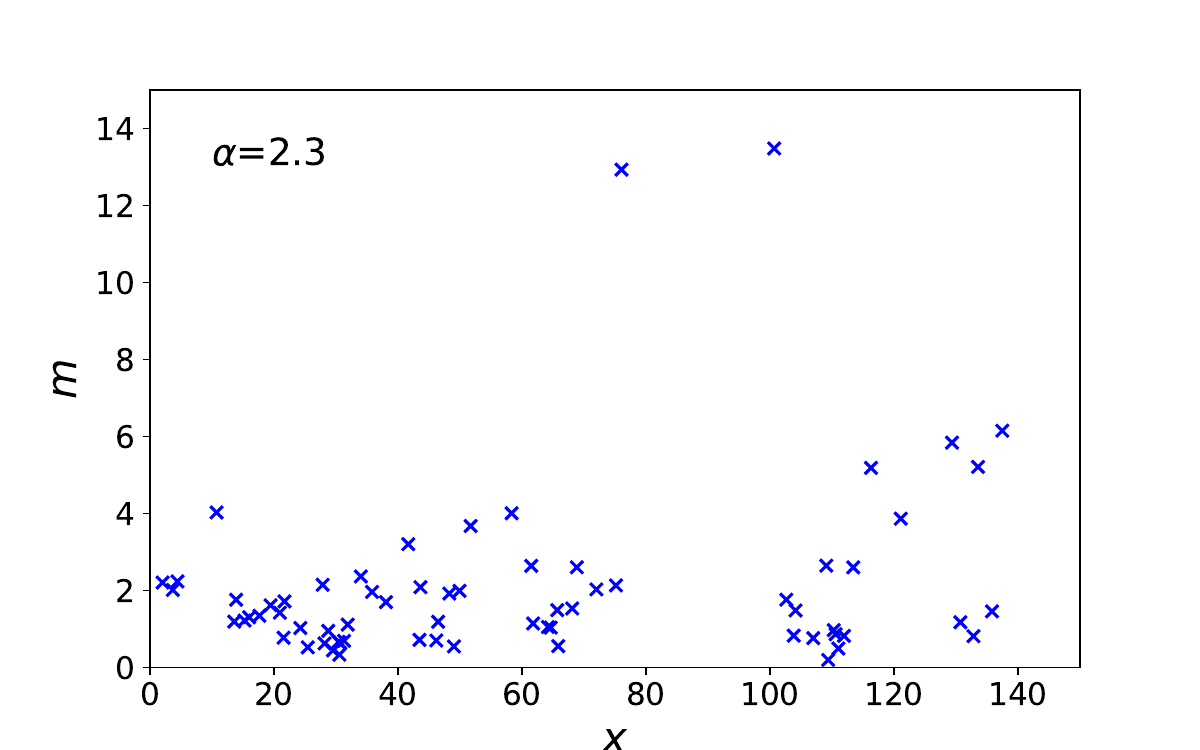}
\\
\includegraphics[height=5.cm,width=0.24\textwidth]{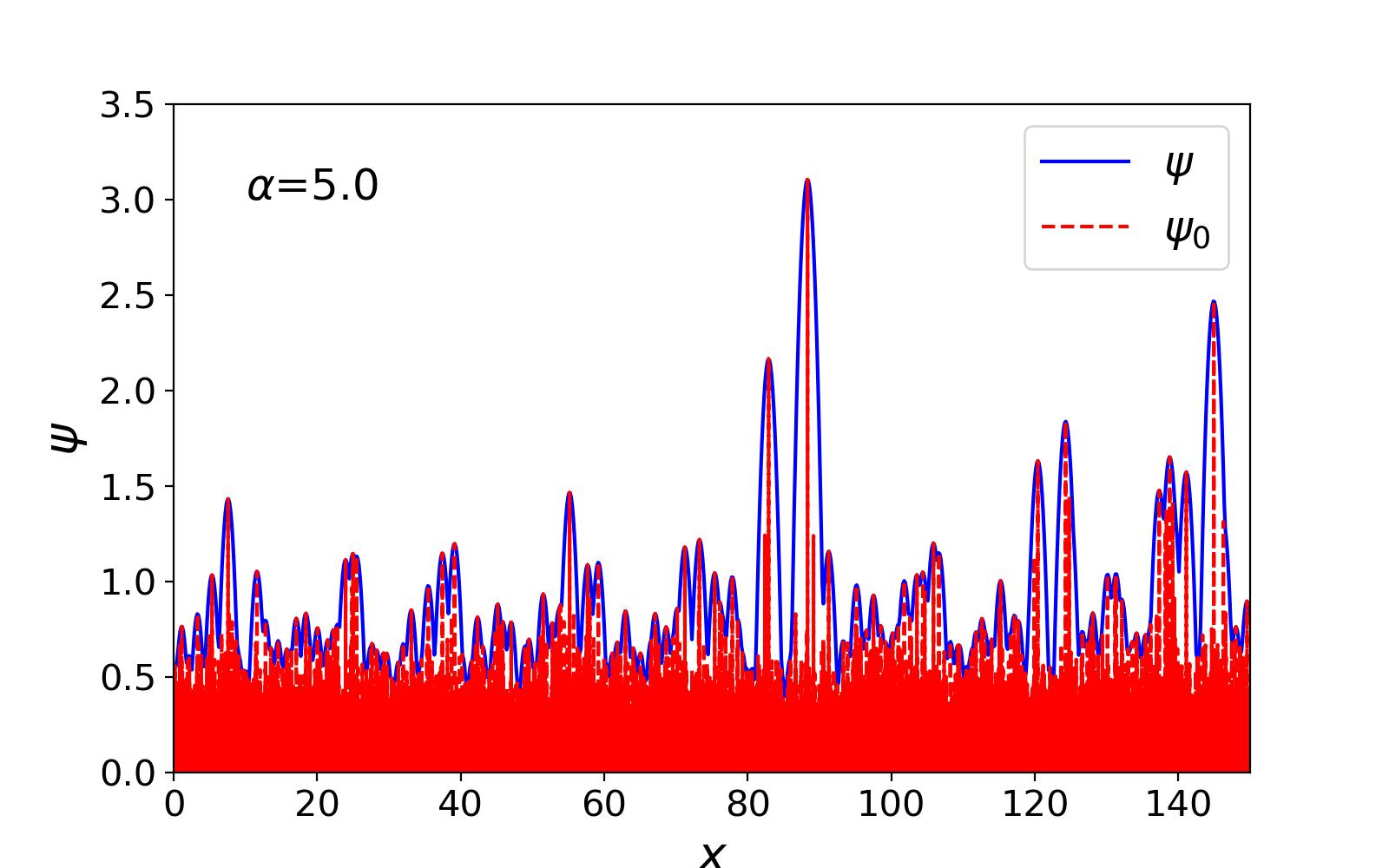}
\includegraphics[height=5.cm,width=0.24\textwidth]{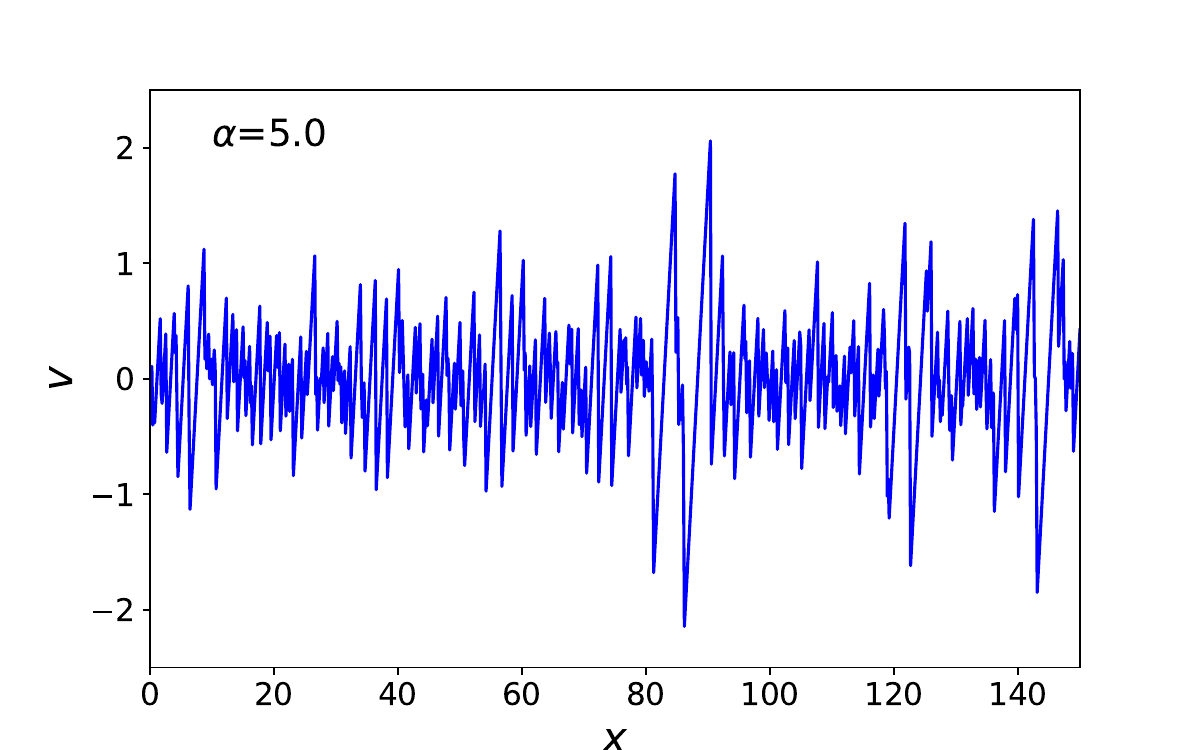}
\includegraphics[height=5.cm,width=0.24\textwidth]{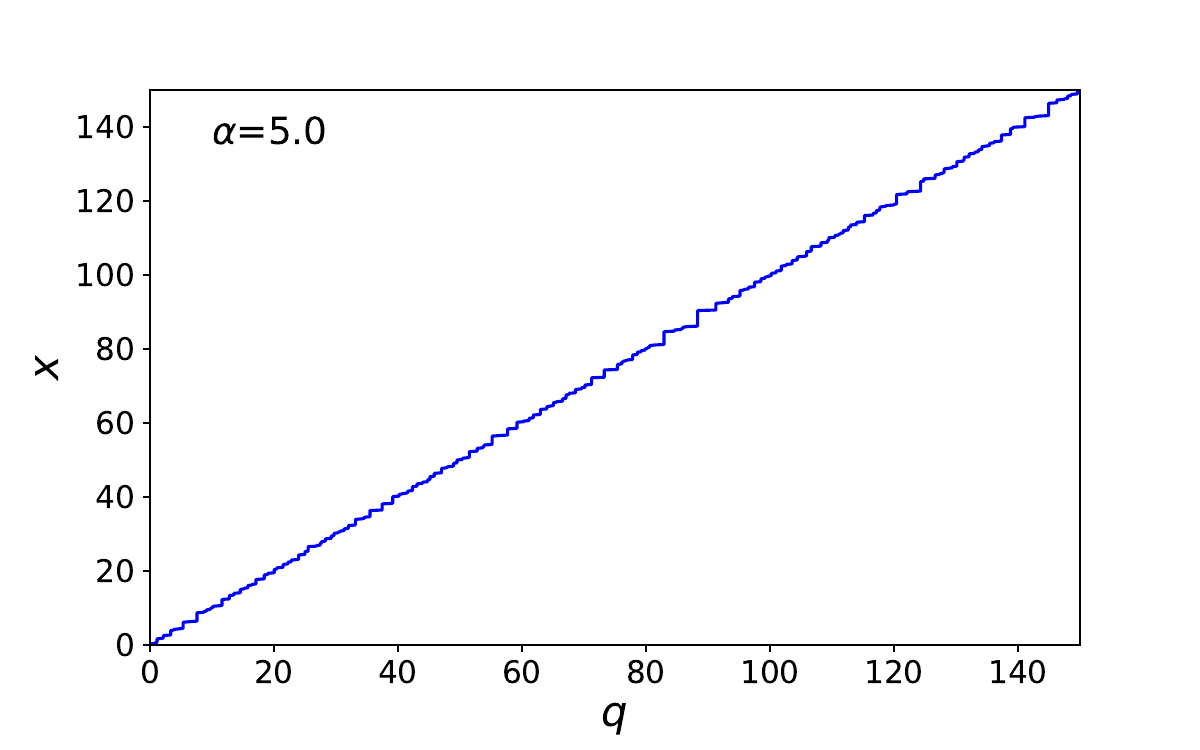}
\includegraphics[height=5.cm,width=0.24\textwidth]{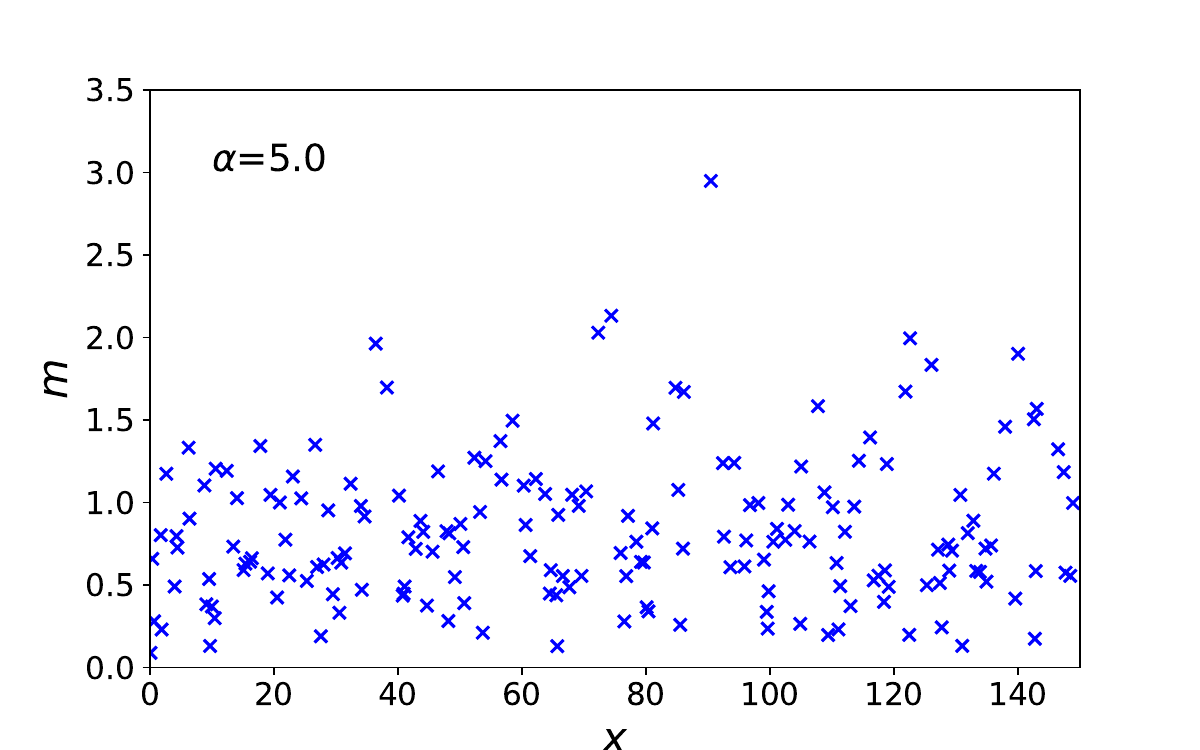}
\caption{A realization of the system for the cases $\alpha=2.3$ (upper row) and $\alpha=5$
(lower row) at time $t=1$.
{\it Left column:} the initial velocity potential $\psi_0(x)$ (red dashed curve) and the final velocity potential
$\psi(x,t)$ (blue solid curve).
{\it Middle left column:} velocity field $v(x,t)$.
{\it Middle right column:} Lagrangian map $x(q,t)$.
{\it Right column:} mass and location of the shocks.
}
\label{fig:realization}
\end{figure}

We show in Fig.~\ref{fig:realization} a numerical realization of the system for the two cases
$\alpha=2.3$ (upper row) and $\alpha=5$ (lower row).
We use the rescaled coordinates (\ref{eq:re-scaling}), which also correspond to the choice $t=1$
and $a=1$.
The initial condition $\psi_0(q)$ is obtained in a straightforward manner from a random generator,
using that the Poisson point process (\ref{eq:lambda-psi}) is homogeneous over the space $(q,y)$ 
with the change of variable $y=\psi_0^{1-\alpha}/(1-\alpha)$. 
The potential $\psi(x,t)$, velocity $v(x,t)$ and inverse Lagrangian map $q(x,t)$ at time $t$
are obtained from the Legendre transform (\ref{eq:H-def}).

The initial velocity potential $\psi_0$ is very singular, as it is defined by a Poisson point process.
For smaller $\alpha$ it shows a heavy power-law tail, which leads to a few very high peaks
(compare the vertical scales between the panels for $\alpha=2.3$ and $\alpha=5$).
The potential $\psi(x,t)$ is made of a collection of downward parabolic arcs, in agreement with
Eq.(\ref{eq:psi-qi}). These arcs peak at the locations of the highest peaks of $\psi_0$.
For lower $\alpha$, because of the greater height of the rare peaks, the latter have a wider region
of influence (i.e., the spatial extent of their parabolic arc is greater).
Therefore, in the nonlinearly evolved field the spatial correlation extends to greater distance
for lower $\alpha$ (even though the initial field $\psi_0$ has no spatial correlations at all).
For large $\alpha$ we can see that the potentials $\psi_0$ and $\psi$ fluctuate within
an increasingly narrow range around $\psi = 1$, as noticed in Sec.~\ref{sec:alpha-infty-rescale}.

The velocity field $v(x,t)$ shows the typical ramps $x/t$ of the Burgers dynamics, between shocks
that are associated with negative jumps.
Following the behavior of the potential $\psi$, for lower $\alpha$ the velocity jumps
and the width of the linear ramps are greater.

The Lagrangian map, $q \mapsto x$, follows the homogeneous medium relation $x=q$ on large
scales. On smaller scales, it shows fluctuations associated with the formation of voids and shocks.
Thus, it is made of a series of horizontal and vertical steps.
Horizontal steps correspond to shocks, where a finite interval $\Delta q$ maps to a unique shock
location $x$, which contains a finite mass (the density field is thus made of a series of Dirac peaks).
Vertical steps correspond to voids, of size $\Delta x$, which are empty of matter and thus correspond
to $\Delta q=0$.
Again, for lower $\alpha$ the steps (and the fluctuations from the mean $\langle x \rangle_q = q$)
are greater.
The inverse Lagrangian map, $x \mapsto q$, can be directly read from the same figure and
it also shows a series of horizontal and vertical steps around the mean $\langle q \rangle_x = x$.

The location $x_s$ and mass $m_s=\Delta q$ of the shocks (shown by the crosses in the right column 
panels) follow the same behaviors. For smaller $\alpha$, the shocks are less numerous and more widely 
separated by the larger voids and their mass distribution shows a heavier power-law tail.

We derive in the next Sections the analytical expressions of the probability distributions
of the displacement, velocity and density fields associated with these dynamics, as well as the
distributions of the voids and shocks.
In agreement with Fig.~\ref{fig:realization}, we find that the system at any time $t>0$ is made
of a series of shocks and voids, with distributions that show power-law tails that decay more
slowly for smaller $\alpha$.

\section{One-point Eulerian distributions}
\label{sec:one-point-Eulerian}

\begin{figure}
\centering
\includegraphics[height=6cm,width=0.33\textwidth]{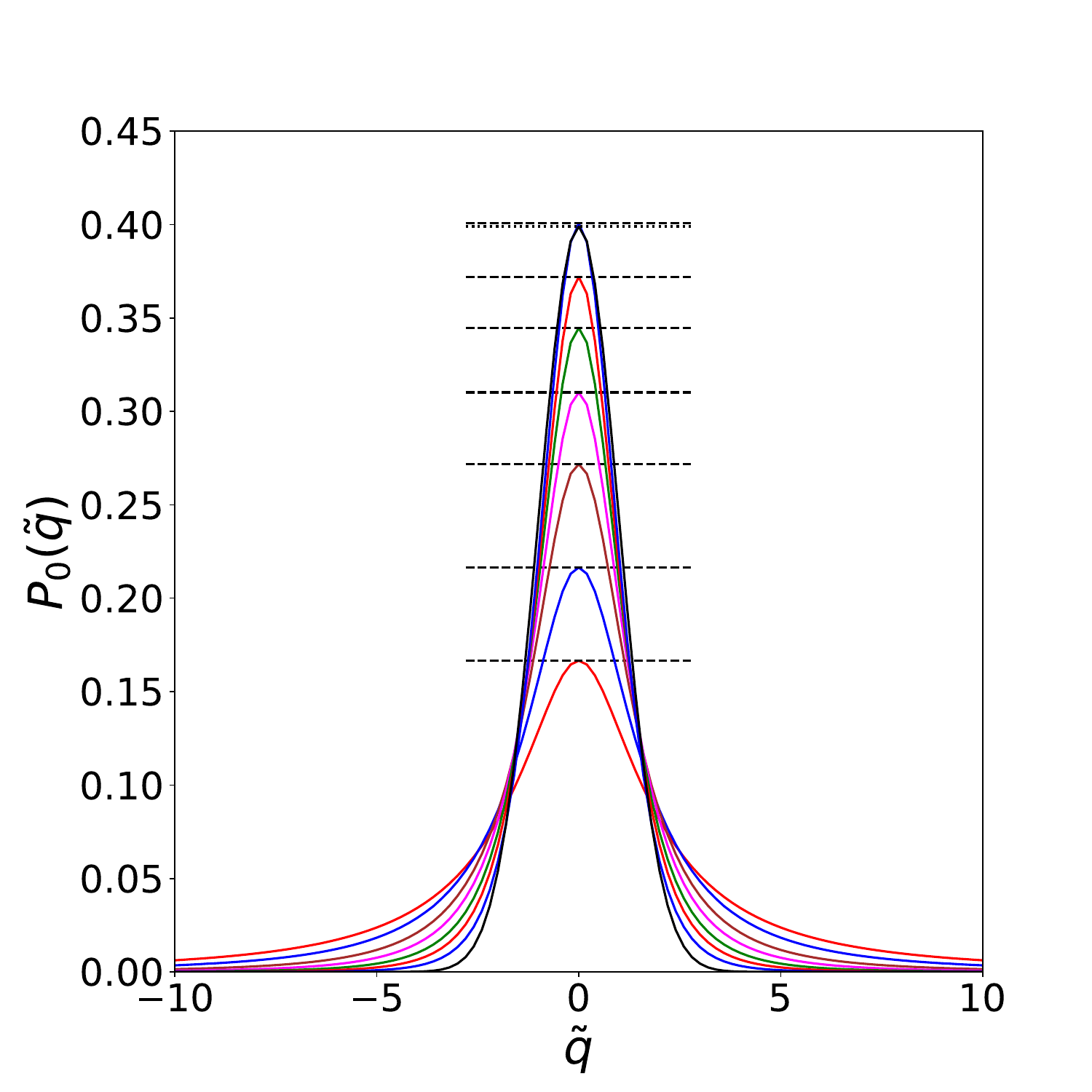}
\includegraphics[height=6cm,width=0.33\textwidth]{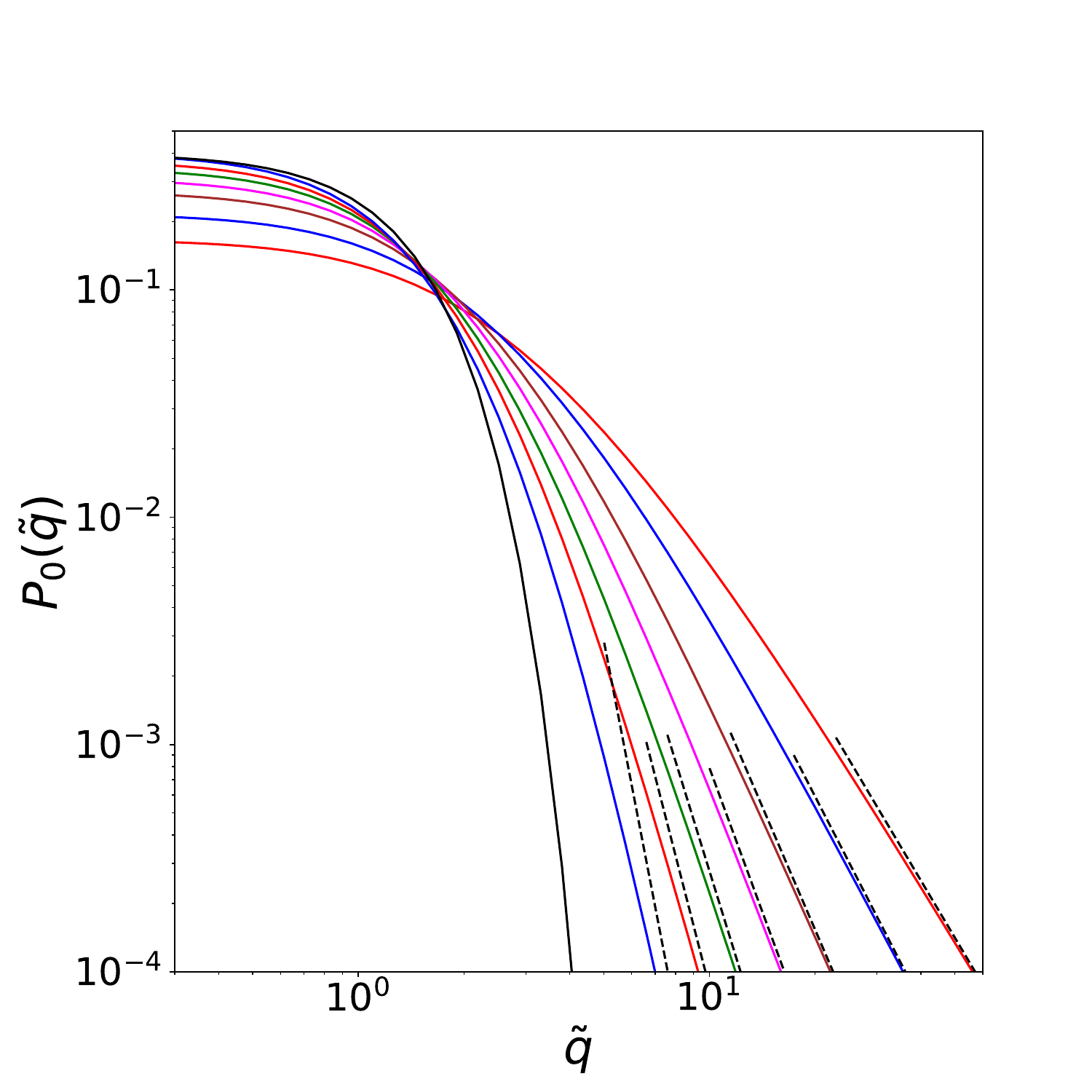}
\includegraphics[height=6cm,width=0.32\textwidth]{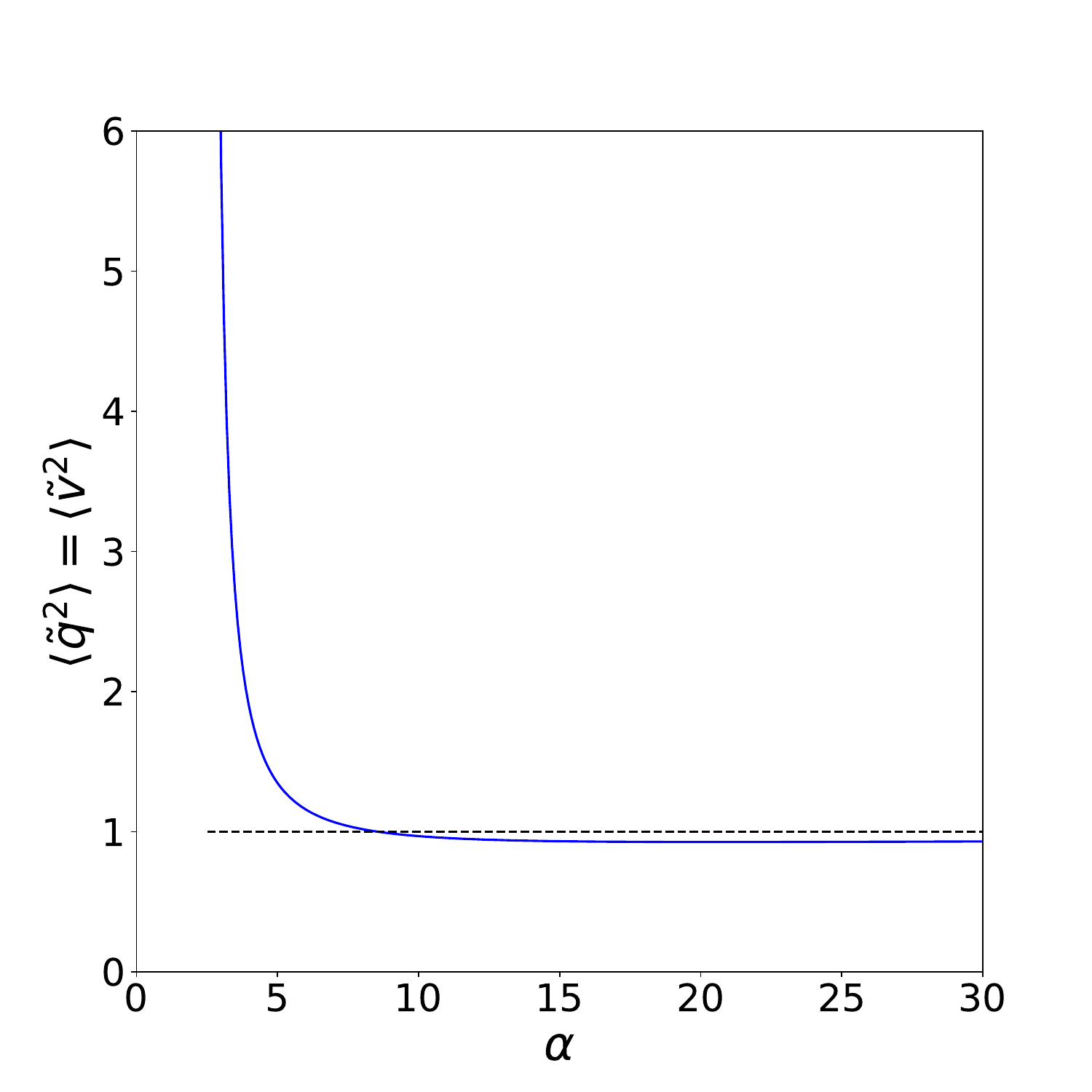}
\caption{
{\it Left panels:} one-point probability distribution $P_0(\tilde q)=P_0(\tilde v)$ of the Lagrangian coordinate
$\tilde q$, or of the velocity $\tilde v$, from Eq.(\ref{eq:P_0-q}). 
We use the rescaled coordinate $\tilde q$ as in (\ref{eq:tilde-def}), to illustrate the convergence to
Eq.(\ref{eq:P0-q-alpha-inf}) in the limit $\alpha\to\infty$.
We display our results on linear scales (left panel) and logarithmic scales (middle panel), for the
cases $\alpha=2.3, 2.5, 2.8, 3.1, 3.5, 4, 5, \infty$. 
The horizontal dashed lines in the left panel are the values $P_0(\tilde q=0)$ of Eq.(\ref{eq:P0_q=0}).
The slope at large $\tilde q$ in the middle panel becomes steeper for larger $\alpha$.
The dashed lines in the middle panel are the asymptotic power-laws (\ref{eq:P0-large-q}).
{\it Right panel:} variance $\langle \tilde q^2 \rangle = \langle \tilde v^2 \rangle$ of the Lagrangian
displacement and of the velocity. 
The horizontal dashed line is the limit $\alpha\to\infty$, from Eq.(\ref{eq:P0-q-alpha-inf}).
}
\label{fig:P0_q}
\end{figure}

We consider in this Section the one-point probability distribution $P_x(q)$ of the Lagrangian
point $q$ for a given Eulerian point $x$. This also gives $P_x(v)$ with $v=x-q$.
Thanks to the statistical invariance by translations, we can focus on $x=0$
as $P_x(q)$ is a function of $q-x$ only and we have $P_x(q) = P_0(q-x)$.
The probability $P_0(<c)$ that the first-contact parabola ${\cal P}_{0,c\star}$ of Eq.(\ref{eq:paraboladef})
has its minimum $c_\star$ lower than $c$ is also the probability that the domain above ${\cal P}_{0,c}$ 
is empty.
From the Poisson intensity (\ref{eq:Lambda-Psi0}), we obtain
\be
P_0(<c) = e^{-\int_{-\infty}^{\infty} dq \int_{c+q^2/2}^{\infty} d\psi \, \psi^{-\alpha} } , \;\;
P_0(c) = \int_{-\infty}^{\infty} dq \, (c+q^2/2)^{-\alpha} 
e^{-\int_{-\infty}^{\infty} dq \int_{c+q^2/2}^{\infty} d\psi \, \psi^{-\alpha} } .
\label{eq:P<c-Pc}
\ee
We clearly have $P_0(<c) \to 1$ for $c \to \infty$ and $P_0(<c) \to 0$ for $c \to 0$, from
Eq.(\ref{eq:Lambda-Parabola}).
On the other hand, the probability $P_0(q,c) dq dc$ that the first contact-point is along the parabola
${\cal P}_{0,c}$ at abscissa $q$ is given by the product of the probabilities that there is one point
at $(q,\psi=c+q^2/2)$ in the range $[dq \times dc]$ and that there are no points above
${\cal P}_{0,c}$,
\be
P_0(q,c) dq dc = P(N_{dq \times dc}=1) \times P_0(<c) = (c+q^2/2)^{-\alpha} dq dc  \times
e^{-\int_{-\infty}^{\infty} dq \int_{c+q^2/2}^{\infty} d\psi \, \psi^{-\alpha} } .
\label{eq:P(q,c)}
\ee
The comparison with Eq.(\ref{eq:P<c-Pc}) shows that this probability is well normalised to unity,
$\int dq P_0(q,c) dq= P_0(c)$ and $\int dc P_0(c) = 1$.
Performing the integrations in the argument of the exponential, we obtain 
as in \cite{Gueudre2014} for the distribution
$P_0(q)$ of the Lagrangian coordinate $q$ associated with the Eulerian position $x=0$,
\be
P_0(q) = \int_0^{\infty} dc \, (c+q^2/2)^{-\alpha} \, e^{- \Lambda_\alpha c^{3/2-\alpha}} .
\label{eq:P_0-q}
\ee
This distribution is even in $q$, it has a finite value at the origin,
\be
P_0(q=0) = \frac{1}{\alpha-1} \Lambda_{\alpha}^{(2-2 \alpha)/(2\alpha-3)} \, 
\Gamma \left( 2 + \frac{1}{2\alpha-3} \right)  ,
\label{eq:P0_q=0}
\ee
and a power-law tail at large distance
\be
|q| \gg 1 : \;\;\; P_0(q) \simeq \frac{1}{\alpha-1} \left( \frac{q^2}{2} \right)^{1-\alpha} .
\label{eq:P0-large-q}
\ee
This power-law tail always decreases faster than $1/|q|$, as $\alpha > 3/2$, and it becomes steeper
for larger $\alpha$.
Since for $x=0$ we have $v=-q$, the velocity probability distribution $P_0(v)=P_0(q=-v)$ is given by the
same expressions.

The moments $\langle q^{2n} \rangle$ are only finite for $n<\alpha-3/2$,
\be
n < \alpha -\frac{3}{2} : \;\;\;
\langle q^{2n} \rangle = \langle v^{2n} \rangle =
\frac{\Gamma(n+1/2) \Gamma[1-2n/(2\alpha-3)] \Gamma(\alpha-n-1/2)}
{\sqrt{\pi}\Gamma(\alpha-1/2)} 2^n \Lambda_{\alpha}^{2n/(2\alpha-3)} .
\label{eq:q2-variance}
\ee
In particular, the variance $\langle q^2 \rangle = \langle v^2 \rangle$ is only finite for $\alpha > 5/2$.

We note that in the case $\alpha=2$, the integral (\ref{eq:P_0-q}) reads
\be
\alpha= 2 : \;\;\; L(t) \propto t , \;\;\;
P_0(q) = \frac{2}{q^3} \left[ q - 2 \pi {\rm Ci}(2\pi/q) \sin(2\pi/q) - \pi \cos(2\pi/q) 
( \pi - 2 {\rm Si}(2\pi/q) ) \right] ,
\ee
where ${\rm Ci}$ and ${\rm Si}$ are the cosine and sine integrals,
whereas in the case $\alpha=5/2$ we obtain
\be
\alpha= 5/2 : L(t) \propto t^{3/4} , \;\;\; 
P_0(q) = \frac{\sqrt{2\pi}}{q^3} U(3/2,0,8\sqrt{2}/(3 q^2))  ,
\ee
where $U$ is Kummer's confluent hypergeometric function.

We show the curves $P_0(\tilde q)$ in Fig.~\ref{fig:P0_q} for the cases 
$\alpha=2.3, 2.5, 2.8, 3.1, 3.5, 4, 5, \infty$, using the rescaled coordinate $\tilde q$ of Eq.(\ref{eq:tilde-def})
to illustrate the convergence to the limit $\alpha\to\infty$.
We shall consider these values of $\alpha$ for all figures in this article, as they span all regimes
associated with the initial conditions (\ref{eq:lambda-psi}).
We collect all the results obtained in the limit $\alpha\to\infty$ in Section~\ref{sec:alpha-infty}
below.

We can check that our numerical computation of Eq.(\ref{eq:P_0-q}) agrees with the analytical
value at the origin (\ref{eq:P0_q=0}) and the asymptotic power-law tail (\ref{eq:P0-large-q}).
While the convergence to $\alpha\to\infty$ appears monotonic at large $\tilde q$, where the power-law tail
becomes steeper with an exponent that goes to infinity as the probability distribution converges
to the Gaussian (\ref{eq:P0-q-alpha-inf}), this is not the case at the origin.
The value $P_0(\tilde q=0)$, in terms of the rescaled coordinate $\tilde q$, first increases
with $\alpha$ to reach a maximum at $\alpha \simeq 11.5$ and then slightly decreases to reach the
asymptotic value $1/\sqrt{2\pi}$.
Nevertheless, the use of the rescaled coordinate $\tilde q$ permits a meaningful comparison between
the different values of $\alpha$.

The variance $\langle \tilde q^2 \rangle = \langle \tilde v^2 \rangle$ becomes close to its limit
(\ref{eq:P0-q-alpha-inf}) for $\alpha\to\infty$ as soon as $\alpha \gtrsim 5$.
It diverges for $\alpha \leq 5/2$.

\section{Two-point Eulerian distributions}
\label{sec:two-point}

\subsection{Two first-contact parabolas}

We consider in this Section two-point Eulerian probability distributions, such as the probability
$P_{x_1,x_2}(q_1,q_2)$ to have the two Lagrangian coordinates $q_1$ and $q_2$ at the
Eulerian points $x_1$ and $x_2$.
As in Sec.~\ref{sec:one-point-Eulerian}, we first consider the probability $P_{x_1,x_2}(<c_1,<c_2)$
that the two first-contacts parabolas ${\cal P}_{x_1,c_{\star 1}}$ and ${\cal P}_{x_2,c_{\star 2}}$
are below those of height $c_1$ and $c_2$.
For the Poisson point process (\ref{eq:Lambda-Psi0}) this is
\be
x_1 < x_2 : \;\;\; P_{x_1,x_2}(<c_1,<c_2) = e^{-\int_{-\infty}^{q_\star} dq 
\int_{c_1+(q-x_1)^2/2}^{\infty} d\psi \, \psi^{-\alpha} - \int_{q_\star}^{\infty} dq 
\int_{c_2+(q-x_2)^2/2}^{\infty} d\psi \, \psi^{-\alpha} } ,
\label{eq:P<c1<c2}
\ee
which is the probability that there are no points above the two parabolas ${\cal P}_{x_1,c_1}$ and 
${\cal P}_{x_2,c_2}$.
Here we introduced the point $(q_\star,\psi_\star)$ which is the intersection of these two parabolas,
\be
\psi_\star = c_1 + \frac{(q_\star-x_1)^2}{2} = c_2 + \frac{(q_\star-x_2)^2}{2} , \;\;\;
q_\star = \frac{x_1+x_2}{2} + \frac{c_2-c_1}{x_2-x_1} , \;\;\;
\psi_\star = \frac{(x_2-x_1)^2}{8} + \frac{(c_2-c_1)^2}{2 (x_2-x_1)^2} + \frac{c_2+c_1}{2} .
\ee
To the left of $q_\star$, ${\cal P}_{x_1,c_1}$ is below ${\cal P}_{x_2,c_2}$, whereas to the right of
$q_\star$, ${\cal P}_{x_2,c_2}$ is below ${\cal P}_{x_1,c_1}$, because we take $x_1 < x_2$.
Taking the derivative of the cumulative distribution (\ref{eq:P<c1<c2}) with respect to $c_1$ and $c_2$,
we obtain the probability distribution $P_{x_1,x_2}(c_1,c_2) dc_1 dc_2$,
\bea
x_1 < x_2 : \;\;\; P_{x_1,x_2}(c_1,c_2) & = & \left[ \frac{\psi_\star^{-\alpha}}{x_2-x_1} 
+ \int_{-\infty}^{q_\star} dq_1 ( c_1 + (q_1-x_1)^2/2)^{-\alpha} 
\int_{q_\star}^{\infty} dq_2 ( c_2 + (q_2-x_2)^2/2)^{-\alpha} \right] \nonumber \\
&& \times e^{-\int_{-\infty}^{q_\star} dq 
\int_{c_1+(q-x_1)^2/2}^{\infty} d\psi \, \psi^{-\alpha} - \int_{q_\star}^{\infty} dq 
\int_{c_2+(q-x_2)^2/2}^{\infty} d\psi \, \psi^{-\alpha} } .
\label{eq:Pc1c2}
\eea
The first term in the bracket corresponds to the case where the two parabolas have a common contact point
with the initial potential $\psi_0(q)$, which is their intersection point $(q_\star,\psi_\star)$.
The second term corresponds to the case where the two contact points $q_1$ and $q_2$ are
distinct and somewhere along the parabolas with $q_1<q_\star<q_2$.
Thus, in agreement with \cite{Gueudre2014},
the probability distribution $P_{x_1,x_2}(q_1,c_1,q_2,c_2) dq_1 dc_1 dq_2 dc_2$ reads
\bea
&& x_1 < x_2 : \;\;\; P_{x_1,x_2}(q_1,c_1,q_2,c_2) = \biggl [ \frac{\psi_\star^{-\alpha}}{x_2-x_1} 
\delta_D(q_1-q_\star) \delta_D(q_2-q_\star) + \theta(q_1<q_\star) 
\left( c_1 + (q_1-x_1)^2/2 \right)^{-\alpha} 
\nonumber \\
&& \times \theta(q_2 > q_\star) \left( c_2 + (q_2-x_2)^2/2 \right)^{-\alpha} \biggl ] 
e^{-\int_{-\infty}^{q_\star} dq 
\int_{c_1+(q-x_1)^2/2}^{\infty} d\psi \, \psi^{-\alpha} - \int_{q_\star}^{\infty} dq 
\int_{c_2+(q-x_2)^2/2}^{\infty} d\psi \, \psi^{-\alpha} } ,
\label{eq:Pq1c1q2c2}
\eea
where $\delta_D$ and $\theta$ are the Dirac distribution and Heaviside function with obvious notations.
Making the change of variables
\be
\bar x  = \frac{x_1+x_2}{2} , \;\;\; x = x_2 - x_1 > 0 , \;\;\;  q_1 = \bar x + q'_1 , \;\;\; q_2 = \bar x + q'_2 ,
\label{eq:qp1-qp2-def}
\ee
and changing integration variables from $(c_1,c_2)$ to $(q_\star',\psi_\star)$, we obtain
\be
P_x(q'_1,q'_2) = \! \int_{-\infty}^{\infty} \!\! dq'_\star \int_{\psi_{\min}(q'_\star)}^{\infty} \!\!\! 
d\psi_\star \, \bigg[ \psi_\star^{-\alpha} \delta_D(q'_1-q'_\star) \delta_D(q'_2-q'_\star) 
+ x \theta(q'_1<q'_\star) \theta(q'_2 > q'_\star) \psi_-(q'_1)^{-\alpha} 
\psi_+(q'_2)^{-\alpha} \bigg] e^{-{\cal I}} ,
\label{eq:Pq1q2}
\ee
where we introduced the functions
\bea
&& \psi_-(q') = \psi_\star + \frac{1}{2} \left[ \left( \frac{x}{2} + q' \right)^2 
- \left ( \frac{x}{2} + q'_\star \right)^2 \right] , \;\;\;\;
\psi_+(q') = \psi_\star + \frac{1}{2} \left[ \left( \frac{x}{2} - q' \right)^2 
- \left ( \frac{x}{2} - q'_\star \right)^2 \right] , \nonumber \\
&& \psi_{\min}(q'_\star) = \frac{1}{2} \left( | q'_\star | + \frac{x}{2} \right)^2  , \;\;\;\;
{\cal I}(\psi_\star,q'_\star) = \frac{1}{\alpha-1} \left[ 
\int_{-\infty}^{q'_\star} dq' \psi_-(q')^{1-\alpha} 
+ \int_{q'_\star}^{\infty} dq' \psi_+(q')^{1-\alpha} \right] . \;\;\;
\eea
This explicitly shows that $P_x(q'_1,q'_2)$ only depends on $x=x_2-x_1$, thanks to the statistical
invariance of the system over translations and the centering of the coordinates (\ref{eq:qp1-qp2-def})
around $(x_1+x_2)/2$.
By parity symmetry we also have ${\cal I}(\psi_\star,-q'_\star)={\cal I}(\psi_\star,q'_\star)$.
For $x=0$ the two parabolas coincide and the quantity ${\cal I}$ is equal to the one found
in Eq.(\ref{eq:P_0-q}) for the one-point distribution,
\be
x=0 : \;\;\; {\cal I}(\psi_\star,q'_\star) = \Lambda_{\alpha} 
\left( \psi_\star - q'_\star \, \! ^{2}/2 \right)^{3/2-\alpha} .
\ee

For future convenience, it is useful to define the quantities ${\cal A}_{\nu}(x,q'_\star)$
and ${\cal R}_{\nu}(x) $,
\be
\nu > 3/2 , \;\; x \geq 0 : \;\; {\cal A}_{\nu}(x,q'_\star) = \int_{\psi_{\min}(q'_\star)}^{\infty} d \psi_\star \, 
\psi_\star^{-\nu} e^{- {\cal I}(\psi_\star,q'_\star) } , 
\;\;\; {\cal R}_{\nu}(x) = \int_{-\infty}^{\infty} d q'_\star \, {\cal A}_{\nu}(x,q'_\star) .
\label{eq:A-nu-def}
\ee
In particular, we have for $x=0$
\be
{\cal R}_{\nu}(0) =  \int_{-\infty}^{\infty} d q'_\star \, \int_0^{\infty} dc \, \left( c 
+ \frac{q'_\star \, ^{\!2}}{2} \right)^{-\nu}  e^{-\Lambda_\alpha c^{3/2-\alpha}} 
= \frac{\sqrt{2\pi} \Gamma(\nu-1/2)}{(\alpha-3/2) \Gamma(\nu)} \,
\Gamma\left(\frac{2\nu-3}{2\alpha-3}\right) \Lambda_{\alpha}^{(3/2-\nu)/(\alpha-3/2)} ,
\label{eq:R-nu-0}
\ee
\be
{\cal R}_{\nu}'(0) = - \sqrt{2\pi} \frac{ \Gamma(\alpha+\nu-3/2) 
\Gamma\left(\frac{2\alpha+2\nu-5}{2\alpha-3}\right) } {(\alpha-1) (\nu-1) (\alpha-3/2) \Gamma(\alpha+\nu-2)} 
\Lambda_{\alpha}^{(5-2\alpha-2\nu)/(2\alpha-3)} ,
\label{eq:Rp-nu-0}
\ee
and for large $x$
\be
x \gg 1 : \;\;\; {\cal R}_{\nu}(x) \simeq
\frac{2^{3\nu-3}}{(\nu-1) (2\nu-3)} x^{3-2\nu} .
\label{eq:R-nu-large-x}
\ee

\subsection{Void probabilities}
\label{sec:Void-probabilities}

\subsubsection{Probability of an empty interval}

\begin{figure}
\centering
\includegraphics[height=7cm,width=0.48\textwidth]{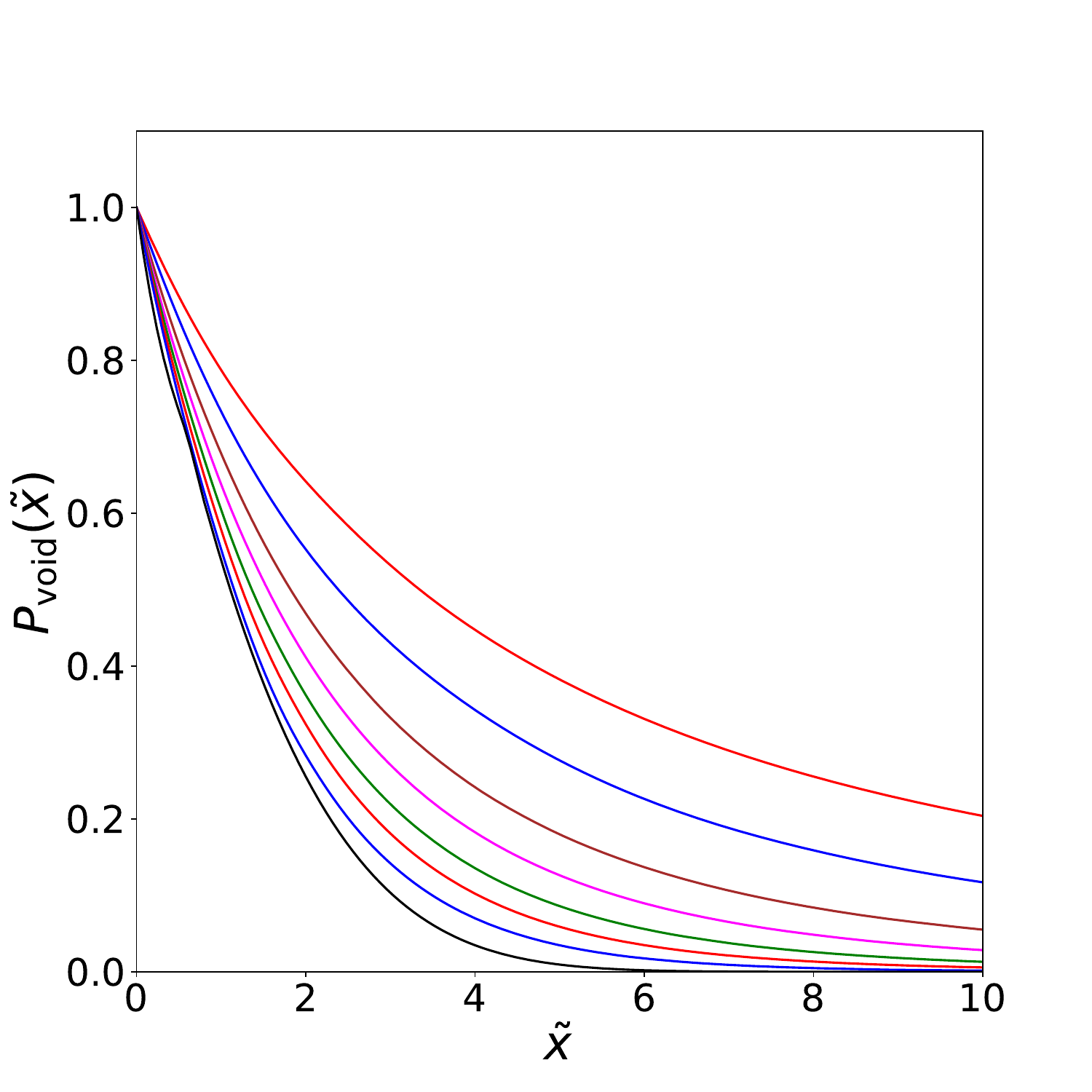}
\includegraphics[height=7cm,width=0.48\textwidth]{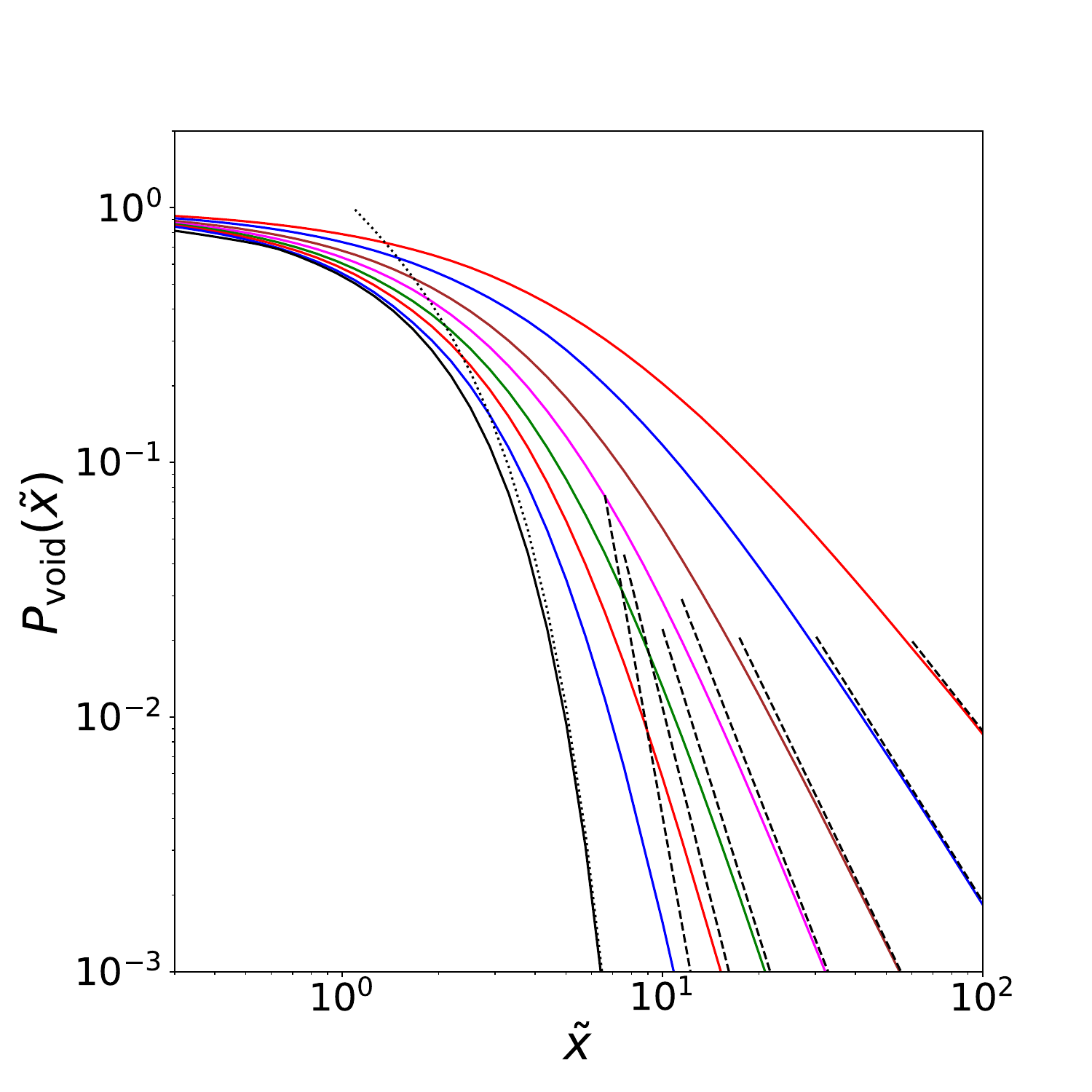}
\caption{
Void probability $P_{\rm void}(\tilde x)$ from Eq.(\ref{eq:Pvoid}), for the cases 
$\alpha=2.3, 2.5, 2.8, 3.1, 3.5, 4, 5, \infty$, as in Fig.~\ref{fig:P0_q}.
Again we use the rescaled coordinate $\tilde x$ to illustrate the convergence to the limit $\alpha\to\infty$.
The slope at large $\tilde x$ in the right panel increases with $\alpha$.
The dashed lines in the right panel are the asymptotic power-laws (\ref{eq:Pvoid-x-0-large-x}),
while the dotted line is the asymptotic result (\ref{eq:Pvoid-alpha-infty-asymp}) for the case
$\alpha=\infty$.
}
\label{fig:Pvoid}
\end{figure}

The overdensity within the Eulerian interval $[x_1,x_2]$ is given by 
\be
\rho_{x_1,x_2} = \frac{q_2-q_1}{x_2-x_1} \geq 0 ,
\ee
where we rescaled the density by the mean density $\bar\rho(t)$ at time $t$, but keep the notation
$\rho$ hereafter.
If the two parabolas have the same contact point, $q_1=q_2=q_\star$, it means that the density vanishes,
$\rho=0$, and the interval $[x_1,x_2]$ is void of matter. Thus, from Eq.(\ref{eq:Pq1q2}) the probability for the 
interval to be empty is
\be
P_{\rm void}(x) = {\cal R}_{\alpha}(x) ,
\label{eq:Pvoid}
\ee
where the function ${\cal R}_{\alpha}(x)$ was defined in Eq.(\ref{eq:A-nu-def}).
From the results (\ref{eq:R-nu-0}) and (\ref{eq:R-nu-large-x}) we obtain
\be
P_{\rm void}(0) = 1  \;\;\; \mbox{and for} \;\;\; x \gg 1 : \;\;\; 
P_{\rm void}(x) \simeq \frac{2^{3(\alpha-1)}}{(\alpha-1) (2\alpha-3)} x^{3-2\alpha} .
\label{eq:Pvoid-x-0-large-x}
\ee
Thus, the void probability goes to unity for $x\to 0$ and decays as a power law for large intervals.
The result $P_{\rm void}(0) = 1$ means that voids cover all the Eulerian space $x$
and matter is concentrated in Dirac density peaks of vanishing width.

We display the void probability $P_{\rm void}(\tilde x)$ in Fig.~\ref{fig:Pvoid}, together with
the asymptotic power-laws (\ref{eq:Pvoid-x-0-large-x}) and the asymptotic result 
(\ref{eq:Pvoid-alpha-infty-asymp}) for $\alpha=\infty$ in the right panel.
Again, the large-distance tails are steeper for larger $\alpha$ and the power-law tail becomes
a Gaussian (with a power-law prefactor) in the limit $\alpha \to \infty$.

\subsubsection{Multiplicity function of voids and distance between shocks}

\begin{figure}
\centering
\includegraphics[height=6.cm,width=0.33\textwidth]{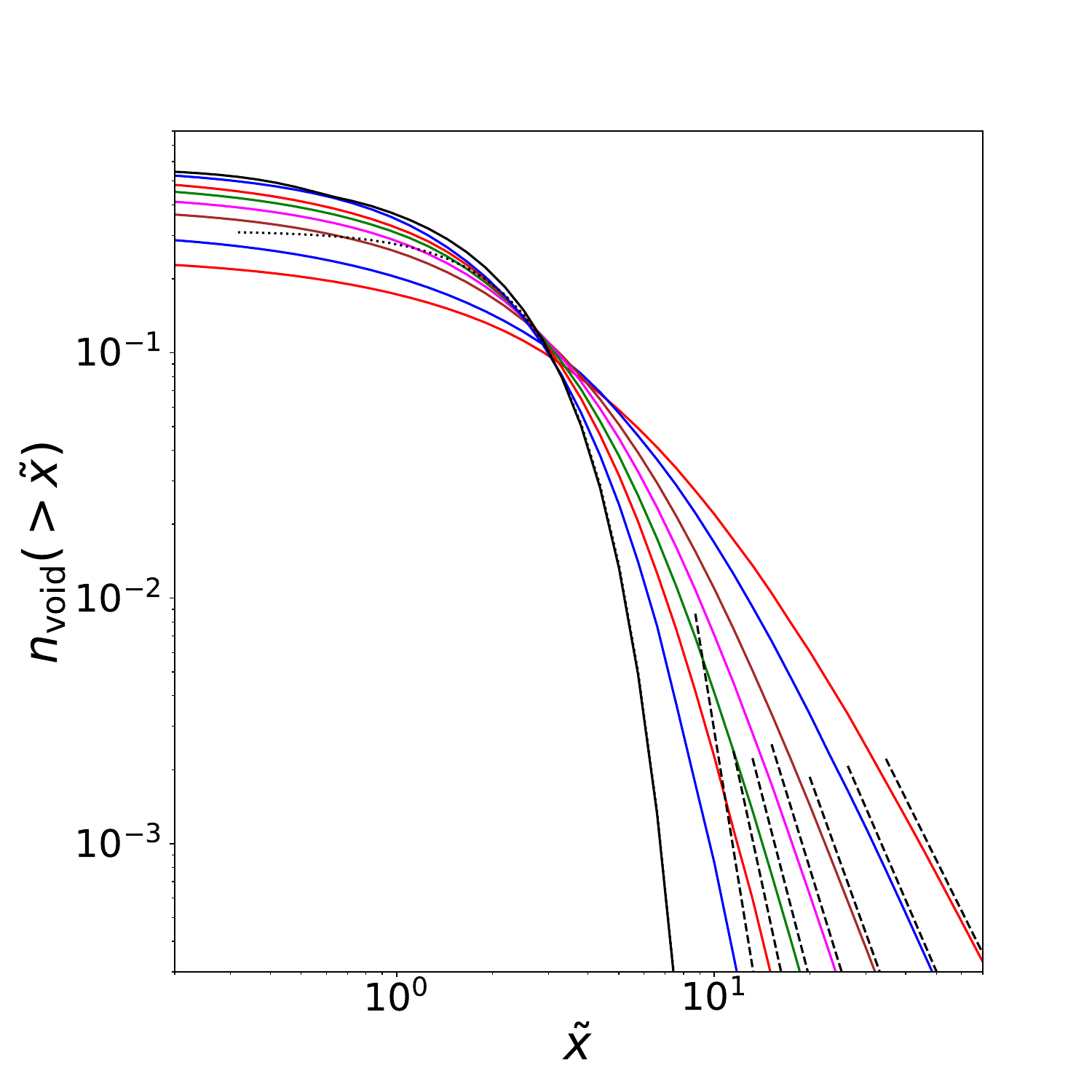}
\includegraphics[height=6.cm,width=0.33\textwidth]{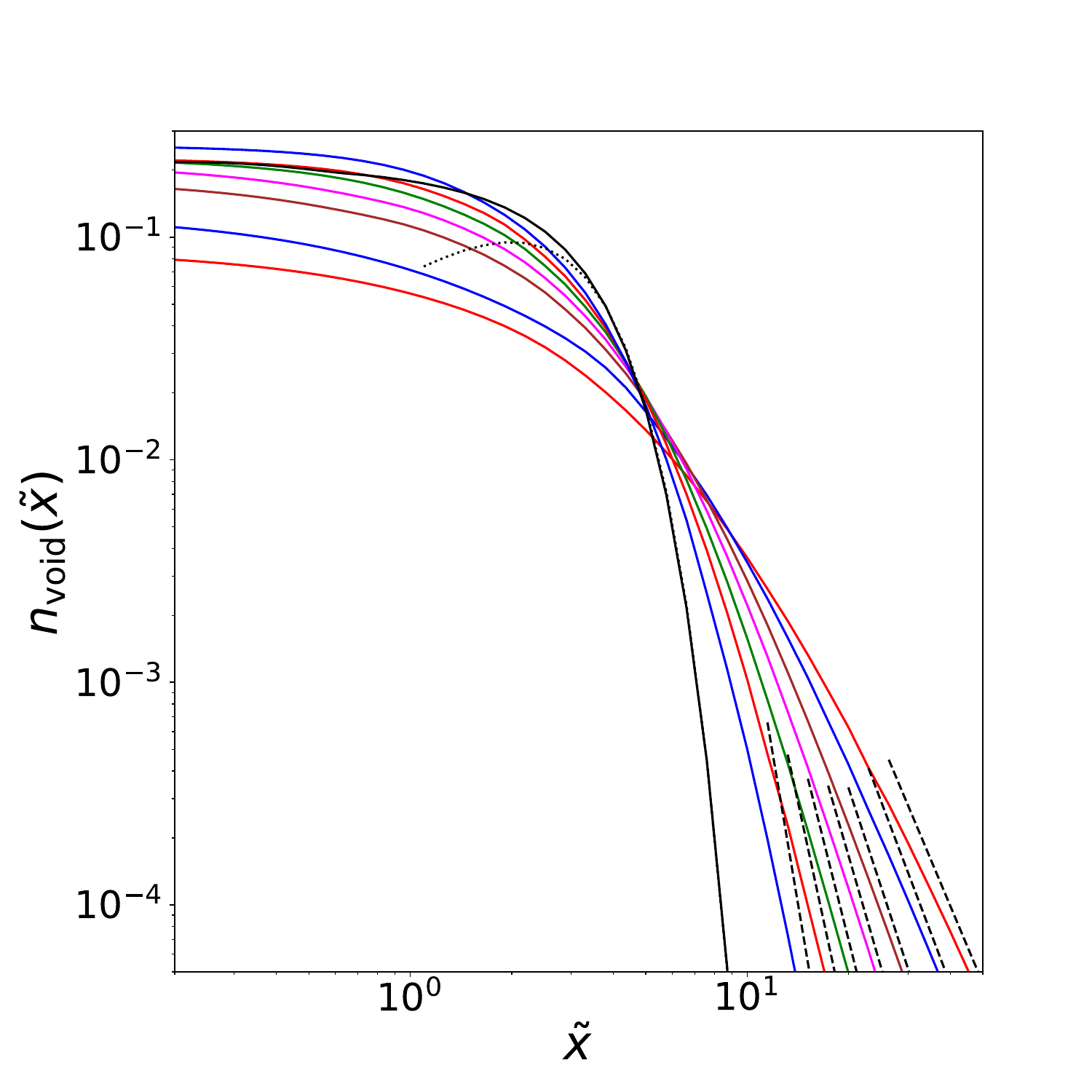}
\includegraphics[height=6.cm,width=0.32\textwidth]{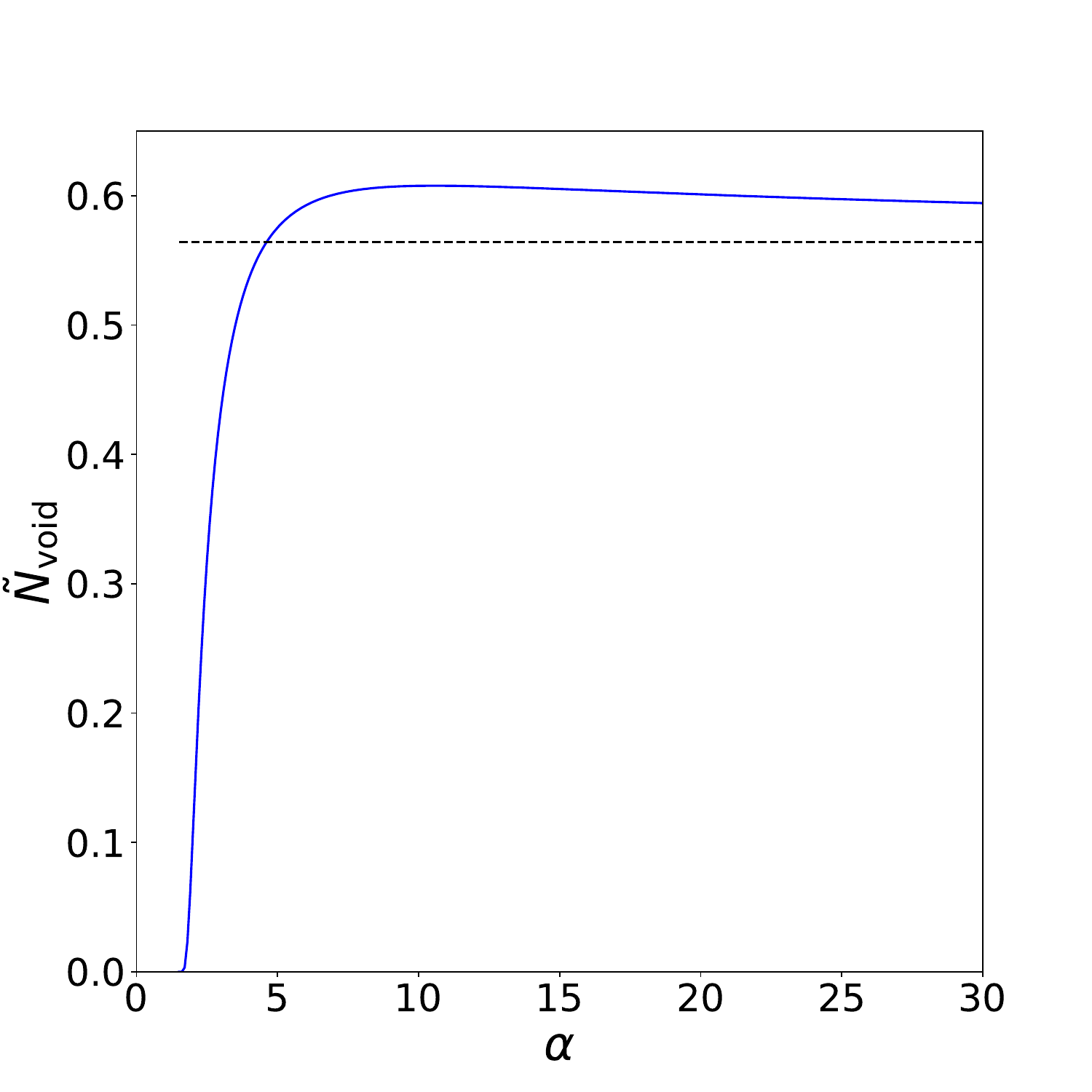}
\caption{
{\it Left panel:} cumulative void multiplicity function $n_{\rm void}(> \tilde x)$ from 
Eq.(\ref{eq:n-void-R-alpha}). 
{\it Middle panel:} void multiplicity function $n_{\rm void}(\tilde x)$ from 
Eq.(\ref{eq:n-void-R-alpha}). 
{\it Right panel:} rescaled number density of voids $\tilde N_{\rm void}$ as a function of $\alpha$.
In the left and middle panels, the dashed lines are the power laws associated with
Eq.(\ref{eq:nvoid-power-law}), whereas the dotted lines are the asymptotic regimes associated with
Eq.(\ref{eq:Pvoid-alpha-infty-asymp}).
In the right panel the horizontal dotted line is the value (\ref{eq:Nvoid-asymp}) in the limit
$\alpha\to\infty$.
}
\label{fig:nvoid}
\end{figure}

Let us define $n_{\rm void}(x) dx$ the number of voids per unit length of size $x$ to $x+dx$.
Then we have 
\be
P_{\rm void}(x) = \int_x^{\infty} dx' n_{\rm void}(x') \, (x'-x) , \;\;\; \mbox{whence} \;\;\;
n_{\rm void}(>x) = - \frac{dP_{\rm void}}{dx} = - {\cal R}'_{\alpha}(x) , \;\;
n_{\rm void}(x) = \frac{d^2P_{\rm void}}{dx^2} = {\cal R}''_{\alpha}(x) .
\label{eq:n-void-R-alpha}
\ee
The results (\ref{eq:R-nu-large-x}) and $P_{\rm void}(0)=1$ give
\be
x \gg 1 : \;\; n_{\rm void}(x) \simeq 2^{3\alpha-2} x^{1-2\alpha} , \;\;\;
\mbox{and} \;\; \int_0^{\infty} dx \, n_{\rm void}(x) \, x = 1 ,
\label{eq:nvoid-power-law}
\ee
which again means that voids cover all the Eulerian space $x$.
From Eq.(\ref{eq:Rp-nu-0}) we obtain the number of voids $N_{\rm void}$ per unit length,
\be
N_{\rm void} = n_{\rm void}(>0) = - {\cal R}'_{\alpha}(0) = \sqrt{2\pi} \frac{ \Gamma(2\alpha-3/2) 
\Gamma\left( \frac{4\alpha-5}{2\alpha-3} \right) } { (\alpha-1)^2 (\alpha-3/2) \Gamma(2\alpha-2) }
\Lambda_{\alpha}^{(5-4\alpha)/(2\alpha-3)} .
\label{eq:Nvoid}
\ee
It shows the asymptotic behaviors
\be
\alpha \to (3/2)^+ : \;\; N_{\rm void} \sim e^{-[2+5 \ln(2)] /[4 (\alpha-3/2)]} \to 0 , \;\;\;
\mbox{and for} \;\;\; \alpha \to \infty : \;\; 
\tilde N_{\rm void} = \frac{N_{\rm void}}{\sqrt{\alpha}} \to \frac{1}{\sqrt{\pi}} .
\label{eq:Nvoid-asymp}
\ee
In the limit $\alpha \to \infty$ we introduced the rescaled number density of voids $\tilde N_{\rm void}$
as in Eq.(\ref{eq:tilde-def}).
This also gives for the void probability the small-scale behavior
\be
x \to 0 : \;\; P_{\rm void}(x) = 1 - N_{\rm void} x + \dots 
\label{eq:Pvoid-x-0}
\ee
We can define the mean void size by
\be
\langle x \rangle_{\rm void} = \frac{ \int_0^{\infty} dx \, n_{\rm void}(x) \, x }{ \int_0^{\infty} dx \, n_{\rm void}(x)}
= \frac{1}{N_{\rm void}} .
\ee
It displays the asymptotic regimes
\be
\alpha \to 3/2 : \;\; \langle x \rangle_{\rm void} \to \infty , \;\;\;
\alpha \to \infty : \;\;  \langle \tilde x \rangle_{\rm void} \to \sqrt{\pi} .
\ee
Thus, we find that for $\alpha \to 3/2$ the mean size of the voids diverges. This agrees with the increase
of typical displacements for lower $\alpha$ already noticed in Section~\ref{sec:one-point-Eulerian} ,
where we found that the variances $\langle q^2 \rangle$ and $\langle v^2 \rangle$ are actually 
infinite for $\alpha < 5/2$. 
We also recover a finite limit in the rescaled coordinates (\ref{eq:tilde-def}) in the limit
$\alpha \to \infty$.
Because the system is made of a series of shocks separated by voids, the void multiplicity function
$n_{\rm void}(x)$ also provides the probability distribution $P(x_s)$ of the distance $x_s$ between
adjacent shocks,
\be
P(x_s) = \frac{n_{\rm void}(x_s)}{N_{\rm void}} = - \frac{{\cal R}''_{\alpha}(x_s)}{{\cal R}'_{\alpha}(0)} .
\label{eq:P-xs}
\ee

We display the void multiplicity functions $n_{\rm void}(>\tilde x)$ and $n_{\rm void}(\tilde x)$
in the left and middle panels in Fig.~\ref{fig:nvoid}, and the mean number of voids per unit length
in the right panel.
Again, we find a monotonic steepening of the far power-law tails and a non-monotonic behavior
at the origin, as also shown in the right panel by $\tilde N_{\rm void}$.
The void multiplicity function $n_{\rm void}(x)$ is finite and nonzero at $x=0$.
Therefore, the number of voids and their mean size are dominated by the typical voids of size
$\tilde x \sim 1$, except for $\alpha \to 3/2$.
In the limit $\alpha \to 3/2$, $N_{\rm void} \to 0$ and $\langle x \rangle_{\rm void} \to \infty$ as
voids become infinitely large and the system is actually ill-defined for $\alpha \leq 3/2$,
as already seen in Section~\ref{sec:Initial-condition}.

\subsection{Two-point velocity correlation and energy power spectrum}
\label{sec:energy-spectrum}

\begin{figure}
\centering
\includegraphics[height=7cm,width=0.48\textwidth]{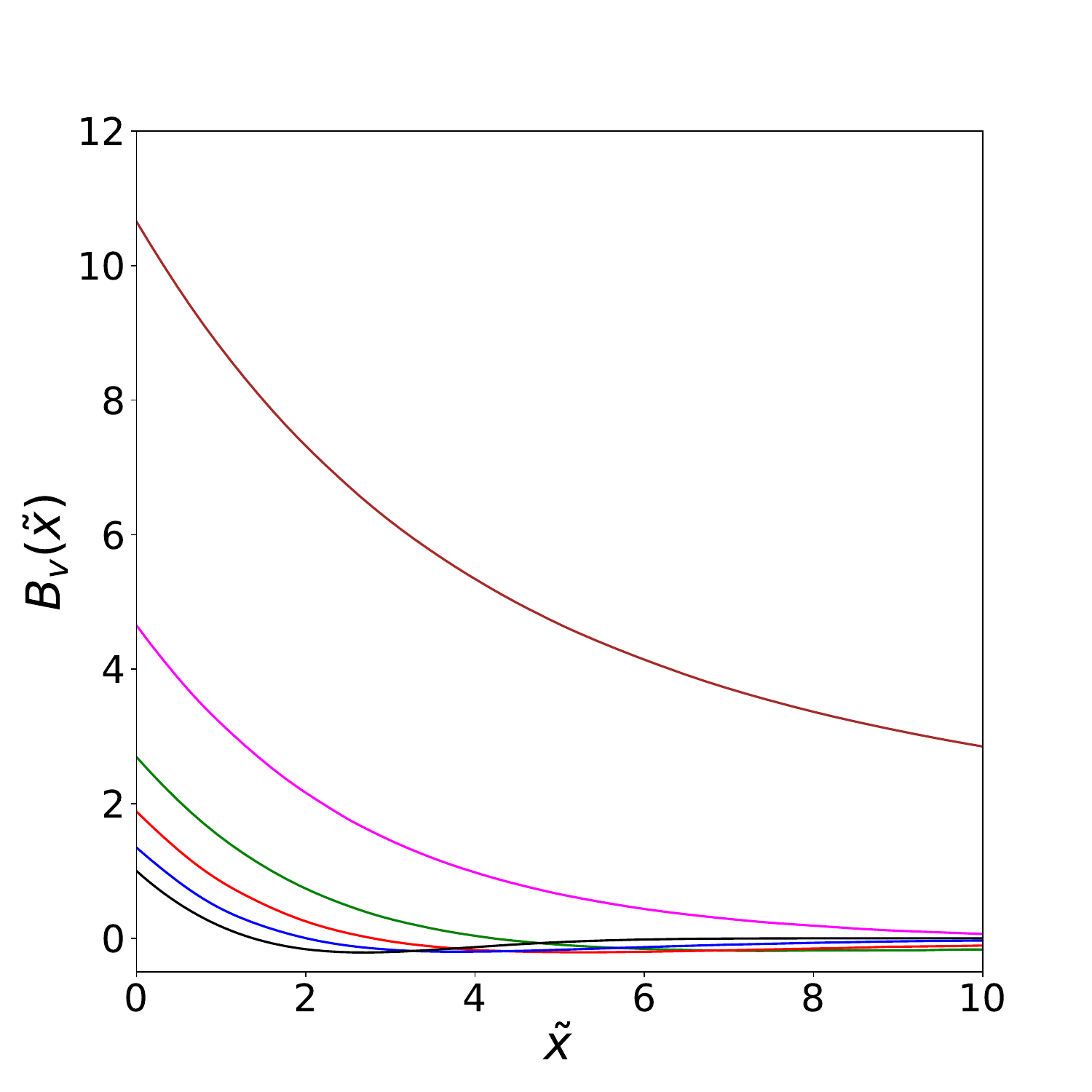}
\includegraphics[height=7cm,width=0.48\textwidth]{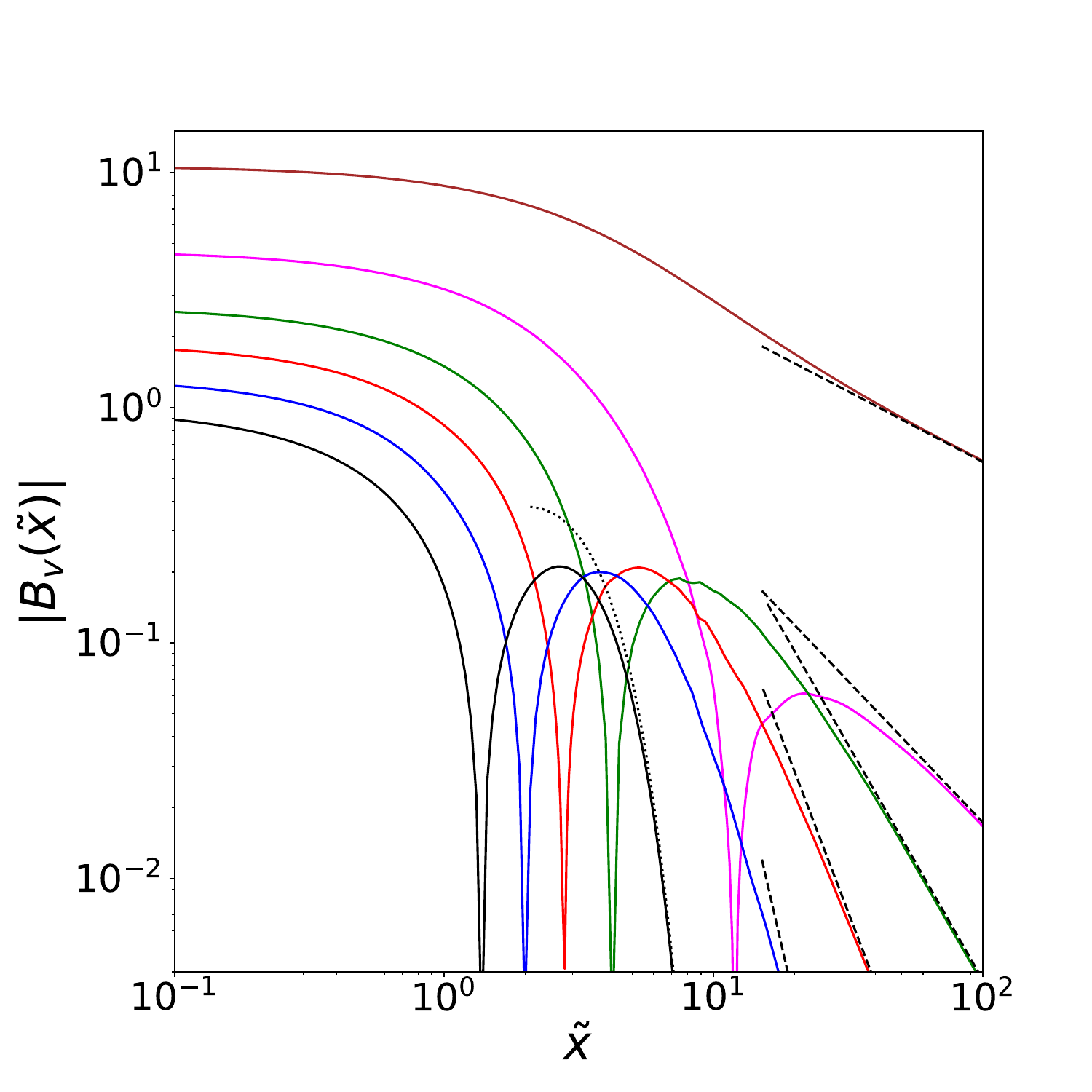}
\caption{
Velocity correlation $B_{\tilde v}(\tilde x)$ for  the cases $\alpha=2.8, 3.1, 3.5, 4, 5, \infty$.
In the right panel the dashed lines are the asymptotic power laws (\ref{eq:Bv-asymp})
while the dotted line is the asymptotic result (\ref{eq:Bv-asymp-alpha-infty})
for $\alpha=\infty$.
}
\label{fig:Bv}
\end{figure}

\begin{figure}
\centering
\includegraphics[height=7cm,width=0.48\textwidth]{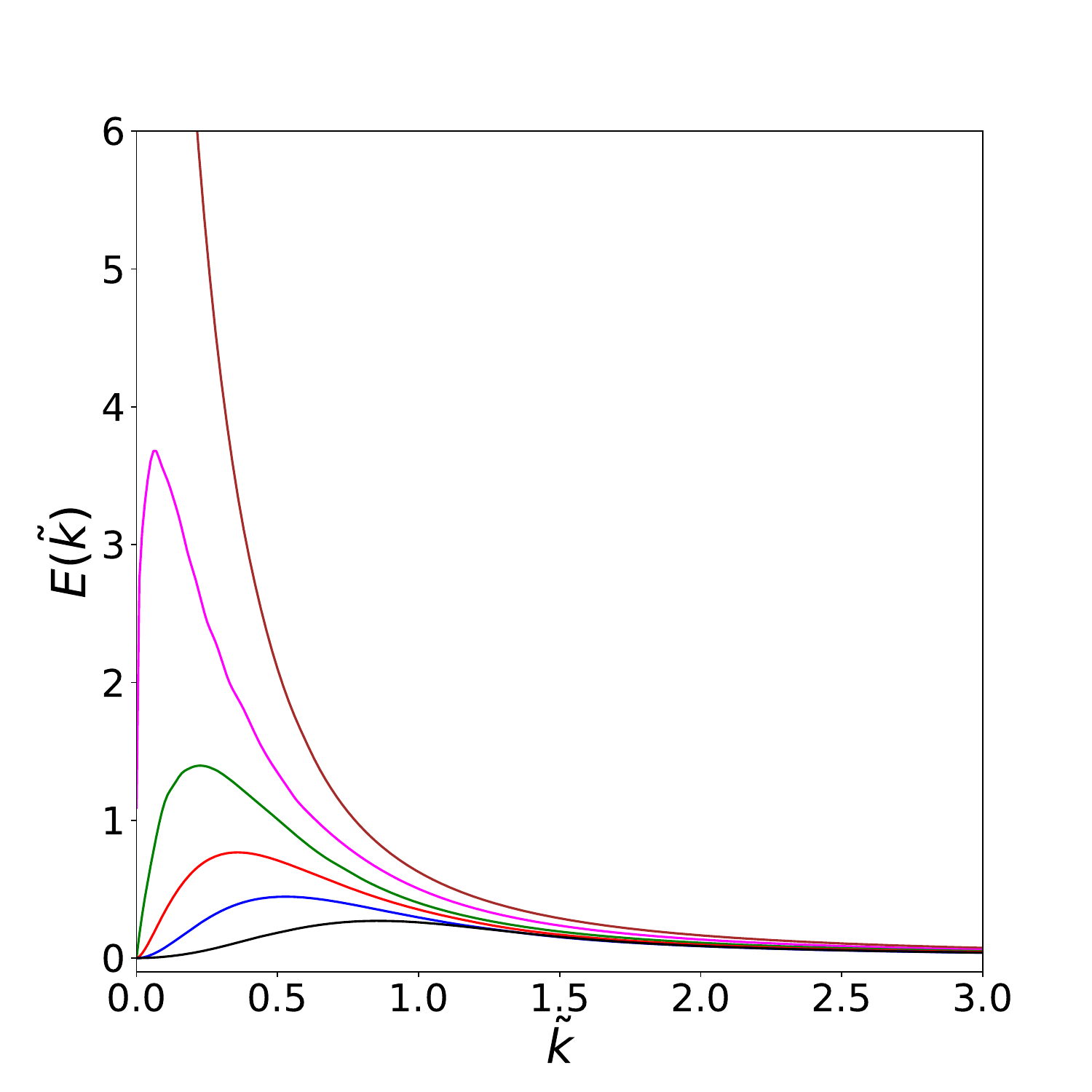}
\includegraphics[height=7cm,width=0.48\textwidth]{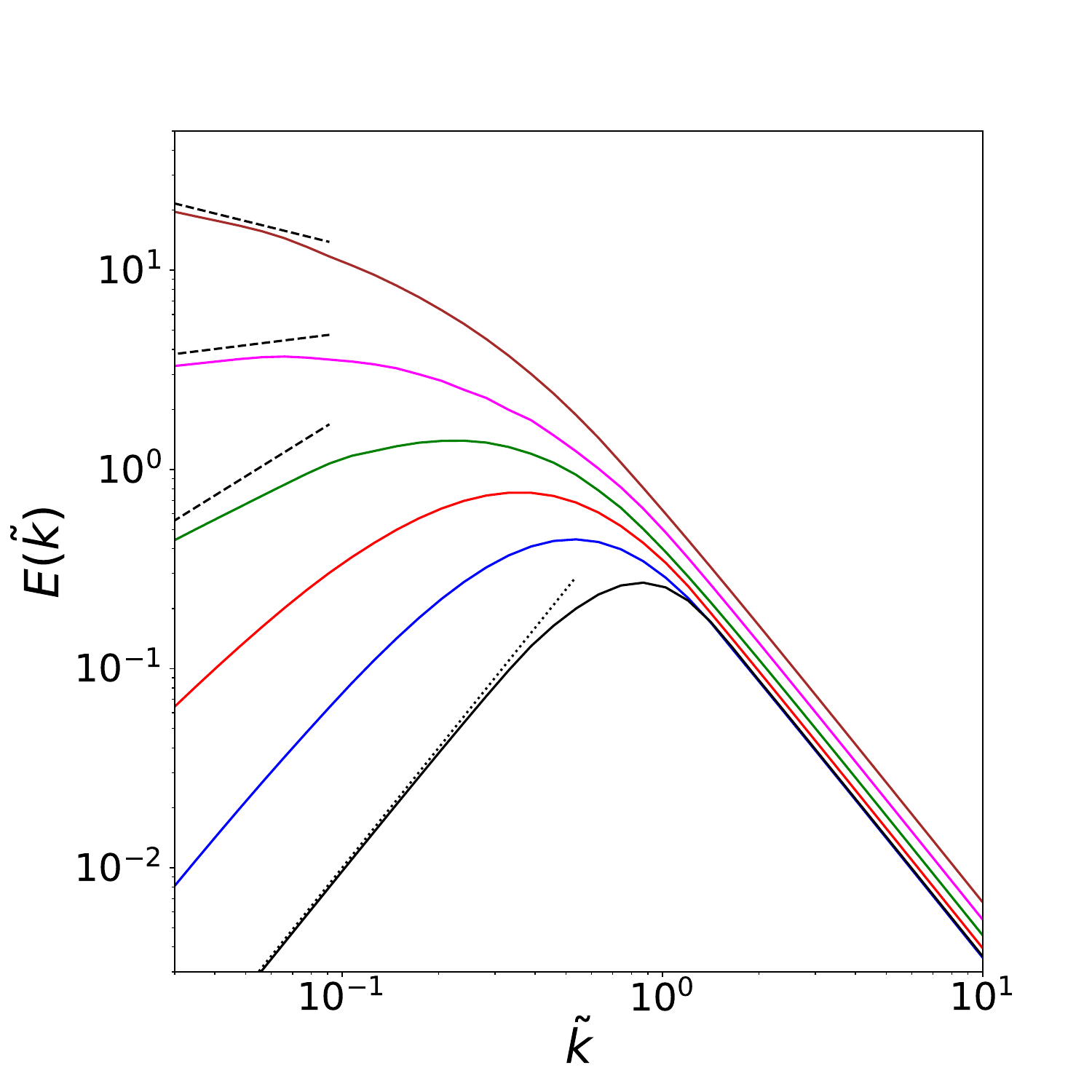}
\caption{
Velocity power spectrum $E(\tilde k)$.
In the right panel the dashed lines are the power laws (\ref{eq:Ek-small-k-power}) for the cases
$\alpha=2.8, 3.1, 3.5$. The dotted line is the power law $E(\tilde k) = \tilde k^2$, which shows the
low-$k$ slope for $\alpha > 4$.
}
\label{fig:Ek}
\end{figure}

Using $v_1=x_1-q_1=-q'_1-x/2$, $v_2=x_2-q_2=-q'_2+x/2$, and the expression (\ref{eq:Pq1q2})
for the probability distribution of $q'_1$ and $q'_2$, we obtain for the velocity correlation
between two points of distance $x=x_2-x_1 \geq 0$,
\be
B_v(x) \equiv \langle v_1 v_2 \rangle_x = \int_{-\infty}^{\infty} dq'_\star 
\int_{\psi_{\min}(q'_\star)}^{\infty} d \psi_\star \left[ \psi_\star^{-\alpha} 
\left( q'_\star \, \! ^2 - \frac{x^2}{4} \right) - \frac{x}{(\alpha-1)^2} \psi_\star^{2-2\alpha} 
\right]  e^{- {\cal I}(\psi_\star,q'_\star) } ,
\label{eq:v1v2-1}
\ee
where we integrated over $q'_1$ and $q'_2$.
We can check, with an integration by parts over $\psi_\star$, that we can write the expression
(\ref{eq:v1v2-1}) as
\be
B_v(x) = \frac{1}{\alpha-1} \int_{-\infty}^{\infty} dq'_\star \left( x \frac{\partial}{\partial x} 
- q'_\star \frac{\partial}{\partial q'_\star} \right) {\cal A}_{\alpha-1} ,
\ee
where ${\cal A}_{\nu}$ was defined in Eq.(\ref{eq:A-nu-def}).
Integrating by parts over $q'_\star$ this gives
\be
\alpha > 5/2, \;\; x \geq 0 : \;\;\; B_v(x) =  \frac{1}{\alpha-1}  \frac{d}{dx} \left[ x {\cal R}_{\alpha-1}(x) \right] , 
\label{eq:Bv-deriv}
\ee
where ${\cal R}_{\nu}$ was defined in Eq.(\ref{eq:A-nu-def}).
This generalizes to the case of finite $\alpha$ the relation (\ref{eq:Ek-inf-def}) obtained
in the Gaussian case with vanishing large-scale power \cite{Gurbatov1991}.
However, for finite $\alpha$ we cannot identify ${\cal R}_{\alpha-1}$ with ${\cal R}_{\alpha}$
and the term in the brackets is not $x P_{\rm void}(x)$ from Eq.(\ref{eq:Pvoid}).

For $\alpha \leq 5/2$ the velocity correlation $B_v(x)$ diverges,
as already seen in (\ref{eq:q2-variance}) for the one-point velocity variance $\langle v^2 \rangle$.
By parity symmetry we have $B_v(-x) = B_v(x) = B_v(|x|)$.
This gives the small-scale and large-scale asymptotics
\be
|x| \ll 1 : B_v(x) = \frac{{\cal R}_{\alpha-1}(0)}{\alpha-1} 
+ \frac{2 {\cal R}'_{\alpha-1}(0)}{\alpha-1} |x| + \dots ,   \;\;\; |x| \gg 1 :  \;\;\;  
B_v(x) \simeq \frac{2^{3\alpha-5} (3-\alpha)}{(\alpha-1) (\alpha-2) (2\alpha-5)} |x|^{5-2\alpha} .
\label{eq:Bv-asymp}
\ee
The non-analytic $|x|$ term, with a negative prefactor, corresponds to shocks. It is associated with
the probability to have at least one shock within the interval $|x|$, which decreases linearly with $|x|$
following the linear slope of the complementary void probability (\ref{eq:Pvoid-x-0}).
A shock at position $x_s$ leads to a discontinuous decrease of the velocity, 
$v(x_s^+)-v(x_s^-)<0$, following the familiar sawtooth pattern of solutions of the Burgers equation,
which is displayed in Fig.~\ref{fig:realization}.
This leads to the negative prefactor of the $|x|$ term in (\ref{eq:Bv-asymp}).

We show the velocity correlation in Fig.~\ref{fig:Bv}, for the cases $\alpha > 5/2$ where it is 
well-defined.
It is always positive at small distance, $|\tilde x| \lesssim 1$, with an amplitude that increases for smaller
$\alpha$. It becomes negative at large distance, $|\tilde x|  \gg 1$, for $\alpha > 3$,
as implied by Eq.(\ref{eq:Bv-asymp}).

The energy spectrum $E(k)$ is the Fourier transform of the correlation $B_v$,
\be
\alpha > 5/2 : \;\;\; E(k) \equiv \int_{-\infty}^{\infty} \frac{dx}{2\pi} B_v(x) e^{ikx} 
= \int_0^{\infty} \frac{dx}{\pi} B_v(x) \cos(kx) ,
\label{eq:Ek-def}
\ee
where we used $B_v(-x) = B_v(x)$.
For $\alpha>3$ we can use Eq.(\ref{eq:Bv-deriv}) to make an integration by parts, which gives
\be
\alpha > 3 : \;\; E(k) = \frac{k}{\pi (\alpha-1)} \int_0^{\infty} dx \, x {\cal R}_{\alpha-1}(x) \sin(kx) .
\label{eq:Ek-sin}
\ee
Using Eq.(\ref{eq:Ek-def}) we obtain the low-wavenumber asymptotics
\be
5/2 < \alpha < 4 , \;\; | k | \ll 1 : \;\; E(k) \simeq \frac{2^{\alpha-1}}{\sqrt{\pi}}  
\frac{\Gamma(4-\alpha)}{(\alpha-1) (\alpha-2) \Gamma(\alpha-3/2)} | k |^{2\alpha-6} ,
\label{eq:Ek-small-k-power}
\ee
and
\be
\alpha > 4 , \;\; | k | \ll 1 : \;\; E(k) \simeq \frac{k^2}{\pi (\alpha-1)} 
\int_0^{\infty} dx \, x^2 {\cal R}_{\alpha-1}(x) \propto k^2 .
\ee
Thus, the energy spectrum shows a power-law at low wavenumbers.
For $5/2 < \alpha < 4$ the exponent increases from $k^{-1}$ to $k^2$ whereas for $\alpha>4$ 
we have a universal $k^2$ tail.
In particular, $E(0) = \infty$ for $\alpha < 3$ and $E(0)=0$ for $\alpha>3$.
The non-analytic behavior at $x=0$ of the correlation $B_v(x)$ in (\ref{eq:Bv-asymp}),
associated with the term $|x|$ due to shocks, leads to the universal $k^{-2}$ decay of the energy 
spectrum at large wavenumbers,
\be
\alpha > 5/2 , \;\; | k | \gg 1 : \;\;\; E(k) \simeq \frac{16}{\sqrt{2\pi}} 
\frac{ \Gamma(2\alpha-5/2) \Gamma[2-1/(2\alpha-3)] } { (\alpha-1) (\alpha-2) \Gamma(2\alpha-1)}
\Lambda_{\alpha}^{(7-4 \alpha)/(2\alpha-3)} k^{-2} .
\label{eq:Ek-large-k-power}
\ee

We show the energy power spectrum in Fig.~\ref{fig:Ek}. We can check that it is always positive
and agrees with the asymptotic regimes (\ref{eq:Ek-small-k-power})-(\ref{eq:Ek-large-k-power}).

\subsection{Lagrangian increment}
\label{sec:Lagrangian-increment}

\begin{figure}
\centering
\includegraphics[height=6.cm,width=0.33\textwidth]{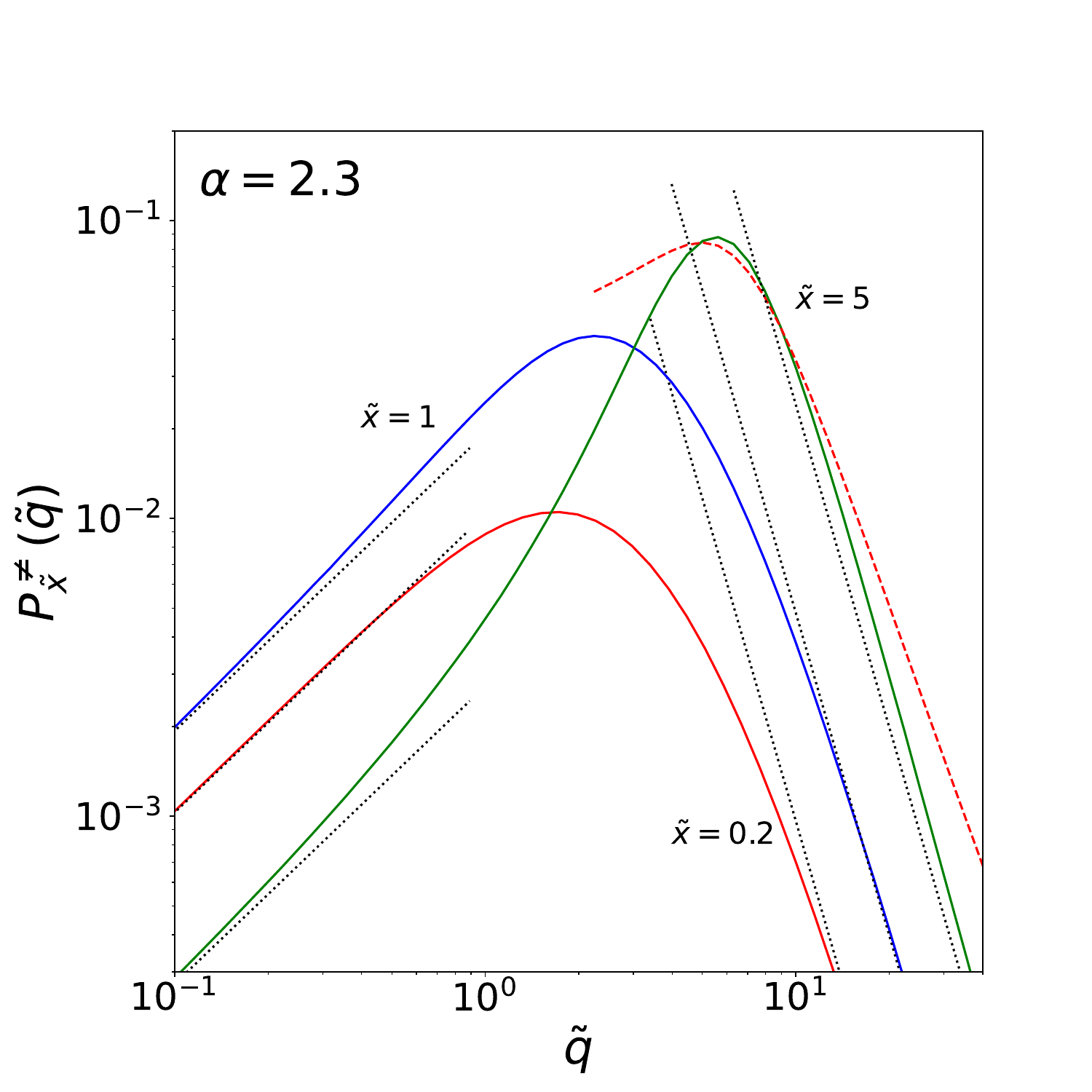}
\includegraphics[height=6.cm,width=0.33\textwidth]{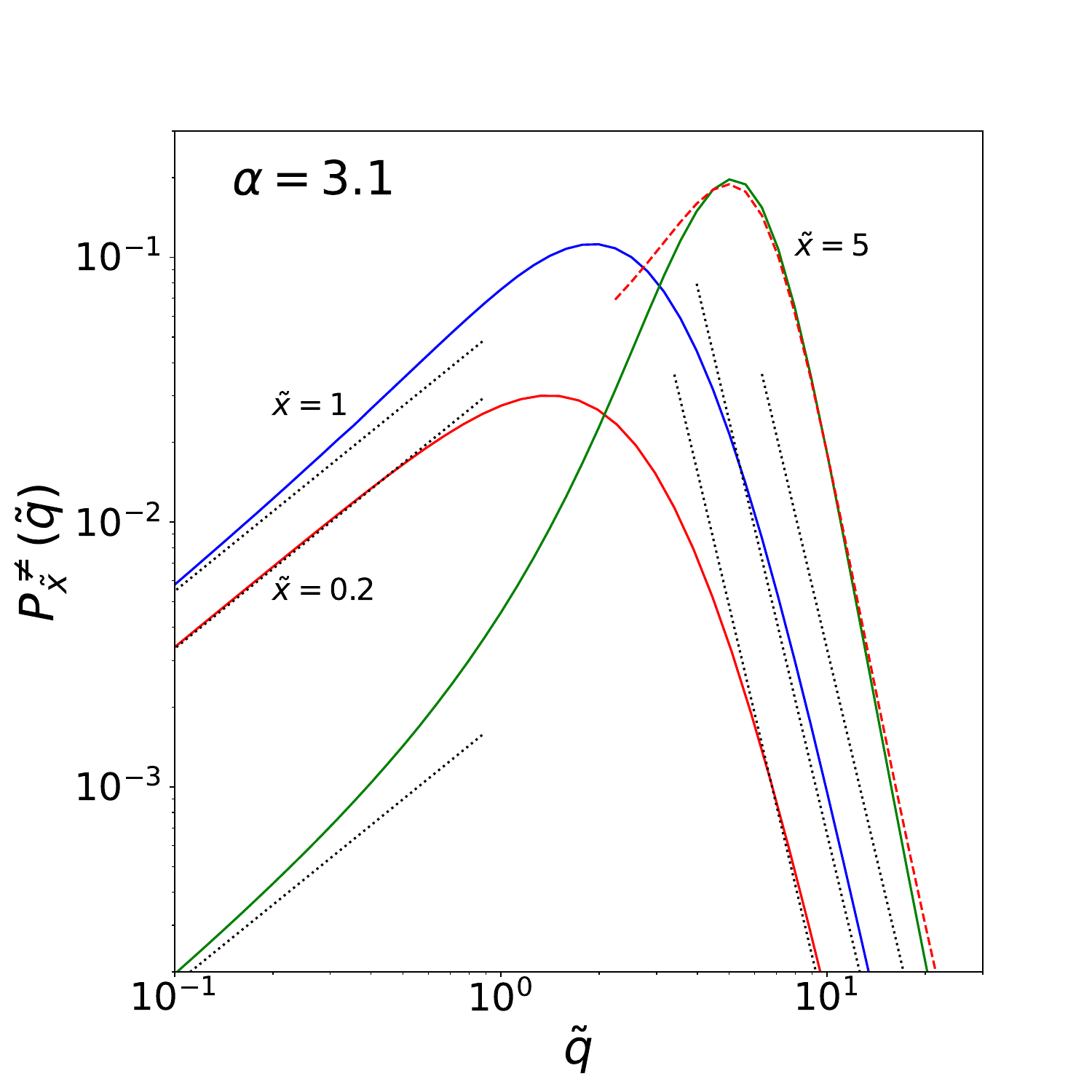}
\includegraphics[height=6.cm,width=0.32\textwidth]{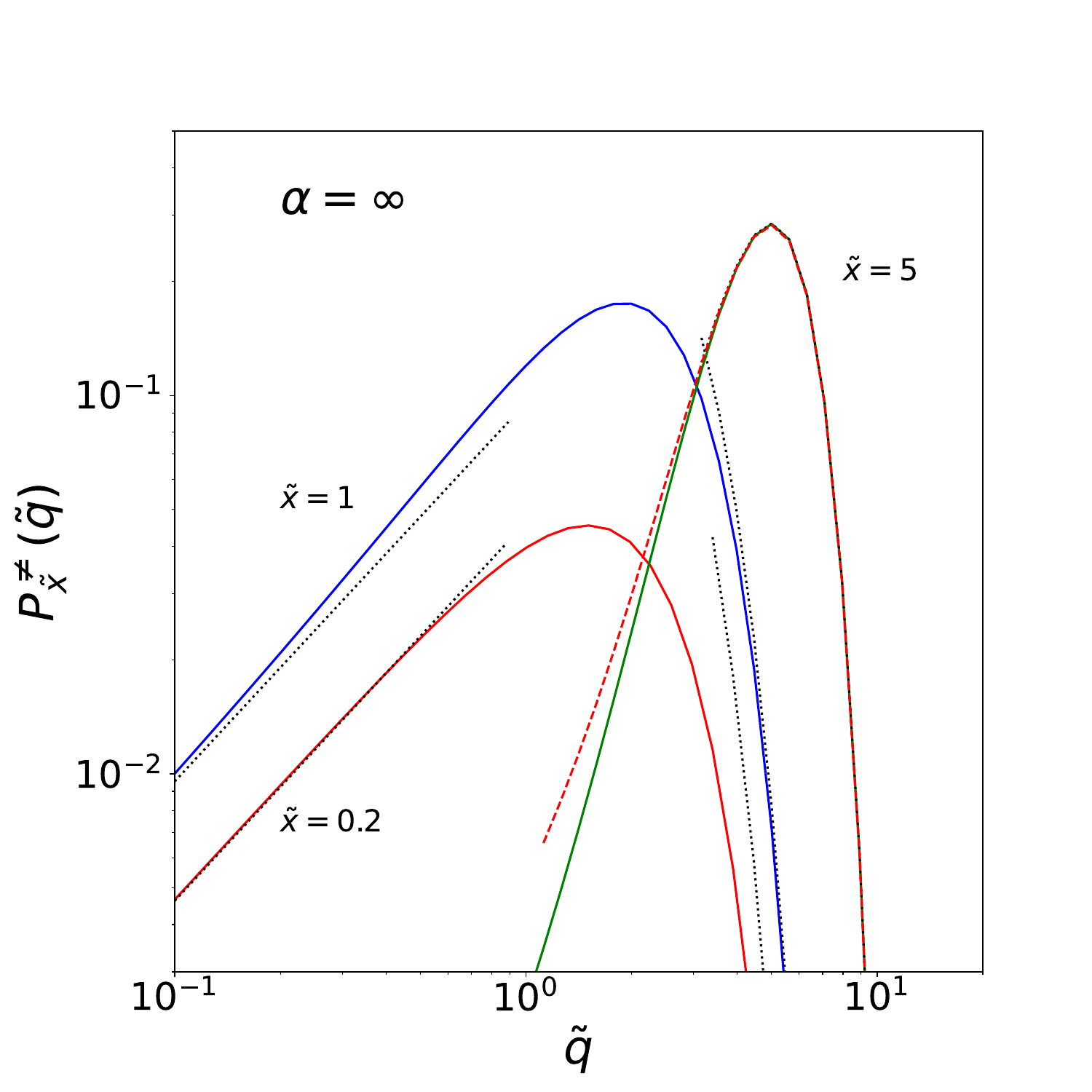}
\caption{
Probability distribution $P_{\tilde x}^{\neq}(\tilde q)$ for the cases $\alpha=2.3$, $3.1$ and
$\infty$, from left to right panel, and for the three scales, $\tilde x= 0.2, 1$ and $5$.
The black dotted lines are the small-$\tilde q$ and large-$\tilde q$ asymptotes (\ref{eq:P-x-q-asymp})
and (\ref{eq:P_x-q-alpha-infty-asymp-q}).
The red dashed lines for $\tilde x = 5$ are the large-separation asymptotes
(\ref{eq:P_x_q-large-peak}) and (\ref{eq:P_x_q-large-peak-alpha-infty}).
}
\label{fig:P_x_q}
\end{figure}

We now consider the distribution of the Lagrangian increment $q=q_2-q_1=q'_2-q'_1$ on the Eulerian 
interval $[x_1,x_2]$, which is given by the integration
\be
x = x_2 - x_1 \geq 0 , \;\; q = q_2 - q_1 \geq 0 : \;\;  P_x(q) = \int_{-\infty}^{\infty} 
dq'_1 dq'_2 \, P_x(q'_1,q'_2) \delta_D(q'_2-q'_1-q) .
\label{eq:P-x-q-inc-def}
\ee
From Eq.(\ref{eq:Pq1q2}) it contains a Dirac term, associated with the case where $q_1=q_2=q_\star$,
and a regular part,
\be
P_x(q) = P_{\rm void}(x) \, \delta_D(q) + P^{\neq}_x(q) ,
\label{eq:P-x-q-split}
\ee
with
\be
P^{\neq}_x(q) = x \int_{-\infty}^{\infty} dq'_\star \int_{\psi_{\min}(q'_\star)}^{\infty} 
\!\! d\psi_\star \, e^{- {\cal I}(\psi_\star,q'_\star)} \int_{q'_\star-q/2}^{q'_\star+q/2} 
d q' \, \psi_-(q'-q/2)^{-\alpha} \, \psi_+(q'+q/2)^{-\alpha}  ,
\label{eq:P-neq-def}
\ee
which also reads as
\be
P^{\neq}_x(q) = \int_0^{\infty} dc_1 dc_2 \, e^{- {\cal I}} \int_{q'_\star-q/2}^{q'_\star+q/2} 
d q' \, \psi_-(q'-q/2)^{-\alpha} \, \psi_+(q'+q/2)^{-\alpha}  .
\label{eq:P-neq-c1c2}
\ee
Throughout this article, the Dirac distribution is such that $\int_0^\infty dq \delta_D(q)=1$,
that is, it gives a unit weight as we integrate over $q\geq 0$ (we might write $\delta_D(q-0^+)$
but we keep simple notations).
The Dirac term is set by the void probability studied in Section~\ref{sec:Void-probabilities},
therefore we focus on the regular term $P^{\neq}_x(q)$ in this Section.
In the limit of small intervals we have from Eq.(\ref{eq:Pvoid-x-0})
\be
x \to 0 : \;\;  \int_0^{\infty} dq \, P^{\neq}_x(q) = 1 - P_{\rm void}(x) = N_{\rm void} x + \dots , 
\;\; x \to \infty : \;\;  \int_0^{\infty} dq \, P^{\neq}_x(q) \to 1 .
\label{eq:Px_q_x0}
\ee
Thus, the total weight of the regular part $P^{\neq}_x(q)$ decreases linearly with $x$ at small $x$,
because $P_{\rm void}$ goes to unity.

Using $q_2'-q_1'= \frac{\partial \psi_+}{\partial q'_2} - \frac{\partial \psi_-}{\partial q'_1} + x$
and integration by parts, we can check that the first moment satisfies
\be
\langle q \rangle_x = x {\cal R}_{\alpha}(x) + x \int_0^{\infty} dq P^{\neq}_x(q) =
x P_{\rm void}(x) + x (1-P_{\rm void}(x) ) = x , 
\label{eq:mean-q-x}
\ee
where we used the normalization to unity of the full probability distribution (\ref{eq:P-x-q-split})
and the result (\ref{eq:Pvoid}).
This means that there is no global contraction or expansion of the system. As the motion until time $t$
occurs on scales of the order of $L(t)$ introduced in (\ref{eq:L-t}), on much larger scales
we have $q \simeq x$, as seen in the right middle panel in Fig.~\ref{fig:realization}.
This implies that for any interval $x$ the mean $\langle q \rangle_x$
is equal to $x$.

From (\ref{eq:P-neq-def}) we obtain the power-law tails at fixed $x$,
\be 
q \to 0 : \;\; P_x^{\neq}(q) = {\cal R}_{2\alpha}(x) \, x q , \;\;\;
q \to \infty : \;\; P_x^{\neq}(q) = 2^{1+\alpha} x \, q^{1-2\alpha} .
\label{eq:P-x-q-asymp}
\ee
At large $x$ and $q$ and fixed $|q-x|$, we obtain from (\ref{eq:P-neq-c1c2})
\be
x \to \infty , \;\; q \to \infty , \;\; |q-x| \sim 1 : \;\;
P_x^{\neq}(q) \simeq f_{\infty}^{\neq}(|q-x|) \;\;\; \mbox{with} \;\;\;
f_{\infty}^{\neq}(q) = \int_{-\infty}^{\infty} dq' P_0(q'+q/2) P_0(q'-q/2) ,
\label{eq:P_x_q-large-peak}
\ee
where $P_0$ is the one-point probability distribution (\ref{eq:P_0-q}).
This is the large-separation limit, $x \gg 1$, where the statistics at locations $x_1$ and $x_2$ become uncorrelated, as Eq.(\ref{eq:P_x_q-large-peak}) also reads
$P_x^{\neq}(q) \simeq \int_{-\infty}^{\infty} dq_1 dq_2 \delta_D(q_2-q_1-q) P_0(q_1-x_1) P_0(q_2-x_2)$.
For small distance $x$ at fixed $q$ we obtain
\be
x \to 0 : \;\; P_x^{\neq}(q) \simeq x \, n_{\rm shock}(q) \;\;\; \mbox{with} \;\;\; 
n_{\rm shock}(q) = q \int_0^{\infty} dc \, e^{-\Lambda_\alpha c^{3/2-\alpha}}
\int_{-\infty}^{\infty} dq' (c+q'^{\,2}/2)^{-\alpha} (c+(q'+q)^2/2)^{-\alpha} ,
\label{eq:P-x-q-small-x}
\ee
where we anticipated that the function $n_{\rm shock}(q)$ defined by this expression is also
the mass function of shocks, as we will derive in Eq.(\ref{eq:nq-shocks}) below.
This means that the probability distribution of the matter content within small intervals,
$x \to 0$, is governed up to order $x$ by the probability to contain zero shock (i.e.,
being empty as described by the contribution $P_{\rm void}(x) \, \delta_D(q)$)
or one shock (as described by the shock mass function $n_{\rm shock}(q)$).
This is because the shocks are discrete and contain all of the matter.
In particular, we can see that the normalizations $\langle q \rangle_x = x$ and
(\ref{eq:nshock-all-mass}) are consistent with the small-$x$ factorized form
(\ref{eq:P-x-q-small-x}).
 
We note that taking the limits $q\to 0$ or $q\to\infty$ in Eq.(\ref{eq:P-x-q-small-x}) recovers
(\ref{eq:P-x-q-asymp}) in the limit $x\to 0$.
Thus the limits $q \to 0$ and $x \to 0$, or $q \to \infty$ and $x \to 0$, commute.

For numerical computations, the expression (\ref{eq:P-neq-def}) is better suited for small distances
$x$, as it already includes the prefactor $x$ that captures the linear decrease of the total
weight (\ref{eq:Px_q_x0}), whereas the expression (\ref{eq:P-neq-c1c2}) is better suited for
large distances, as the heights of the two first-contact parabolas $c_1$ and $c_2$ become
independent.

We show the regular part $P^{\neq}_{\tilde x}(\tilde q)$ in Fig.~\ref{fig:P_x_q} for the three cases
$\alpha=2.3$, $3.1$ and $\infty$, and for the three scales $\tilde x= 0.2, 1$ and $5$.
We obtain a similar behavior for all $\alpha$ (including the cases not shown in the figure).
At large distances, $\tilde x \gg 1$, the total weight of the regular part 
$P^{\neq}_{\tilde x}(\tilde q)$ goes to unity and the distribution is peaked around its mean 
$\langle \tilde q \rangle = \tilde x$.
As explained above, because the typical displacement of fluids elements at time $t$ is
set by the scale $L(t)$ of Eq.(\ref{eq:L-t}), smoothed on much larger scales $x \gg L(t)$
the system has hardly moved at all and $x/q \to 1$.
The expression (\ref{eq:P_x_q-large-peak}) provides an increasingly good approximation
of the distribution around its peak and over a broader domain.
However, the far tails remain described by the power laws (\ref{eq:P-x-q-asymp}).

On small scales, $\tilde x \ll 1$, the total weight of the regular part 
$P^{\neq}_{\tilde x}(\tilde q)$ decreases linearly with $\tilde x$ as in (\ref{eq:Px_q_x0})
while its peak remains at $\tilde q \sim 1$.
This is consistent with the normalization of the full distribution $P_{\tilde x}(\tilde q)$ to unity and of the
mean to $\langle \tilde q \rangle = \tilde x$.
Then, the location of the peak at $\tilde q \sim 1$ is decorrelated with the length $\tilde x$
and is simply set by the typical displacement scale $L(t)$, which corresponds to $\tilde x \sim 1$
in the dimensionless units (\ref{eq:re-scaling}).
Thus, as seen in Eq.(\ref{eq:P-x-q-small-x}), for small intervals $\tilde x \ll 1$ the regular
part $P^{\neq}_{\tilde x}(\tilde q)$ converges to a fixed shape $n_{\rm shock}(\tilde q)$
with an amplitude that decreases with $\tilde x$.

\subsection{Density field}
\label{sec:Density-field}

\begin{figure}
\centering
\includegraphics[height=6.cm,width=0.33\textwidth]{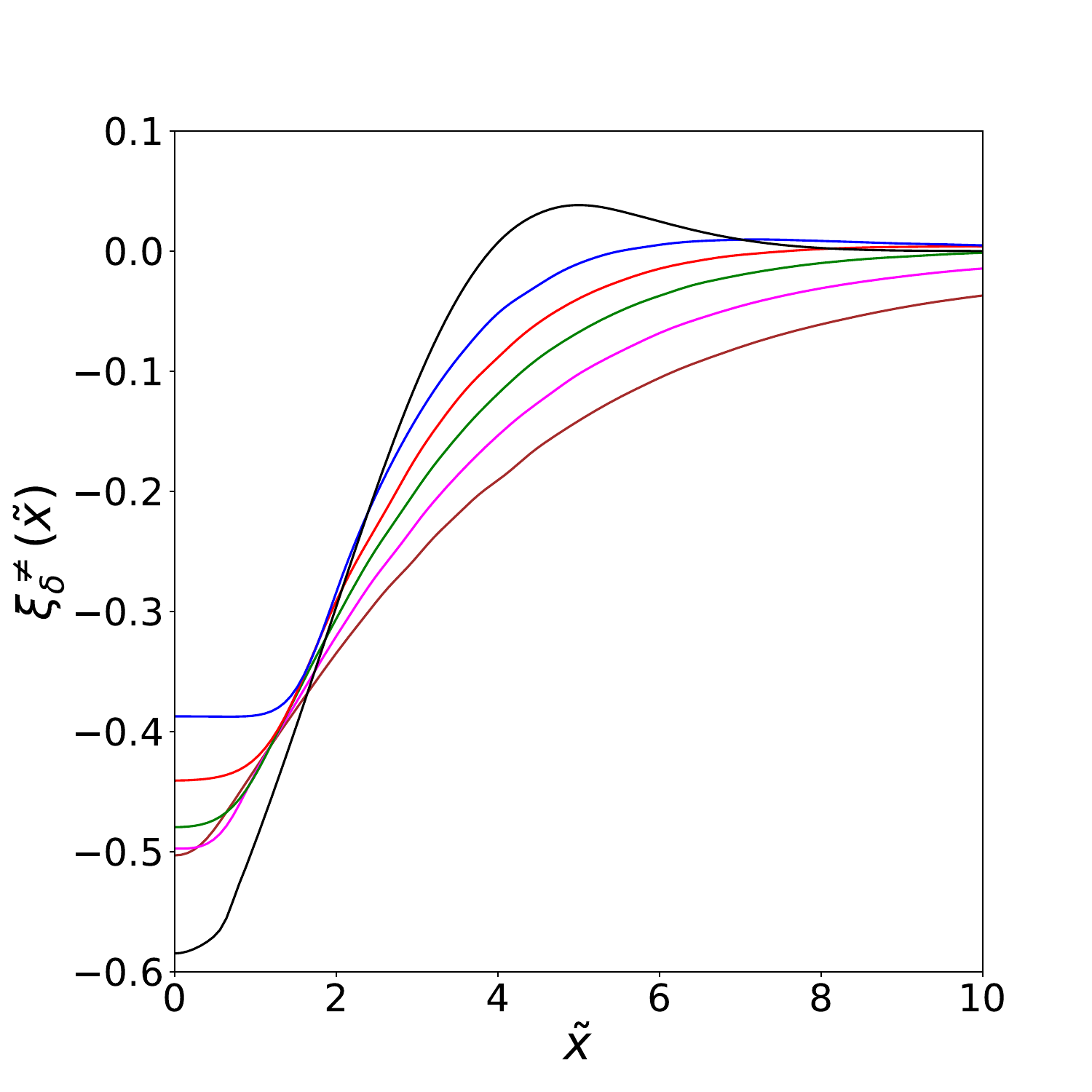}
\includegraphics[height=6.cm,width=0.33\textwidth]{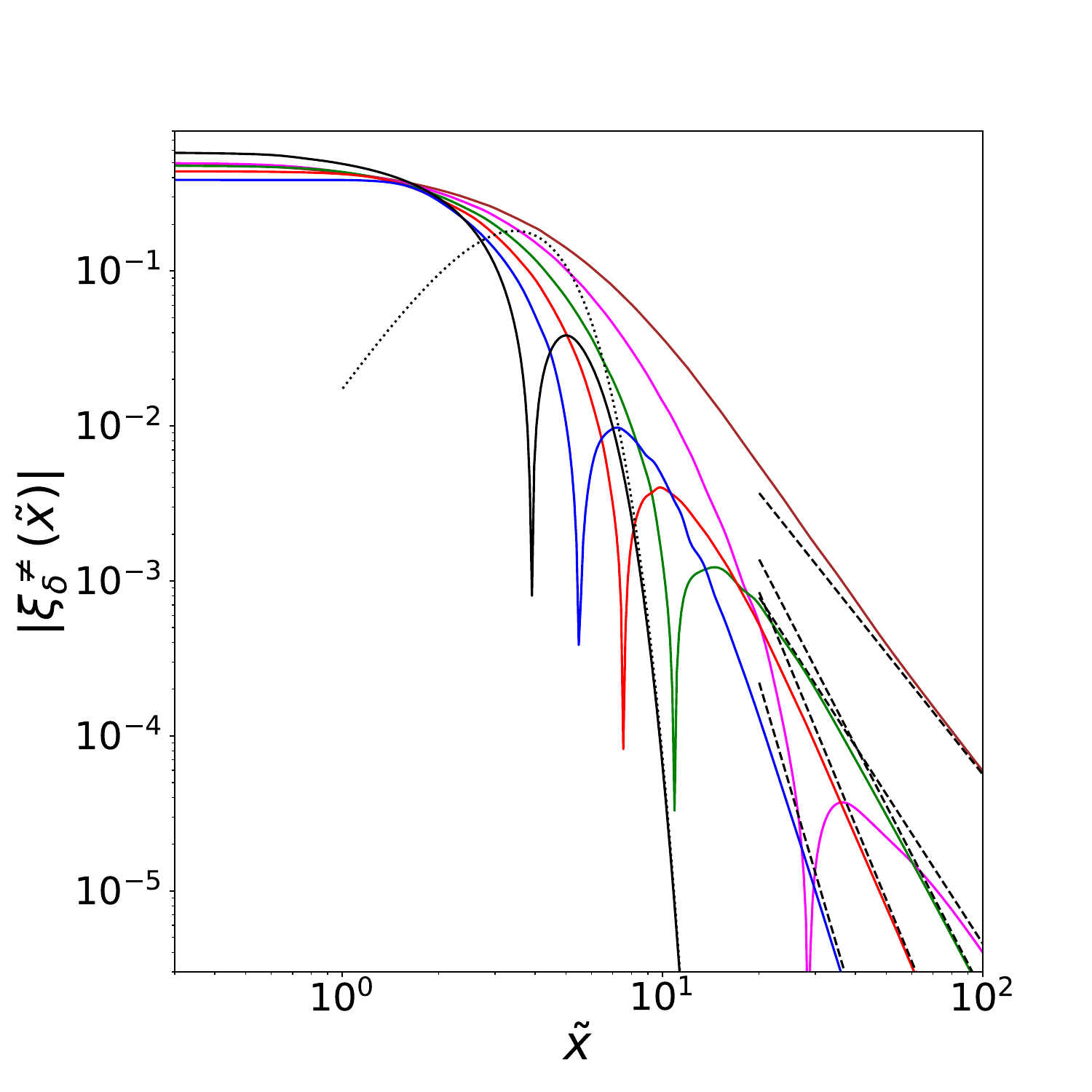}
\includegraphics[height=6.cm,width=0.32\textwidth]{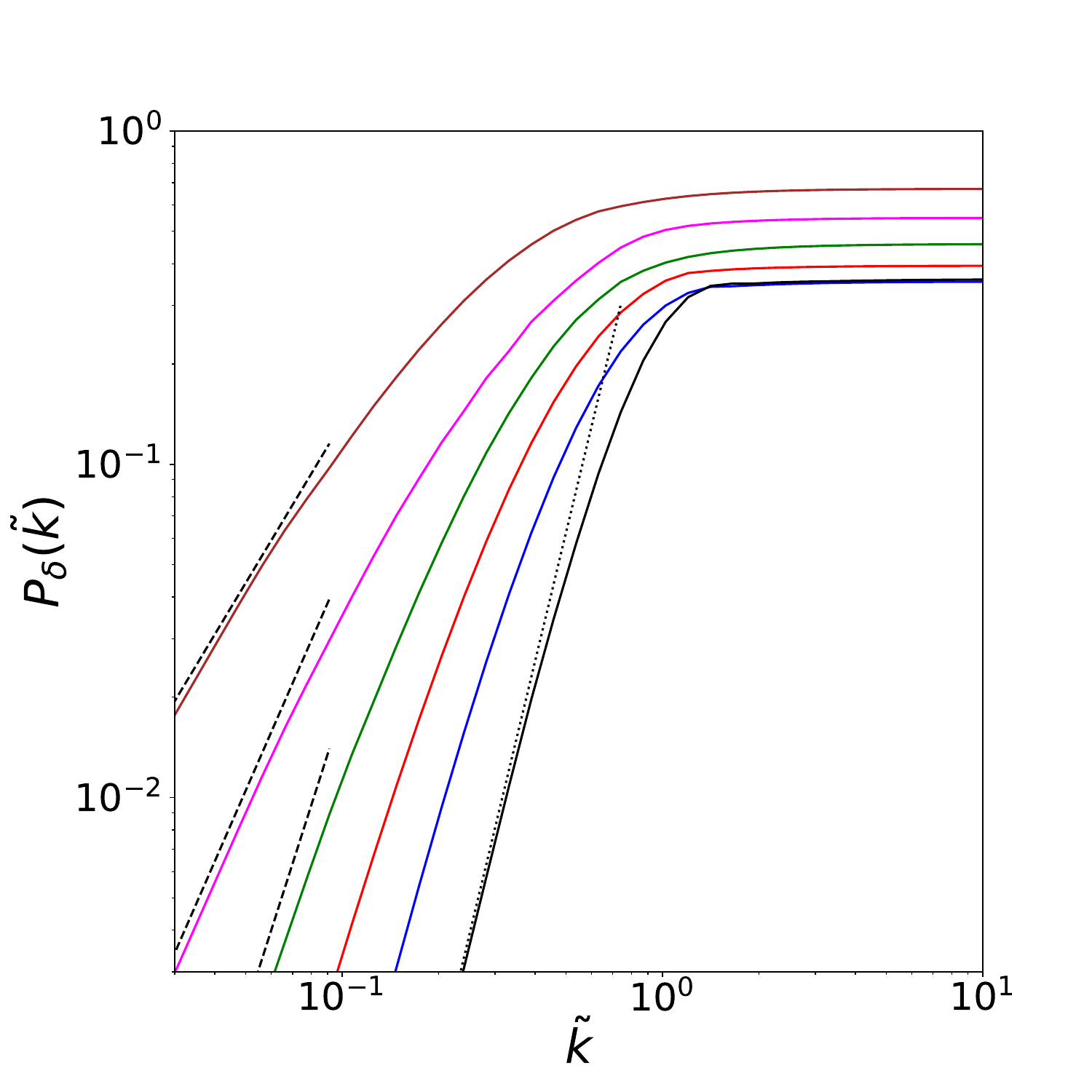}
\caption{
{\it Left and middle panels:} density correlation function $\xi_{\delta}^{\neq}(\tilde x)$ 
on linear and logarithmic scales for $\tilde x>0$.
{\it Right panel:} density power spectrum $P_{\delta}(\tilde k)$.
}
\label{fig:xi}
\end{figure}

\begin{figure}
\centering
\includegraphics[height=6.cm,width=0.33\textwidth]{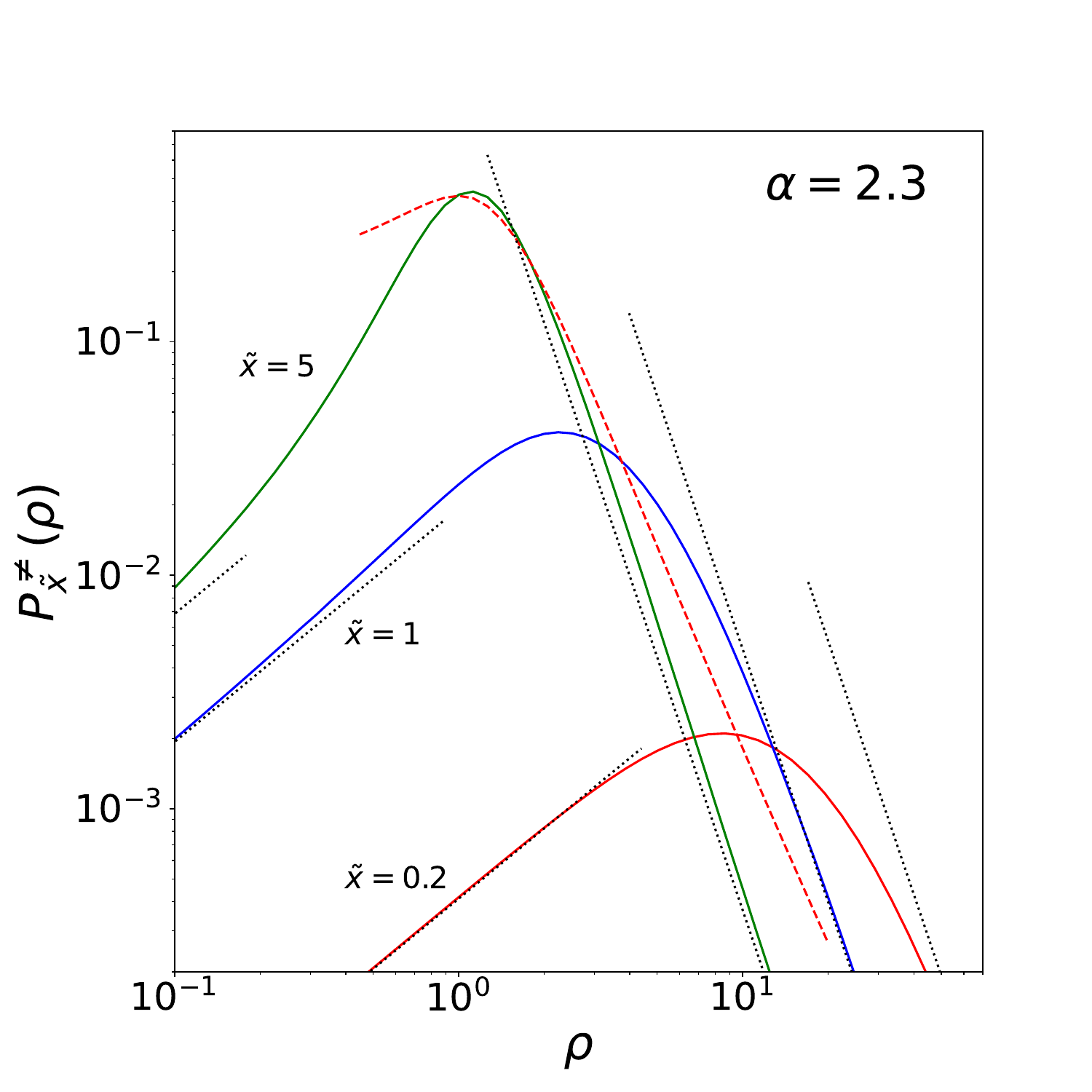}
\includegraphics[height=6.cm,width=0.33\textwidth]{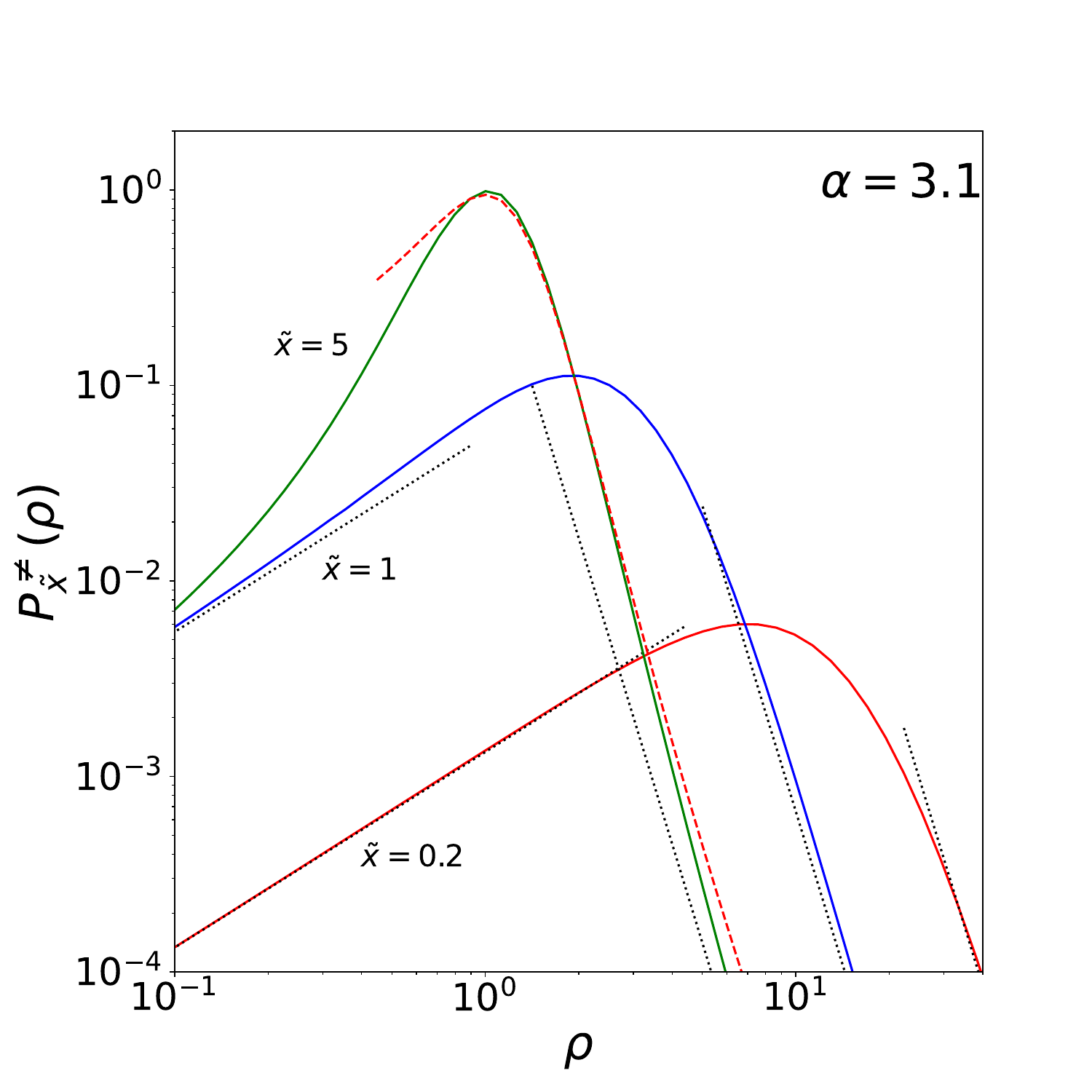}
\includegraphics[height=6.cm,width=0.32\textwidth]{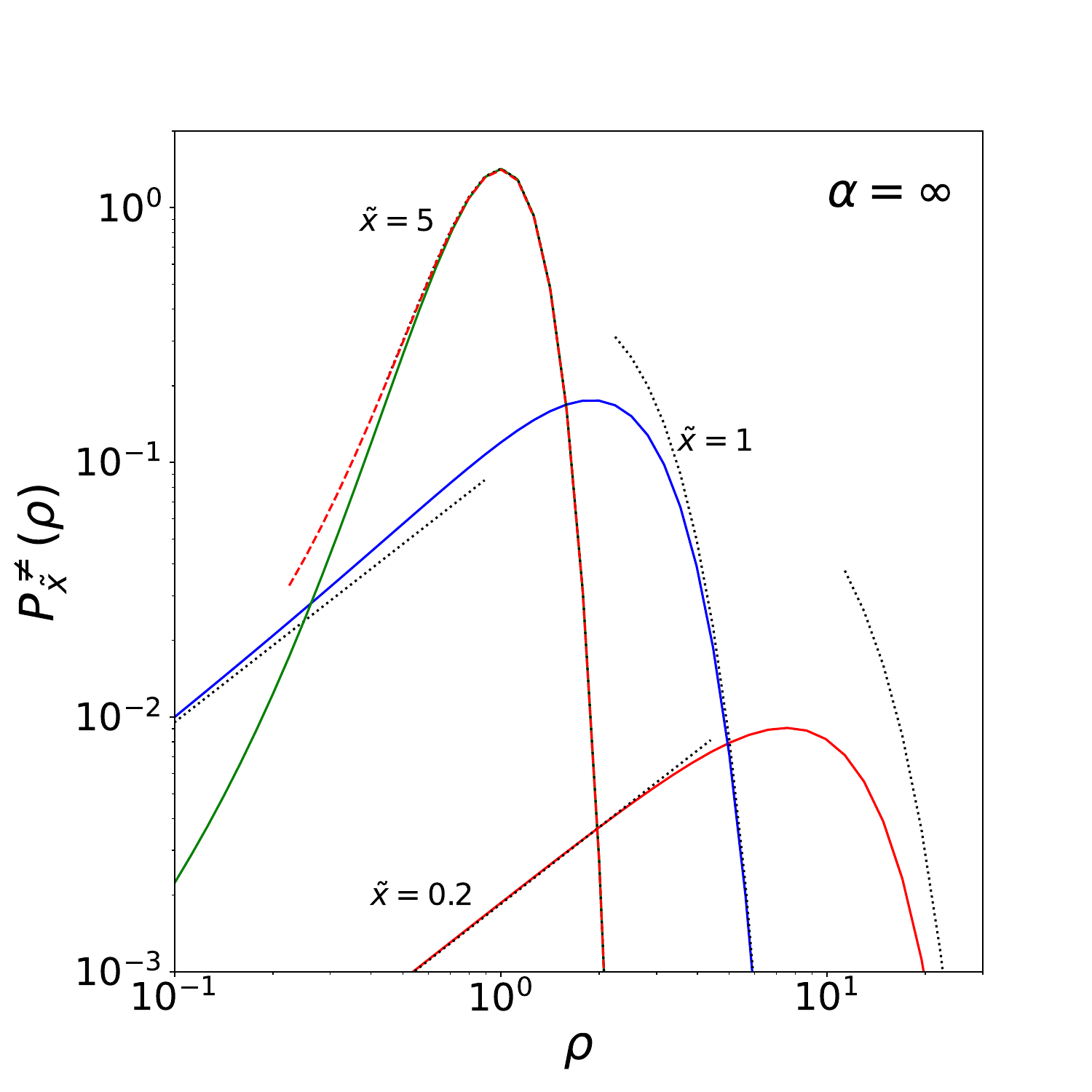}
\caption{
Probability distribution $P_{\tilde x}^{\neq}(\rho)$ of the overdensity $\rho$ for the cases $\alpha=2.3$, 
$3.1$ and $\infty$, from left to right panel, and for the three scales, $\tilde x= 0.2, 1$ and $5$.
The black dotted lines are the small-$\rho$ and large-$\rho$ asymptotes (\ref{eq:P-x-q-asymp})
and (\ref{eq:P_x-q-alpha-infty-asymp-q}).
The red dashed lines for $\tilde x = 5$ are the large-separation asymptotes
(\ref{eq:P_x_q-large-peak}) and (\ref{eq:P_x_q-large-peak-alpha-infty}).
}
\label{fig:P_x_rho}
\end{figure}

The conservation of matter, also encoded by the continuity equation, implies that the density field
$\rho(x)$ (normalized to the mean density of the system) and the density contrast $\delta=\rho-1$
are given by
\be
\rho(x) = \frac{dq}{dx} , \;\;\; \delta(x) = - \frac{dv}{dx} , \;\; \mbox{and} \;\; \langle \rho \rangle =1 , \;\;
\;\; \langle \delta \rangle =0 ,
\ee
in terms of the Eulerian map $q(x)$ and of the velocity field $v(x)$, where we used $v(x)=x-q(x)$.
Defining the density correlation $\xi_\delta(x) = \langle \delta(x_1) \delta(x_1+x) \rangle$ and the
power spectrum $P_\delta(k)$, we obtain
\be
\xi_\delta(x) = - B_v''(x) , \;\; P_\delta(k) = k^2 E(k) ,
\ee
where $B_v(x)$ and $E(k)$ are the velocity correlation and energy spectrum introduced
in Section~\ref{sec:energy-spectrum}.
The results (\ref{eq:Bv-deriv}) and (\ref{eq:Bv-asymp}) imply
\be
\xi_\delta(x) = \xi_0 \, \delta_D(x) + \xi_{\delta}^{\neq}(x) , \;\; \mbox{with} \;\;
\;\;\; x>0 : \; \xi_{\delta}^{\neq}(x) =  - \frac{1}{\alpha-1}  \frac{d^3}{dx^3} 
\left[ x {\cal R}_{\alpha-1}(x) \right] , 
\label{eq:xi-delta-deriv}
\ee
\be
\xi_0 = -\frac{4}{\alpha-1} {\cal R}'_{\alpha-1}(0) = \frac{8 \sqrt{2\pi} \Gamma(2\alpha-5/2) 
\Gamma\left( \frac{4\alpha-7}{2\alpha-3}\right)}{(\alpha-1)^2 (\alpha-2) \Gamma(2\alpha-2)}
\Lambda_{\alpha}^{(7-4\alpha)/(2\alpha-3)} > 0 ,
\label{eq:xi-0-def}
\ee
where the Dirac term associated with shocks comes from the absolute value term in
Eq.(\ref{eq:Bv-asymp}) and $\xi_{\delta}^{\neq}(x)$ is a finite even function. 

This provides at once the expressions and the properties of $\xi(x)$ and $P_\delta(k)$,
with
\be
\alpha > 5/2 , \;\; | x | \gg 1 : \xi_\delta(x) \simeq 2^{3\alpha-4} \frac{\alpha-3}{\alpha-1} |x|^{3-2\alpha} ,
\label{eq:xi-large-x}
\ee
and
\be
5/2 < \alpha < 4 , \;\; | k | \ll 1 : P_\delta(k) \propto |k|^{2\alpha-4} ; \;\;
\alpha > 4 , \;\; | k | \ll 1 : P_\delta(k) \propto |k|^{4}  ;
\label{eq:P-delta-low-k}
\ee
as well as
\be
\alpha > 5/2 , \;\; | k | \gg 1 : \;\;\; P_\delta(k) \simeq \frac{16}{\sqrt{2\pi}} 
\frac{ \Gamma(2\alpha-5/2) \Gamma[2-1/(2\alpha-3)] } { (\alpha-1) (\alpha-2) \Gamma(2\alpha-1)}
\Lambda_{\alpha}^{(7-4\alpha)/(2\alpha-3)}  .
\ee
At high wavenumbers we recover a white-noise density power spectrum because of the shocks.
We show the density correlation and the power spectrum in Fig.~\ref{fig:xi}.
The density correlation is negative at small distances, $|\tilde x| \lesssim 1$. However, we note that
the Dirac term $\xi_0 \delta_D(x)$ is positive, see Eq.(\ref{eq:xi-0-def}).
Therefore, the density correlation is positive at infinitesimally small distance, $x=0$, because
the shocks are high (infinite) positive overdensities of vanishing width.
For $0 < \tilde x \lesssim 1$, the density correlation is negative, which expresses the facts that
locally matter has fallen into the shocks and that shocks are isolated, with a void probability
$P_{\rm void}(x)$ that goes to unity at small distance.
At large distance, $|\tilde x| \gg 1$, $\xi_{\delta}$ changes sign and becomes positive for
$\alpha > 3$, see Eq.(\ref{eq:xi-large-x}).
This may be related to the fact that for smaller $\alpha$ the power-law tails are heavier, so that 
the negative regime extends up to large scales, whereas for larger $\alpha$ the steeper decrease
of the correlations allows the local exclusion effect to be more efficiently screened and there appears
a positive correlation due to large-scale effects (a positive fluctuation on a long wavelength
$\lambda \gg x$ increases the probability of mass concentrations at both $x_1$ and $x_2$).
The right panel clearly shows the convergence to a constant high-$\tilde k$ tail for the density
power spectrum in all cases, while the low-$\tilde k$ tail agrees with the asymptotic results
(\ref{eq:P-delta-low-k}).

The mean density $\rho$ within the Eulerian interval $[x_1,x_2]$ of length $x$ is given by
\be
\rho = \frac{q_2-q_1}{x_2-x_1} = \frac{q}{x} , \;\; \mbox{and} \;\; 
P_x(\rho) = P_{\rm void}(x) \delta_D(\rho) + P^{\neq}_x(\rho) \;\; \mbox{with} \;\;
P^{\neq}_x(\rho) = x P^{\neq}_x(q) , 
\ee
where $q=q_2-q_1$ is the Lagrangian increment as in Section~\ref{sec:Lagrangian-increment}.
This gives the low and high density asymptotics
\be
\rho \to 0 : \;\; P^{\neq}_x(\rho) = {\cal R}_{2\alpha}(x) x^3 \rho , \;\;\;
\rho \to \infty : \;\; P^{\neq}_x(\rho) = 2^{1+\alpha} x^{3-2\alpha} \rho^{1-2\alpha} ,
\ee
whereas we have the small and large scale behaviors
\be
x \to 0 : \;\; P^{\neq}_x(\rho) = x^2 n_{\rm shock}(x \rho) , \;\;\;
x \to \infty , \;\; | \delta | \propto 1/x : \;\;  P^{\neq}_x(\delta) \simeq x f_{\infty}^{\neq}( x | \delta | ) .
\label{eq:P_x-rho_small-large-scale}
\ee
On large scales we recover an homogeneous system with small density fluctuations
and the universal behavior 
\be
x \to \infty : \;\; \langle \delta^2 \rangle_x \propto 1/x^2 ,
\ee
whereas on small scales the density distribution is highly inhomogeneous, as it is dominated by the
voids and the shocks.

We show the regular part $P^{\neq}_{\tilde x}(\rho)$ in Fig.~\ref{fig:P_x_rho} for the three cases
$\alpha=2.3$, $3.1$ and $\infty$, and for the three scales $\tilde x= 0.2, 1$ and $5$,
as in Fig.~\ref{fig:P_x_q}. We can clearly see the convergence to an homogeneous system
on large scales, with an increasingly narrow peak around the mean $\langle \rho \rangle = 1$.
On small scales, the total weight of $P^{\neq}_{\tilde x}(\rho)$ decreases linearly with $\tilde x$,
as for $P^{\neq}_{\tilde x}(\tilde q)$ in Fig.~\ref{fig:P_x_q}, because most intervals are empty.
The peak of the regular function $P^{\neq}_{\tilde x}(\rho)$ now grows as 
$\rho_{\rm peak} \propto 1/\tilde x$, in agreement with the factorized form 
(\ref{eq:P_x-rho_small-large-scale}).

\section{Higher-order distributions}
\label{sec:higher-order}

We now turn to higher-order distributions, where explicit expressions can also be derived 
(see for instance \cite{Molchanov1995} for the Gaussian case).
In fact, as for the two-point distribution (\ref{eq:Pq1c1q2c2}), the problem simplifies if we consider
together the two variables $(q_i,c_i)$ associated with an Eulerian point $x_i$.
Indeed, thanks to the uncorrelated nature of the Poisson point process and to the ordering
$q_1 \leq q_2 \leq \cdots \leq q_n$ for $x_1 \leq x_2 \leq \cdots \leq x_n$,
we can write for the $n$-point distribution
\be
x_1 \leq x_2 \leq \cdots \leq x_n : \;\;\; P_{x_1 ,\cdots,x_n}(q_1,c_1; \cdots ;  q_n,c_n) 
= P_{x_1}(q_1,c_1) \prod_{i=2}^n P_{x_i,x_{i-1}}(q_i,c_i | q_{i-1},c_{i-1}) ,
\label{eq:Pn-prod}
\ee
where $P_{x_1}(q_1,c_1) = P_0(q_1-x_1,c_1)$ is the one-point distribution obtained in
Section~\ref{sec:one-point-Eulerian} and $P_{x_i,x_{i-1}}(q_i,c_i | q_{i-1},c_{i-1})$ is the conditional
probability given by
\bea
P_{x_2,x_1}(q_2,c_2 | q_1,c_1) & = & \left[ \delta_D(q_2-q_1) \delta_D(c_2-c_{12}) + \theta(q_2>q_{12})
\theta(c_2>c_{12}) ( c_2 + (q_2-x_2)^2/2)^{-\alpha} \right] \nonumber \\
&& \times e^{ - \int_{q_{12}}^{\infty} dq \int_{c_2+(q-x_2)^2/2}^{c_1+(q-x_1)^2/2} d\psi \, \psi^{-\alpha} } .
\label{eq:P21-conditional}
\eea
We can check that this expression agrees with Eq.(\ref{eq:Pq1c1q2c2}) for the two-point distribution.
Here we introduced $c_{12}$ as the height of the parabola ${\cal P}_{x_2,c_2}$ that
intersects the previous parabola ${\cal P}_{x_1,c_1}$ at the position $q_1$, and we changed
notation from $q_\star$ to $q_{12}$ for the intersection of two parabolas for arbitrary $c_2$,
\be
c_{12} = c_1 + \frac{ (q_1-x_1)^2 - (q_1-x_2)^2}{2} , \;\;\;
q_{12}(c_2) = \frac{x_1+x_2}{2} + \frac{c_2-c_1}{x_2-x_1} .
\ee
If $c_{12} \leq 0$ the first term in Eq.(\ref{eq:P21-conditional}) and the Heaviside factor
$\theta(c_2>c_{12})$ are removed.
We can again check the normalization $\int dc_2 dq_2 P_{x_2,x_1}(q_2,c_2 | q_1,c_1) =1$.

The factorization (\ref{eq:Pn-prod}) holds because one can easily check that
$q_{13} \geq q_{12}$, considering for instance the three-point distribution.
This implies that the intersection of parabolas ${\cal P}_1$ and ${\cal P}_3$ is irrelevant
as it occurs in the domain where ${\cal P}_2$ has already taken over ${\cal P}_1$,
${\cal P}_2 \leq {\cal P}_1$.
Therefore, the parabolic arcs follow the same ordering $({\cal P}_1, \cdots , {\cal P}_n)$
as the points $(x_1,x_2,\cdots,x_n)$, over the domains 
$-\infty < q_{12} \leq q_{23} \leq \cdots \leq q_{n-1,n} < \infty$.
This implies that when we add a new point $x_n$ in the $n$-point distribution (\ref{eq:Pn-prod}),
we only need to consider the previous parabola ${\cal P}_{x_{n-1},c_{n-1}}$ and its contact
point $q_{n-1}$.
Then, as for the two-point distribution (\ref{eq:Pc1c2}), we must separate whether the new contact 
point $q_n$ is identical to the previous contact point $q_{n-1}$ (the first term in Eq.(\ref{eq:Pn-prod})
where the two parabola have the same contact point which is also their intersection), to avoid
putting a double Poisson weight on this contact point, or it is located further to the right (the second term).
We finally add the exponential term associated with the additional constraint that the domain in the
$(q,\psi)$ plane below the previous parabola ${\cal P}_{n-1}$ and above the new parabola
${\cal P}_{n}$ must be empty.

Therefore, thanks to the uncorrelated nature of the Poisson process, the many-points 
correlation functions satisfy the Markovian factorization (\ref{eq:Pn-prod}).
It is interesting to compare with Brownian and white-noise initial conditions.
There, the $n$-point distributions satisfy the simpler factorization 
\cite{Valageas2009a,Valageas2009b}
\be
\mbox{Brownian or white noise} : \;\;  x_1 \leq \cdots \leq x_n : \;\;\; P_{x_1,\cdots,x_n}(q_1 , \cdots , q_n)
=  P_{x_1}(q_1) \prod_{i=2}^n P_{x_i,x_{i-1}}(q_i | q_{i-1}) .
\label{eq:Pn-prod-Brownian-white-noise}
\ee
Moreover, in the Brownian case the Lagrangian increments are independent and we have
$P_{x_2,x_1}(q_2 | q_1) = P_{x_2-x_1}(q_2-q_1)$.
In the case of the Poisson point process, the factorization of the distributions 
$P_{x_1,\cdots,x_n}(q_1 , \cdots , q_n)$ is lost, but it is recovered when we consider the
pair of variables $q_i$ and $c_i$, as in (\ref{eq:Pn-prod}).
One may wonder whether there are many instances where factorization can be recovered
by adding auxiliary variables and if this can be a useful manner to expand the number of
solvable cases.

\section{Lagrangian distributions or particle displacements}
\label{sec:Lagrangian}

In this Section we now turn to Lagrangian distributions, that is, the statistics of the Eulerian
positions $x$ for given Lagrangian coordinates $q$ of the particles.
This corresponds to the statistics of the displacements $x-q$ of the particles labeled by their
initial positions $q$.

\subsection{One-point Lagrangian distribution}
\label{sec:one-point-Lagrangian}

Because particles do not cross each other, the Lagrangian probability $P_q(\geq x)$ for the particle
$q$ to be located to the right of position $x$ is equal to the Eulerian probability $P_x(\leq q)$
for the Lagrangian coordinate $q(x)$ found at position $x$ to be smaller than or equal to $q$.
Taking the derivative of this equality with respect to $x$ and using 
$P_x(q)=P_0(q-x)$ as in Section~\ref{sec:one-point-Eulerian}, we obtain
\be
P_q(x) = P_0(x-q) = P_x(q) .
\ee
Thus, the one-point Lagrangian and Eulerian distributions are identical and the properties
of $P_q(x)$ can be read from the results obtained in Section~\ref{sec:one-point-Eulerian}.

\subsection{Two-point Lagrangian distribution}
\label{sec:two-point-Lagrangian}

\begin{figure}
\centering
\includegraphics[height=6.cm,width=0.33\textwidth]{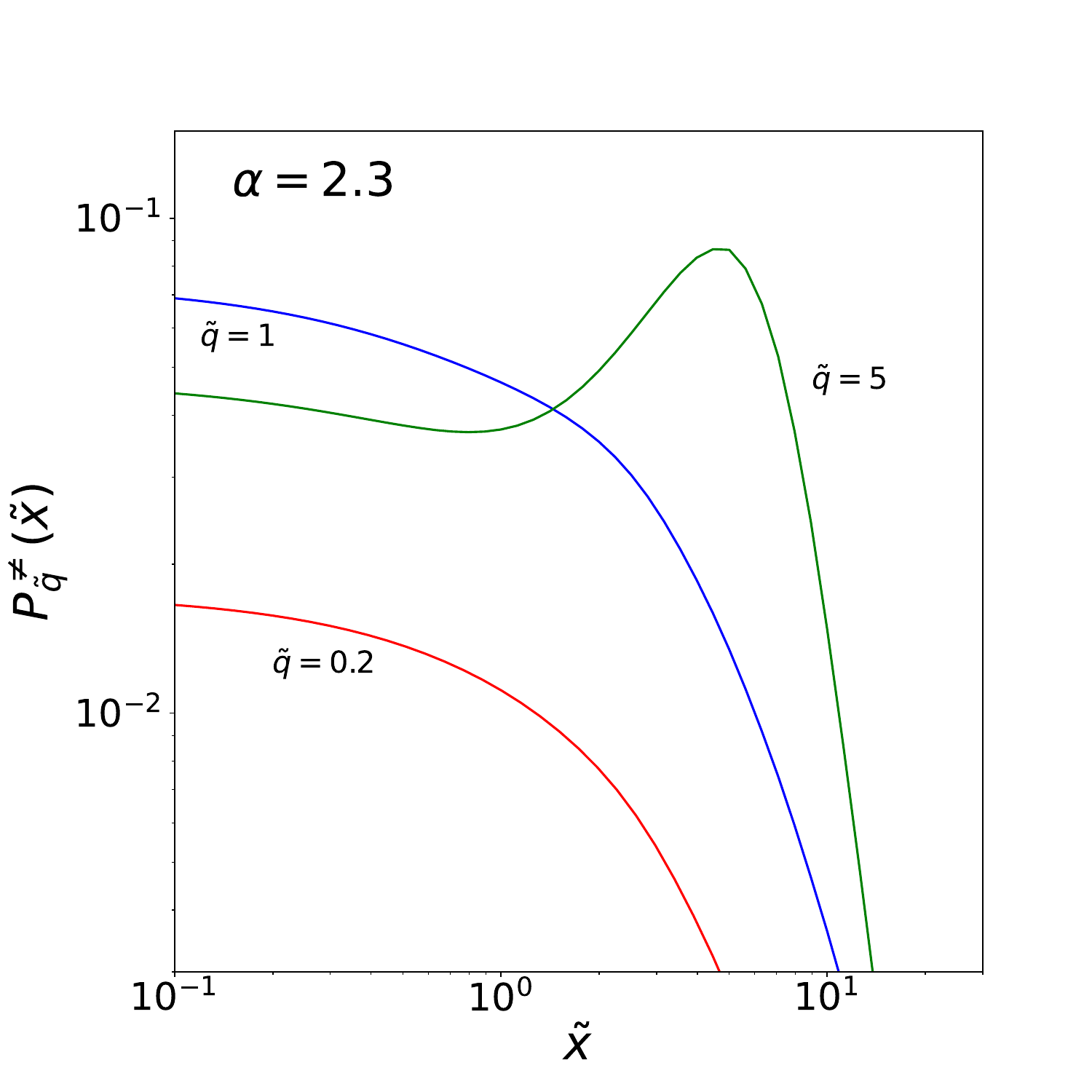}
\includegraphics[height=6.cm,width=0.33\textwidth]{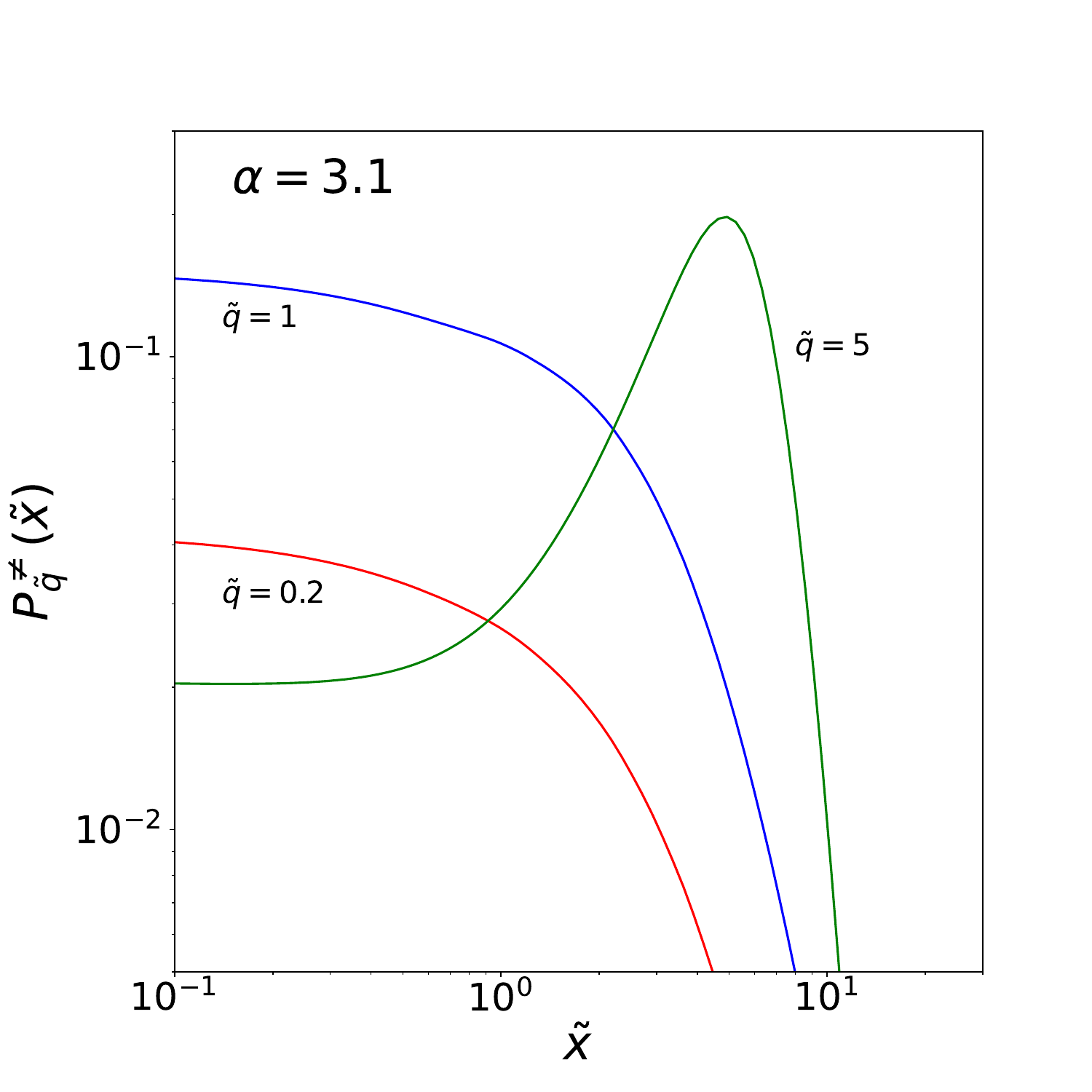}
\includegraphics[height=6.cm,width=0.32\textwidth]{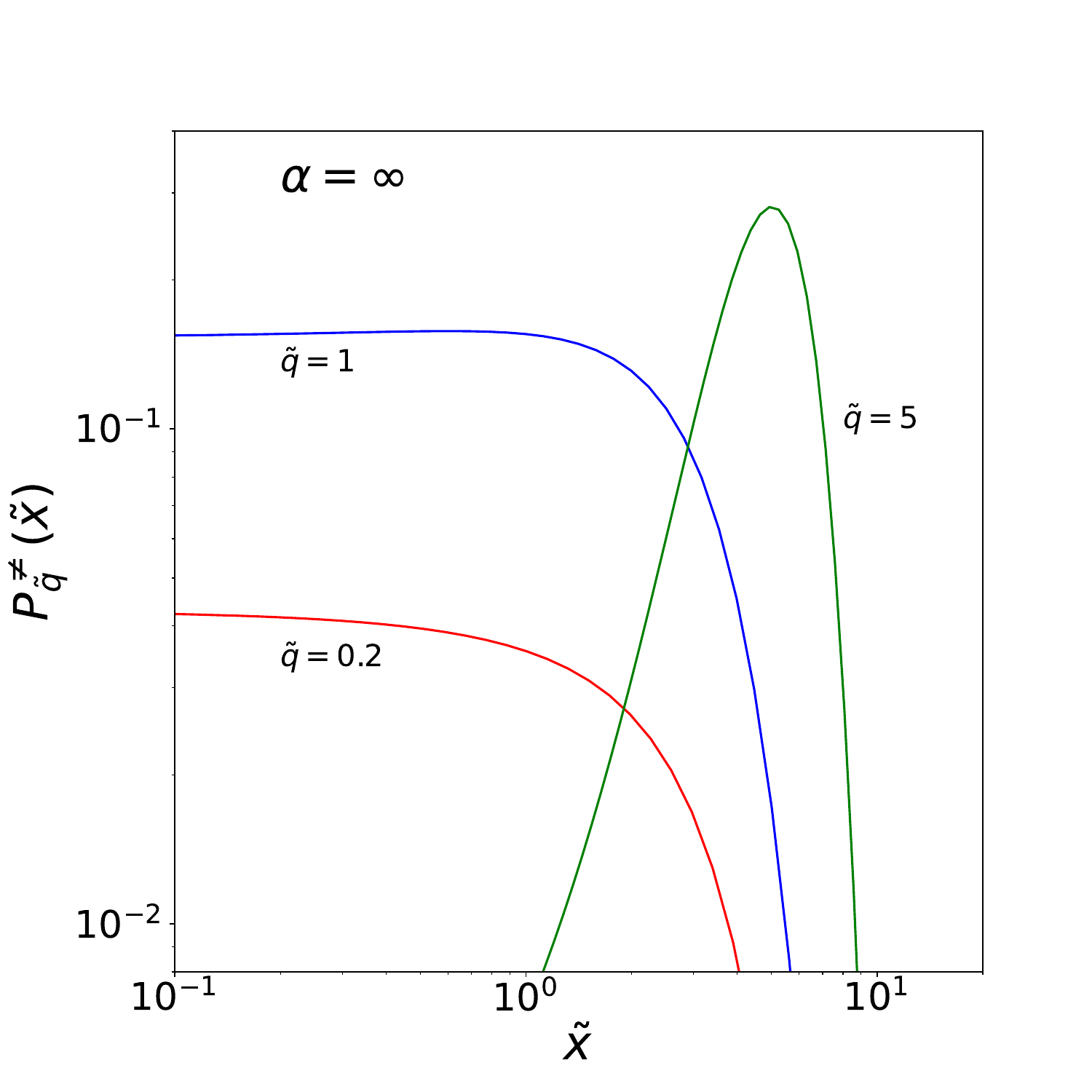}
\caption{
Probability distribution $P_{\tilde q}^{\neq}(\tilde x)$ for the cases
$\alpha=2.3$, $3.1$ and $\infty$, from left to right panel, and for the three scales, 
$\tilde q= 0.2, 1$ and $5$.
}
\label{fig:P_q_x}
\end{figure}

As for the one-point distributions, we can relate the two-point Lagrangian and Eulerian probabilities
by
\be
P_{q_1,q_2}(\geq x_1,\leq x_2) = P_{x_1,x_2}(\leq q_1,\geq q_2) , \;\;\;
P_{q_1,q_2}(x_1,x_2) = - \frac{\partial^2}{\partial x_1 \partial x_2} \int_{-\infty}^{q_1} dq_1''
\int_{q_2}^{\infty} dq_2'' \, P_{x_1,x_2}(q_1'',q_2'') .
\label{eq:P-twopoint-Lag}
\ee
Using the fact that $P_{x_1,x_2}(q_1,q_2) = P_x(q_1-\bar x,q_2-\bar x)$ thanks to statistical
homogeneity, as seen in Eq.(\ref{eq:Pq1q2}), and changing variables from $(x_1,x_2)$ to
$(x,\bar x)$, we obtain
\be
P_{q_1,q_2}(x_1,x_2) = \left( \frac{\partial^2}{\partial x^2} - \frac{1}{4} \frac{\partial^2}{\partial \bar x^2}
\right) \int_{-\infty}^0 dq_1'  \int_{q}^{\infty} dq_2' \, P_x(q_1'+q_1-\bar x,q_2'+q_1-\bar x) ,
\ee
where $q=q_2-q_1$.
As in Section~\ref{sec:Lagrangian-increment}, we focus on the probability
distribution of the Eulerian increment $x$,
\be
P_q(x) = \int dx_1 dx_2 \, P_{q_1,q_2}(x_1,x_2) \, \delta_D(x_2-x_1-x) .
\ee
Changing again variables from $(x_1,x_2)$ to $(x,\bar x)$ and using the definition (\ref{eq:P-x-q-inc-def}),
we obtain the relation between the distributions of the Lagrangian and Eulerian increments,
\be
q > 0 : \;\;\; P_q(x) = P_{\rm shock}(q) \, \delta_D(x) + P_q^{\neq}(x) \;\;\; \mbox{with} \;\;\;
P_q^{\neq}(x) = \frac{\partial^2}{\partial x^2} \int_q^{\infty} dq' \, P_x^{\neq}(q') (q'-q) ,
\label{eq:P-q-x-def}
\ee
where we used that for $q>0$ the Dirac term in the distribution (\ref{eq:P-x-q-split}) does not
contribute and we added the contribution $P_{\rm shock}(q) \delta_D(x)$, associated
with the probability that the interval $[q_1,q_2]$ belongs to a single shock, which is not included
in the regular term $P_q^{\neq}(x)$.
Writing the integral over $q$ as $\int_q^{\infty}=\int_0^{\infty}-\int_0^q$ and using 
Eqs.(\ref{eq:P-x-q-split}) and (\ref{eq:mean-q-x}) in the first integral, we can also write
\be
P_q^{\neq}(x) = q \, n_{\rm void}(x) + 
\frac{\partial^2}{\partial x^2} \int_0^q dq' \, P_x^{\neq}(q') (q-q') , \;\; \mbox{whence} \;\;
\langle x \rangle_q = q ,
\label{eq:P-q-x-void}
\ee
where the second equality easily follows from integrations by parts and the result $P_{\rm void}(0)=1$
in (\ref{eq:Pvoid-x-0-large-x}).
Again, this means that there is no global expansion or contraction of the system. Particles
move on scales of the order of $L(t)$ introduced in (\ref{eq:L-t}), so that in the limit
$q\to\infty$ the relative amplitude of the displacement is negligible and $x/q \to 1$,
which implies that for any interval $q$ the mean $\langle x \rangle_q$ is equal to $q$.

For small Lagrangian interval $q$ at fixed $x$ we obtain
\be
q \to 0 : \;\;\; P_q^{\neq}(x) \simeq q \, n_{\rm void}(x) .
\label{eq:Pq-x-small-q}
\ee
Thus, we obtain a factorized form similar to Eq.(\ref{eq:P-x-q-small-x}) found for the Eulerian
distribution $P_x^{\neq}(q)$ at small $x$.
This now means that for small Lagrangian mass intervals, $q \to 0$, the probability distribution 
of the Eulerian distance $x$ is governed up to order $q$ by the probability to have merged within
a single shock (as described by the contribution $P_{\rm shock}(q) \, \delta_D(x)$) or to contain
one void (as described by the void multiplicity function $n_{\rm void}(x)$).
This is because the voids are discrete (in Lagrangian space) and contain all of the volume.
Again, we can see that the normalizations $\langle x \rangle_q = q$ and
(\ref{eq:nvoid-power-law}) are consistent with the small-$q$ factorized form (\ref{eq:Pq-x-small-q}).

On the other hand, on large scales the Eulerian distribution $P_x^{\neq}(q)$ takes the form
$P_x^{\neq}(q) \simeq f_\infty^{\neq}(|q-x|)$ from Eq.(\ref{eq:P_x_q-large-peak}).
Substituting into Eq.(\ref{eq:P-q-x-def}) gives
\be
q \to \infty , \;\; x \to \infty , \;\; |x-q| \sim 1 : \;\;
P_q^{\neq}(x) \simeq f_{\infty}^{\neq}(|x-q|) \simeq P_x^{\neq}(x) .
\label{eq:P_q_x-large-peak}
\ee
As expected, this again gives a distribution that peaks around the mean $\langle x \rangle_q=q$,
with a fixed width that is set by the typical displacement length of the particles.

We show the regular part $P^{\neq}_{\tilde q}(\tilde x)$ in Fig.~\ref{fig:P_q_x} for the three cases
$\alpha=2.3$, $3.1$ and $\infty$, and for the three scales $\tilde q= 0.2, 1$ and $5$.
For large $\tilde q$ the numerical computation of Eq.(\ref{eq:P-q-x-def}) is not so easy
and it is plagued by compensations between large contributions. Therefore, in the left two panels
of Fig.~\ref{fig:P_q_x} the curve $P^{\neq}_{\tilde q}(\tilde x)$ for the case $\tilde q =5$
corresponds to the simple approximation 
$P^{\neq}_{\tilde q}(\tilde x) \simeq P^{\neq}_{\tilde q}(0) e^{-\tilde x} 
+ P^{\neq}_{\tilde x}(\tilde q)$.
It correctly describes the peak around $\tilde x \simeq q$ and the low-$x$ value, but it may
not reproduce very accurately the intermediate region $1 \lesssim \tilde x \lesssim \tilde q$.
In the right panel we compute numerically the exact expression (\ref{eq:P-q-x-def}) as it is more stable
for $\alpha \to \infty$ because it only involves a double integral, using Eq.(\ref{eq:P_x-q-alpha-infty}).
For $\tilde q =0.2$ and $1$ in all panels, we compute numerically the expression (\ref{eq:P-q-x-void}), which is robust because it is dominated by the first contribution $q \, n_{\rm void}(x)$.
The computation of $P^{\neq}_{q}(0)$ simplifies as two integrations can be performed
analytically so that the numerical computation is reduced to a double integral.
In particular, the resulting expression shows that
\be
P^{\neq}_{q}(0) = \infty \;\;\; \mbox{for} \;\;\; \alpha \leq 2 .
\ee

As seen in Fig.~\ref{fig:P_q_x}, in a fashion similar to the the Eulerian distribution 
$P^{\neq}_{\tilde x}(\tilde q)$ shown in Fig.~\ref{fig:P_x_q}, for large mass intervals, 
$\tilde q \gg 1$, the total weight of the regular part $P^{\neq}_{\tilde q}(\tilde x)$ goes to unity
and the distribution is peaked around its mean $\langle \tilde x \rangle = \tilde q$.
For small mass intervals, $\tilde q \ll 1$, the total weight of the regular part 
$P^{\neq}_{\tilde q}(\tilde x)$ decreases linearly with $\tilde q$ while its characteristic scale
remains at $\tilde x \sim 1$.
It is set by typical size of the voids in the dimensionless units (\ref{eq:re-scaling}).
The main difference between the Eulerian and Lagrangian distributions $P^{\neq}_{\tilde x}(\tilde q)$
and $P^{\neq}_{\tilde q}(\tilde x)$ is that whereas $P^{\neq}_{\tilde x}(\tilde q)$ vanishes linearly
with $\tilde q$ at small $\tilde q$, see Eq.(\ref{eq:P-x-q-asymp}), $P^{\neq}_{\tilde q}(0)$ is nonzero for
$\alpha>2$ and diverges for $\alpha \leq 2$.

\subsection{Multiplicity function of shocks}
\label{sec:shocks}

\begin{figure}
\centering
\includegraphics[height=7cm,width=0.48\textwidth]{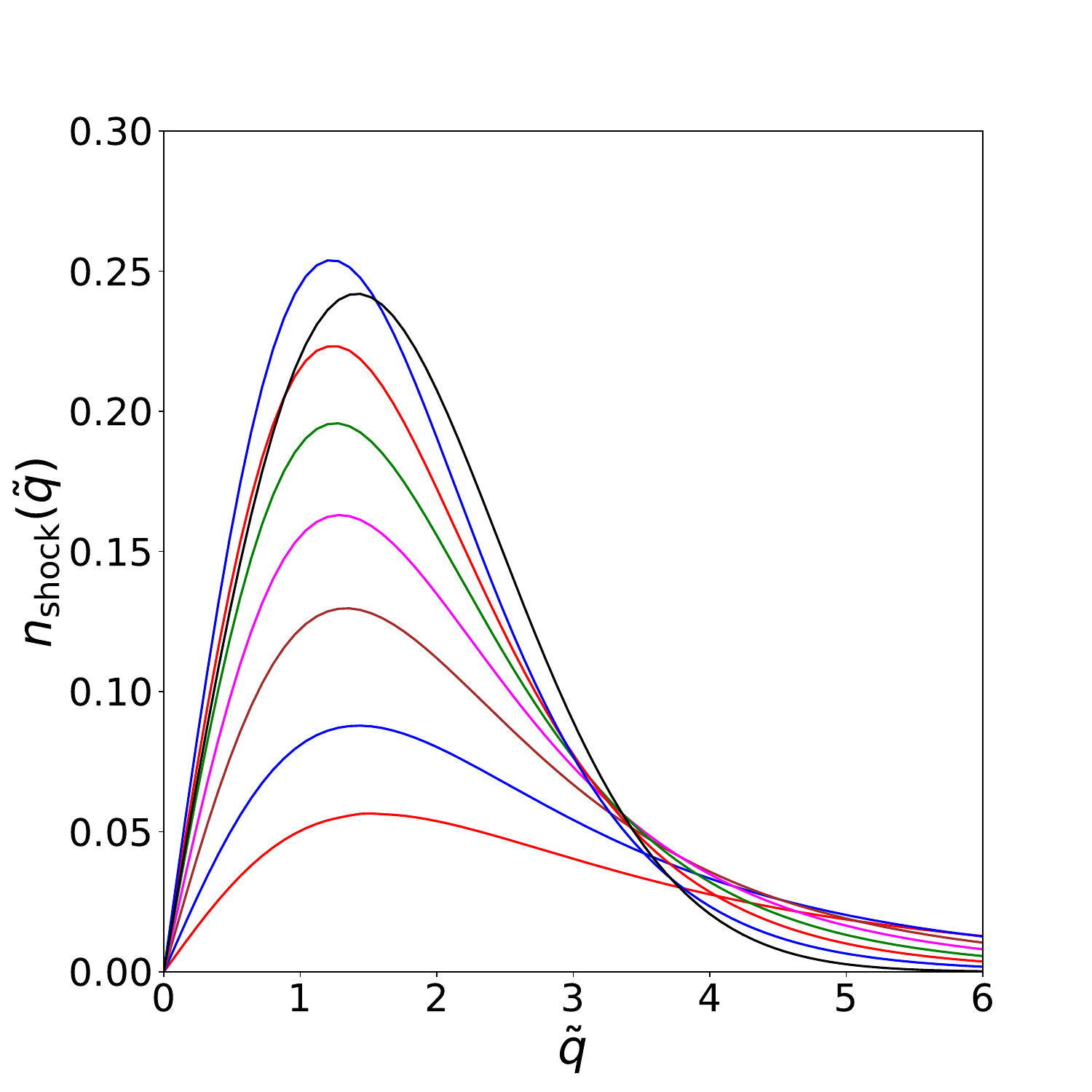}
\includegraphics[height=7cm,width=0.48\textwidth]{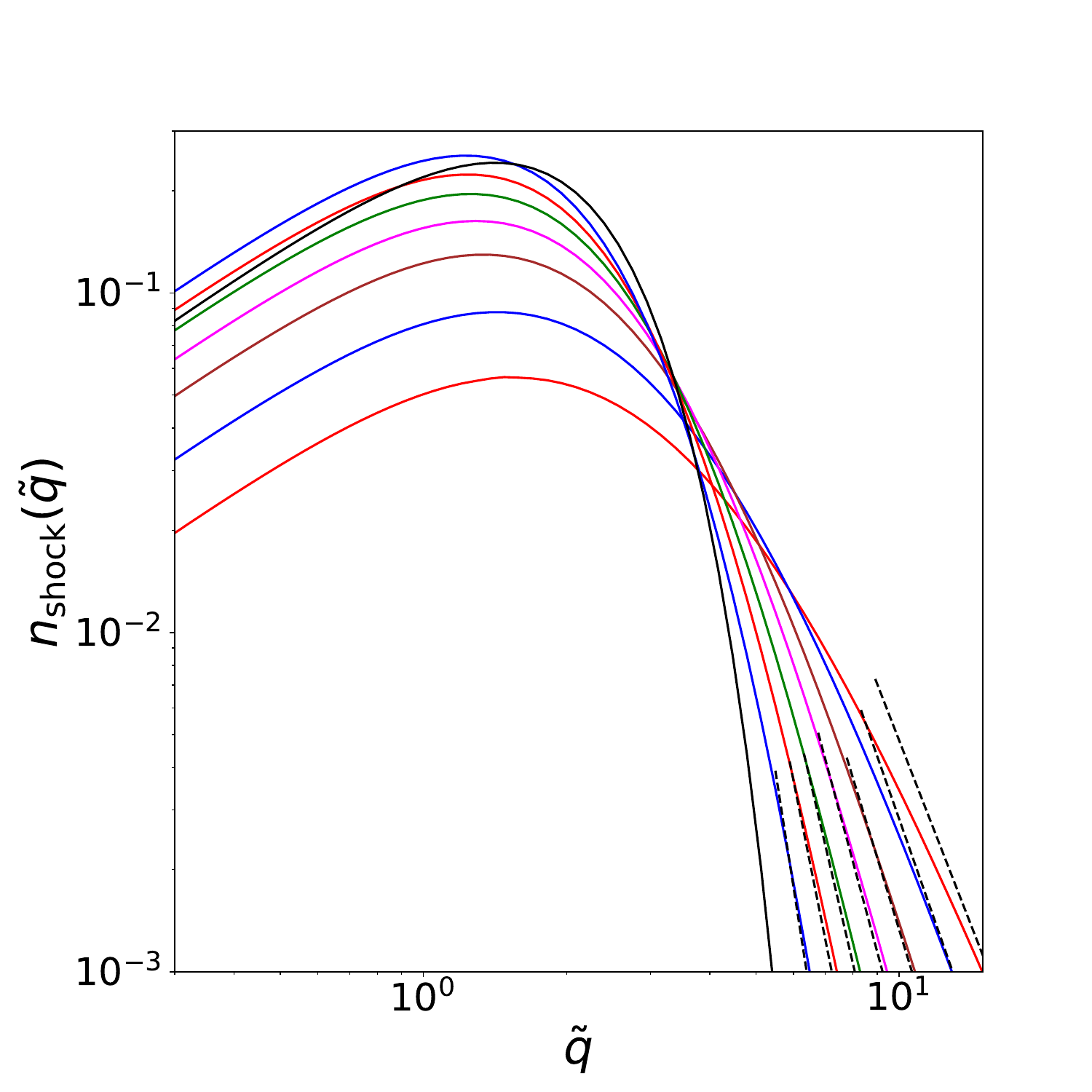}
\caption{
Shock multiplicity function from Eq.(\ref{eq:nq-shocks}) for the cases 
$\alpha=2.3, 2.5, 2.8, 3.1, 3.5, 4, 5, \infty$.
}
\label{fig:nshock}
\end{figure}

The probability $P_{\rm shock}(q)$ for the Lagrangian interval $q$ to belong to a single shock
is obtained from Eq.(\ref{eq:P-q-x-def}) by the normalization to unity of the full probability
distribution $P_q(x)$,
\be
P_{\rm shock}(q) = 1 - \int_0^{\infty} dx P_q^{\neq}(x) = 1 - N_{\rm void} \, q 
+ \left. \frac{\partial}{\partial x} \right|_{x=0} \int_0^q dq' \, P_x^{\neq}(q') (q-q') , 
\ee
where we used Eqs.(\ref{eq:P-q-x-void}) and (\ref{eq:Nvoid}).
On the other hand, $P_{\rm shock}(q)$ is related to the multiplicity function of shocks 
$n_{\rm shock}(q) dq$ per unit length by
\be
P_{\rm shock}(q) = \int_q^{\infty} dq' n_{\rm shock}(q') \, (q'-q) , \;\;\; \mbox{whence} \;\;
n_{\rm shock}(q) = \frac{d^2 P_{\rm shock}}{dq^2} = \left. \frac{\partial}{\partial x} \right|_{x=0} P_x^{\neq}(q) .
\ee
This gives
\be
\int_0^{\infty} dq \, n_{\rm shock}(q) \, q = P_{\rm shock}(0) = 1 ,
\label{eq:nshock-all-mass}
\ee
which means that all the matter is contained in the shocks.
Using the expression (\ref{eq:P-neq-def}) or directly Eq.(\ref{eq:P-x-q-small-x}),
we obtain
\be
n_{\rm shock}(q) =  q \int_0^{\infty} dc \, e^{- \Lambda_{\alpha} c^{3/2-\alpha} }
\int_{-\infty}^{\infty} dq' \left( c + (q'-q/2')^2/2 \right)^{-\alpha} 
\left( c + (q'+q/2)^2/2 \right)^{-\alpha}  ,
\label{eq:nq-shocks}
\ee
and we recover as announced in Section~\ref{sec:Lagrangian-increment} that
the expression (\ref{eq:P-x-q-small-x}) was also the
mass function of the shocks.
This expression was already derived in \cite{Gueudre2014} and
can be directly obtained by writing that the fraction of mass per unit length within
shocks of mass $q$, $q \, n_{\rm shock}(q) dq$, is given by the probability that the Lagrangian point 
$q_0=0$ (or any other point by statistical homogeneity) is located between two simultaneous points
of first contact, $(q_1,\psi_1)$ and $(q_2,\psi_2)$, of a parabola $\cP_{x,c}$,
\be
q \, n_{\rm shock}(q) = \int_{-\infty}^0 dq_1 \int_0^{\infty} dq_2 \, \delta_D(q_2-q_1-q) 
\int d\psi_1 d\psi_2 \, \psi_1^{-\alpha} \psi_2^{-\alpha}
e^{- \Lambda_{\alpha} c^{3/2-\alpha} } ,
\ee
where we used $q_1 \leq 0$, $q_2=q_1+q \geq 0$, and we have
$\psi_1 = c + (q_1-x)^2/2$, $\psi_2 = c + (q_2-x)^2/2$, which determines the parameters
$x$ and $c$ of the first-contact parabola.
Changing integration variables from $(\psi_1,\psi_2)$ to $(x,c)$ and integrating over $q_1$
we recover Eq.(\ref{eq:nq-shocks}).
This gives the asymptotic behaviors
\be
q \to 0 : \;\;\; n_{\rm shock}(q) \simeq  {\cal R}_{2\alpha}(0) \, q , \;\;\;
q \to \infty : \;\;\; n_{\rm shock}(q) \simeq 2^{1+\alpha} q^{1-2\alpha} .
\label{eq:nshock-asymptotic}
\ee
This implies for the probability of a shock,
\be
q \to 0 : \;\;\; P_{\rm shock}(q) \simeq 1 - N_{\rm void} \, q , \;\;\;
q \to \infty : \;\;\; P_{\rm shock}(q) \simeq  \frac{2^{\alpha}}{(\alpha-1) (2\alpha-3)} q^{3-2\alpha} .
\ee

We show the shock multiplicity function in Fig.~\ref{fig:nshock}.
Again the large-mass tail is steeper for larger $\alpha$ and converges to a Gaussian falloff
in the limit $\alpha \to \infty$ as in Eq.(\ref{eq:nshock-alpha-infty}).
It goes to zero linearly in $\tilde q$, as in (\ref{eq:nshock-asymptotic}), and
it peaks at the typical mass scale $\tilde q \sim 1$ in the rescaled units.

\subsection{Higher-order distributions}

As for the two-point distribution (\ref{eq:P-twopoint-Lag}), we can relate the higher-order Lagrangian
and Eulerian distributions as
\be
P_{q_1,\cdots,q_n}( \geq x_1,\cdots, \geq x_n) = P_{x_1, \cdots , x_n}( \leq q_1, \cdots , \leq q_n) .
\ee
However, because of the complicated structure of the Eulerian distributions (\ref{eq:Pn-prod}),
with the auxiliary variables $c_i$ and correlated increments, this does not lead to simple
factorized expressions for the Lagrangian distributions
$P_{q_1,\cdots,q_n}(x_1,\cdots, x_n)$.

\section{Limit $\alpha \to \infty$}
\label{sec:alpha-infty}

We consider in this Section the limit $\alpha \to \infty$ of the system defined in 
Section~\ref{sec:Initial-condition}. As noticed in Section~\ref{sec:alpha-infty-rescale}, in this limit
the slope of the Poisson intensity (\ref{eq:lambda-psi}) becomes infinitely steep so that all first-contact
parabolas have $c \simeq 1$, whereas the typical displacements scale with a factor $1/\sqrt{\alpha}$.
Therefore, to obtain the limit $\alpha \to \infty$ of the probability distributions obtained in the previous
Sections we make the changes of variable (\ref{eq:tilde-def}) and $c=1+u/\alpha$ in the dummy
integration variables.
As announced in Section~\ref{sec:alpha-infty-rescale}, we recover the classical results obtained
at late times for Gaussian initial conditions with vanishing large-scale power 
\cite{Kida1979,Gurbatov1981,Gurbatov1991}, $E_0(k) \propto k^n$ with $n>1$ \cite{Gurbatov1997}.

We first consider the Eulerian distributions.
Making these changes of variable in Eq.(\ref{eq:P_0-q}), 
we obtain in the limit  $\alpha \to \infty$ for the probability distribution $P_0(\tilde q) d\tilde q$
the finite result
\be
\alpha \to \infty : \;\;\; P_0(\tilde q) = \frac{1}{\sqrt{2\pi}} e^{-\tilde q^2/2} , \;\;\;
\langle \tilde q^2 \rangle = 1 .
\label{eq:P0-q-alpha-inf}
\ee
Thus, in the limit  $\alpha \to \infty$ the power-law tail (\ref{eq:P0-large-q}) steepens to 
a Gaussian tail and we recover the Gaussian velocity distribution that is also obtained
at late times for Gaussian initial conditions with vanishing large-scale power
\cite{Kida1979,Gurbatov1981}.
As we shall see below, this convergence to a Gaussian cutoff also applies to the tails of other 
probability distributions, although the Gaussian falloff can be multiplied by a power-law prefactor.
From Eq.(\ref{eq:Pvoid}), we obtain for the void probability distribution
\be
P_{\rm void}(\tilde x) = \int_{-\infty}^{\infty} \frac{d \tilde{q}'_\star}
{{\cal J}(\tilde{q}'_\star,\tilde x)} ,
\ee
where we defined
\be
{\cal J}(\tilde{q}'_\star,\tilde x) = \sqrt{\frac{\pi}{2}} \left[ e^{(\tilde{q}'_\star+\tilde x/2)^2/2}
{\rm erfc}\left( - \frac{\tilde{q}'_\star+\tilde x/2}{\sqrt{2}} \right) + e^{(\tilde{q}'_\star-\tilde x/2)^2/2}
{\rm erfc}\left( \frac{\tilde{q}'_\star-\tilde x/2}{\sqrt{2}} \right) \right] , \;\;\;
{\cal J}(\tilde{q}'_\star,0) = \sqrt{2\pi} e^{ \tilde{q}'_\star \!^2/2} .
\label{eq:J-qs-x-def}
\ee
This again agrees with the case of Gaussian initial conditions without large-scale power
\cite{Kida1979,Gurbatov1991}.
This gives the asymptotic behaviors
\be
\tilde x \ll 1 : \;\; P_{\rm void}(\tilde x) = 1 - \frac{\tilde x}{\sqrt{\pi}} + \dots , \;\;\; \mbox{and for} \;\;
\tilde x \gg 1 : \;\; P_{\rm void}(\tilde x) \simeq \frac{\sqrt{\pi}}{\sqrt{2} \tilde x} \, e^{-\tilde x^2/8} , 
\;\; n_{\rm void}(\tilde x) \simeq \frac{\sqrt{\pi} \tilde x}{16 \sqrt{2}} e^{-\tilde x^2/8} ,
\label{eq:Pvoid-alpha-infty-asymp}
\ee
where we again find a Gaussian falloff, with power-law prefactors.
Because in the limit $\alpha \to \infty$ there is no difference between ${\cal R}_{\alpha}$ and
${\cal R}_{\alpha-1}$, Eq.(\ref{eq:Bv-deriv}) leads to
\be
\tilde x \geq 0 : \; B_{\tilde v}(\tilde x) = \frac{d}{d\tilde x} \left[ \tilde x P_{\rm void}(\tilde x) \right] ,
\;\;\;  E(\tilde k) = \int_0^{\infty} \frac{d\tilde x}{\pi} B_{\tilde v}(\tilde x) \cos(\tilde k \tilde x) 
= \frac{\tilde k}{\pi} \int_0^{\infty} d\tilde x \, \tilde x P_{\rm void}(\tilde x) 
\sin(\tilde k \tilde x) ,
\label{eq:Ek-inf-def}
\ee
where we introduced the rescaled quantities $\tilde v= \sqrt{\alpha} v$ and 
$\tilde k = k/\sqrt{\alpha}$.
This again agrees with the case of Gaussian initial conditions without large-scale power
\cite{Gurbatov1981,Gurbatov1991}.
This gives the large-scale behaviors
\be
| \tilde x | \gg 1 : \;\; B_{\tilde v}(\tilde x) \simeq - \frac{\sqrt{2\pi}}{8} \tilde x \, 
e^{-\tilde x^2/8} , \;\;\;
| \tilde k | \ll 1 : \;\; E(\tilde k) \simeq \frac{\tilde k^2}{\pi} \int_0^{\infty} d\tilde x \, 
\tilde x^2 P_{\rm void}(\tilde x) \propto \tilde k^2 ,
\label{eq:Bv-asymp-alpha-infty}
\ee
and the small-scale asymptotics
\be
| \tilde x | \ll 1 : \;\; B_{\tilde v}(\tilde x) = 1 - \frac{2 |\tilde x|}{\sqrt{\pi}} + \dots ,
\;\; | \tilde k | \gg 1 : \;\; E(\tilde k) \simeq \frac{2}{\pi^{3/2} \tilde k^2} .
\ee

From Eq. (\ref{eq:P-neq-def}), the probability distribution of the Lagrangian increment becomes
\be
P_{\tilde x}^{\neq}(\tilde q) = \frac{\sqrt{\pi}}{2} \tilde x e^{\tilde x \tilde q/2 - \tilde q^2/4} 
\int_{-\infty}^{\infty} d\tilde q'_\star \frac{e^{\tilde q_\star' \! ^2}}{{\cal J}(\tilde q '_\star, \tilde x)^2} 
\left[ {\rm erfc}(\tilde q'_\star-\tilde q/2) - {\rm erfc}(\tilde q'_\star+\tilde q/2) \right] ,
\label{eq:P_x-q-alpha-infty}
\ee
where the function ${\cal J}$ was introduced in Eq.(\ref{eq:J-qs-x-def}).
This gives the small-$\tilde q$ and large-$\tilde q$ behaviors
\be
\tilde q \to 0 : \;\; P_{\tilde x}^{\neq}(\tilde q) = \tilde x \tilde q \int_{-\infty}^{\infty} 
\frac{d\tilde q'_\star}{{\cal J}^2} , \;\;\;\;
\tilde q \to \infty : \;\; P_{\tilde x}^{\neq}(\tilde q) = \sqrt{\pi} \tilde x e^{-(\tilde q - \tilde x)^2/4 + \tilde x^2/4} 
\int_{-\infty}^{\infty} d\tilde q'_\star \frac{e^{\tilde q_\star' \! ^2}}{{\cal J}^2} .
\label{eq:P_x-q-alpha-infty-asymp-q}
\ee
We recover the linear slope at low $\tilde q$ of Eq.(\ref{eq:P-x-q-asymp}),
as this linear exponent does not depend on $\alpha$, while the large-distance power-law falloff
again turns into a Gaussian.
At large distances $\tilde x$ and $\tilde q$ but fixed $\tilde q - \tilde x$, we obtain from either
(\ref{eq:P_x-q-alpha-infty-asymp-q}) or (\ref{eq:P_x_q-large-peak}) the Gaussian peak
\be
\tilde x \to \infty , \;\; \tilde q \to \infty , \;\; | \tilde q - \tilde x | \sim 1 : \;\;
P_{\tilde x}^{\neq}(\tilde q) \simeq \frac{1}{2\sqrt{\pi}} e^{-(\tilde q - \tilde x)^2/4} .
\label{eq:P_x_q-large-peak-alpha-infty}
\ee
whereas for small intervals $\tilde x$ we obtain
\be
\tilde x \to 0 : \;\; P_{\tilde x}^{\neq}(\tilde q) = \frac{\tilde x \tilde q}{2\sqrt{\pi}} 
e^{-\tilde q^2/4} .
\label{eq:Px-q-small-x-alpha-infty}
\ee

For the density correlation and power spectrum we obtain
\be
\xi_{\delta}(\tilde x) = \frac{4}{\sqrt{\pi}} \delta_D(\tilde x) + \xi^{\neq}_{\delta}(\tilde x) , \;\; \mbox{with for} \;
x > 0 : \; \xi^{\neq}_{\delta}(\tilde x) = - \frac{d^3}{d\tilde x^3} \left[ \tilde x P_{\rm void}(\tilde x) \right] ,
\ee
and
\be
| \tilde x | \gg 1 : \;\; \xi_{\delta}(\tilde x) \simeq \frac{\sqrt{2\pi}}{128} \tilde x^3 
e^{-\tilde x^2/8} , \;\;\; | \tilde k | \ll 1 : \;\; P_{\delta}(\tilde k) \propto \tilde k^4 , \;\;\;
| \tilde k | \gg 1 : \;\; P_{\delta}(\tilde k) \simeq \frac{2}{\pi^{3/2}} .
\ee
The probability distribution of the density $\rho=\tilde q/\tilde x$ is again obtained from the
probability distribution of the Lagrangian increment $P_{\tilde x}(\tilde q)$.
From Eq.(\ref{eq:P_x-q-alpha-infty-asymp-q}), this gives a linear slope for $\rho \to 0$
and a Gaussian cutoff for $\rho \to \infty$.
On the other hand, from (\ref{eq:P_x_q-large-peak-alpha-infty}) we obtain a Gaussian distribution
for the density contrast on large scales,
\be
\tilde x \to \infty , \;\; | \delta | \propto 1/\tilde x : \;\; P_{\tilde x}^{\neq}(\delta) \simeq 
\frac{\tilde x}{2\sqrt{\pi}} e^{- \tilde x^2 \delta^2/4} , \;\; \mbox{and for} \; \tilde x \to \infty : \; 
\langle \delta^2 \rangle_{\tilde x} = 2 / \tilde x^2 .
\label{eq:P_x_rho-large-peak-alpha-infty}
\ee

We now consider the Lagrangian distributions.
The relations (\ref{eq:P-q-x-def})-(\ref{eq:P-q-x-void}) still apply, in terms of the rescaled
coordinates $\tilde q$ and $\tilde x$.
The second derivative $\partial^2 P_{\tilde x}^{\neq}/\partial\tilde x^2$ can be computed from
Eq.(\ref{eq:P_x-q-alpha-infty}).
This gives the small and large distance behaviors
\be
\tilde q \to 0 : \;\; P_{\tilde q}^{\neq}(\tilde x) \simeq \tilde q \, n_{\rm void}(\tilde x) , \;\;\;
\tilde q \to \infty , \;\; \tilde x \to \infty , \;\; | \tilde x - \tilde q | \sim 1 : \;\;
P_{\tilde q}^{\neq}(\tilde x) \simeq \frac{1}{2\sqrt{\pi}} e^{-(\tilde x - \tilde q)^2/4} .
\label{eq:P_q_x-asymp-alpha-infty}
\ee
\
The multiplicity function of shocks and the shock probability read as
\be
n_{\rm shock}(\tilde q) = \frac{\tilde q}{2\sqrt{\pi}} e^{-\tilde q^2/4} , \;\;\;
P_{\rm shock}(\tilde q) = {\rm erfc}(\tilde q/2) ,
\label{eq:nshock-alpha-infty}
\ee
in agreement with Eq.(\ref{eq:Px-q-small-x-alpha-infty}).
Again this result agrees with the case of Gaussian initial conditions without large-scale power
\cite{Kida1979,Gurbatov1991}.

\section{Conclusion}
\label{sec:conclusion}

In this paper, we have revisited the case where the initial velocity potential of the
Burgers equation is given by a Poisson point process.
For a power-law Poisson intensity, the dynamics are statistically self-similar and fully
controlled by a single exponent $\alpha$, which determines the heaviness of the tails in both
the initial and final (i.e. at any time $t>0$) probability distributions. 
When $\alpha$ approaches its lower bound, $\alpha \to 3/2$, the dynamics are dominated by very rare
but extremely high initial peaks, leading to widely separated shocks, large voids, and slowly 
decaying correlations reminiscent of strongly intermittent turbulence.
Increasing $\alpha$ suppresses these extremes, giving a denser shock network and steeper correlation 
falloffs. In the limit $\alpha \to \infty$, the power-law tails steepen to Gaussian falloffs
(with power-law multiplicative factors) and we recover the spatial distributions obtained
in the classical study \cite{Kida1979} at late times for Gaussian initial conditions with
vanishing large-scale power.

For all values of $\alpha$, all mass is concentrated in shocks, while the Eulerian space
is filled by voids.
This dominance of shocks and voids produces characteristic mixed statistical signatures: probability
distributions consist of a Dirac contribution (for empty voids or vanishing-size shocks)
plus a regular contribution with power-law tails.
Velocity correlations exhibit a cusp nonanalyticity at the origin, associated with the finite
density of shocks. This leads to the universal $k^{-2}$ asymptote for the energy power spectrum
at high wavenumbers \cite{Frisch2001}. 
Because of the power-law tails, velocity and displacement moments diverge beyond a finite order,
when rare events dominate, revealing a transition between typical and extreme-event controlled regimes.
 
The analytical solvability of this model enables us to derive exact expressions for the
one- and two-point probability distributions of the velocity and displacement, as well as for
the multiplicity functions of shocks and voids.
This work presents an extension of the family of solvable self-similar Burgers systems and
provides an explicit example where broad-tailed initial disorder shapes the nonlinear structure 
formation and leads to heavy-tailed statistics.
It offers a simple benchmark for studying universality classes in nonlinear dynamics
or aggregation phenomena \cite{Gurbatov1999} and for testing approximation schemes
\cite{Kraichnan1968}.

This work could be extended to higher dimensions \cite{Gueudre2014}. 
Although this is straightforward on the conceptual level, as the combination of the geometrical
interpretation and of the Poisson point process still allows explicit derivations, numerical computations
of the higher-dimensional integrals may be more intricate.
Another direction would be to include external random forcing and study the interplay between
the initial conditions and this new source of stochasticity.
These extensions are left for future works.



 



\bibliography{ref}

\begin{thebibliography}{37}%
\makeatletter
\providecommand \@ifxundefined [1]{%
 \@ifx{#1\undefined}
}%
\providecommand \@ifnum [1]{%
 \ifnum #1\expandafter \@firstoftwo
 \else \expandafter \@secondoftwo
 \fi
}%
\providecommand \@ifx [1]{%
 \ifx #1\expandafter \@firstoftwo
 \else \expandafter \@secondoftwo
 \fi
}%
\providecommand \natexlab [1]{#1}%
\providecommand \enquote  [1]{``#1''}%
\providecommand \bibnamefont  [1]{#1}%
\providecommand \bibfnamefont [1]{#1}%
\providecommand \citenamefont [1]{#1}%
\providecommand \href@noop [0]{\@secondoftwo}%
\providecommand \href [0]{\begingroup \@sanitize@url \@href}%
\providecommand \@href[1]{\@@startlink{#1}\@@href}%
\providecommand \@@href[1]{\endgroup#1\@@endlink}%
\providecommand \@sanitize@url [0]{\catcode `\\12\catcode `\$12\catcode
  `\&12\catcode `\#12\catcode `\^12\catcode `\_12\catcode `\%12\relax}%
\providecommand \@@startlink[1]{}%
\providecommand \@@endlink[0]{}%
\providecommand \url  [0]{\begingroup\@sanitize@url \@url }%
\providecommand \@url [1]{\endgroup\@href {#1}{\urlprefix }}%
\providecommand \urlprefix  [0]{URL }%
\providecommand \Eprint [0]{\href }%
\providecommand \doibase [0]{https://doi.org/}%
\providecommand \selectlanguage [0]{\@gobble}%
\providecommand \bibinfo  [0]{\@secondoftwo}%
\providecommand \bibfield  [0]{\@secondoftwo}%
\providecommand \translation [1]{[#1]}%
\providecommand \BibitemOpen [0]{}%
\providecommand \bibitemStop [0]{}%
\providecommand \bibitemNoStop [0]{.\EOS\space}%
\providecommand \EOS [0]{\spacefactor3000\relax}%
\providecommand \BibitemShut  [1]{\csname bibitem#1\endcsname}%
\let\auto@bib@innerbib\@empty
\bibitem [{\citenamefont {Burgers}(1974)}]{Burgers1974}%
  \BibitemOpen
  \bibfield  {author} {\bibinfo {author} {\bibfnamefont {J.~M.}\ \bibnamefont
  {Burgers}},\ }\href {https://doi.org/10.1007/978-94-010-1745-9} {\emph
  {\bibinfo {title} {The Nonlinear Diffusion Equation}}}\ (\bibinfo
  {publisher} {Springer Netherlands},\ \bibinfo {year} {1974})\BibitemShut
  {NoStop}%
\bibitem [{\citenamefont {Hopf}(1950)}]{Hopf1950}%
  \BibitemOpen
  \bibfield  {author} {\bibinfo {author} {\bibfnamefont {E.}~\bibnamefont
  {Hopf}},\ }\bibfield  {title} {\bibinfo {title} {{The partial differential
  equation ut + uux = $\mu$xx}},\ }\href
  {https://doi.org/10.1002/CPA.3160030302} {\bibfield  {journal} {\bibinfo
  {journal} {Communications on Pure and Applied Mathematics}\ }\textbf
  {\bibinfo {volume} {3}},\ \bibinfo {pages} {201} (\bibinfo {year}
  {1950})}\BibitemShut {NoStop}%
\bibitem [{\citenamefont {Cole}(1951)}]{Cole1951}%
  \BibitemOpen
  \bibfield  {author} {\bibinfo {author} {\bibfnamefont {J.~D.}\ \bibnamefont
  {Cole}},\ }\bibfield  {title} {\bibinfo {title} {{On a quasi-linear parabolic
  equation occurring in aerodynamics}},\ }\href
  {https://doi.org/10.1090/QAM/42889} {\bibfield  {journal} {\bibinfo
  {journal} {Quarterly of Applied Mathematics}\ }\textbf {\bibinfo {volume}
  {9}},\ \bibinfo {pages} {225} (\bibinfo {year} {1951})}\BibitemShut {NoStop}%
\bibitem [{\citenamefont {Gurbatov}\ \emph {et~al.}(1981)\citenamefont
  {Gurbatov}, \citenamefont {Saichev}, \citenamefont {Gurbatov},\ and\
  \citenamefont {Saichev}}]{Gurbatov1981}%
  \BibitemOpen
  \bibfield  {author} {\bibinfo {author} {\bibfnamefont {S.~N.}\ \bibnamefont
  {Gurbatov}}, \bibinfo {author} {\bibfnamefont {A.~I.}\ \bibnamefont
  {Saichev}}, \bibinfo {author} {\bibfnamefont {S.~N.}\ \bibnamefont
  {Gurbatov}},\ and\ \bibinfo {author} {\bibfnamefont {A.~I.}\ \bibnamefont
  {Saichev}},\ }\bibfield  {title} {\bibinfo {title} {{Degeneracy of
  one-dimensional acoustic turbulence under large Reynolds numbers}},\ }\href
  {https://ui.adsabs.harvard.edu/abs/1981ZhETF..80..689G/abstract} {\bibfield
  {journal} {\bibinfo  {journal} {ZhETF}\ }\textbf {\bibinfo {volume} {80}},\
  \bibinfo {pages} {689} (\bibinfo {year} {1981})}\BibitemShut {NoStop}%
\bibitem [{\citenamefont {Whitham}(1999)}]{Whitham1999}%
  \BibitemOpen
  \bibfield  {author} {\bibinfo {author} {\bibfnamefont {G.~B.}\ \bibnamefont
  {Whitham}},\ }\href {https://doi.org/10.1002/9781118032954} {\emph {\bibinfo
  {title} {{Linear and Nonlinear Waves}}}}\ (\bibinfo  {publisher} {wiley},\
  \bibinfo {year} {1999})\BibitemShut {NoStop}%
\bibitem [{\citenamefont {Frisch}\ and\ \citenamefont
  {Bec}(2001)}]{Frisch2001}%
  \BibitemOpen
  \bibfield  {author} {\bibinfo {author} {\bibfnamefont {U.}~\bibnamefont
  {Frisch}}\ and\ \bibinfo {author} {\bibfnamefont {J.}~\bibnamefont {Bec}},\
  }\href {https://doi.org/10.1007/3-540-45674-0_7} {\emph {\bibinfo {title}
  {{New trends in turbulence Turbulence: nouveaux aspects}}}}\ (\bibinfo
  {publisher} {Springer, Berlin, Heidelberg},\ \bibinfo {year} {2001})\ pp.\
  \bibinfo {pages} {341--383},\ \Eprint {https://arxiv.org/abs/0012033}
  {arXiv:0012033 [nlin]} \BibitemShut {NoStop}%
\bibitem [{\citenamefont {Bec}\ and\ \citenamefont {Khanin}(2007)}]{Bec2007}%
  \BibitemOpen
  \bibfield  {author} {\bibinfo {author} {\bibfnamefont {J.}~\bibnamefont
  {Bec}}\ and\ \bibinfo {author} {\bibfnamefont {K.}~\bibnamefont {Khanin}},\
  }\bibfield  {title} {\bibinfo {title} {{Burgers turbulence}},\ }\href
  {https://doi.org/10.1016/j.physrep.2007.04.002} {\bibfield  {journal}
  {\bibinfo  {journal} {Physics Reports}\ }\textbf {\bibinfo {volume} {447}},\
  \bibinfo {pages} {1} (\bibinfo {year} {2007})}\BibitemShut {NoStop}%
\bibitem [{\citenamefont {Kraichnan}(1968)}]{Kraichnan1968}%
  \BibitemOpen
  \bibfield  {author} {\bibinfo {author} {\bibfnamefont {R.~H.}\ \bibnamefont
  {Kraichnan}},\ }\bibfield  {title} {\bibinfo {title} {{Lagrangian‐History
  Statistical Theory for Burgers' Equation}},\ }\href
  {https://doi.org/10.1063/1.1691900} {\bibfield  {journal} {\bibinfo
  {journal} {The Physics of Fluids}\ }\textbf {\bibinfo {volume} {11}},\
  \bibinfo {pages} {265} (\bibinfo {year} {1968})}\BibitemShut {NoStop}%
\bibitem [{\citenamefont {Kida}(1979)}]{Kida1979}%
  \BibitemOpen
  \bibfield  {author} {\bibinfo {author} {\bibfnamefont {S.}~\bibnamefont
  {Kida}},\ }\bibfield  {title} {\bibinfo {title} {{Asymptotic properties of
  Burgers turbulence}},\ }\href {https://doi.org/10.1017/S0022112079001932}
  {\bibfield  {journal} {\bibinfo  {journal} {Journal of Fluid Mechanics}\
  }\textbf {\bibinfo {volume} {93}},\ \bibinfo {pages} {337} (\bibinfo {year}
  {1979})}\BibitemShut {NoStop}%
\bibitem [{\citenamefont {Frachebourg}\ \emph {et~al.}(2000)\citenamefont
  {Frachebourg}, \citenamefont {Martin},\ and\ \citenamefont
  {Piasecki}}]{Frachebourg2000}%
  \BibitemOpen
  \bibfield  {author} {\bibinfo {author} {\bibfnamefont {L.}~\bibnamefont
  {Frachebourg}}, \bibinfo {author} {\bibfnamefont {P.~A.}\ \bibnamefont
  {Martin}},\ and\ \bibinfo {author} {\bibfnamefont {J.}~\bibnamefont
  {Piasecki}},\ }\bibfield  {title} {\bibinfo {title} {{Ballistic aggregation:
  a solvable model of irreversible many particles dynamics}},\ }\href
  {https://doi.org/10.1016/S0378-4371(99)00585-3} {\bibfield  {journal}
  {\bibinfo  {journal} {Physica A: Statistical Mechanics and its Applications}\
  }\textbf {\bibinfo {volume} {279}},\ \bibinfo {pages} {69} (\bibinfo {year}
  {2000})}\BibitemShut {NoStop}%
\bibitem [{\citenamefont {Valageas}(2009{\natexlab{a}})}]{Valageas2009}%
  \BibitemOpen
  \bibfield  {author} {\bibinfo {author} {\bibfnamefont {P.}~\bibnamefont
  {Valageas}},\ }\bibfield  {title} {\bibinfo {title} {{Ballistic aggregation
  for one-sided Brownian initial velocity}},\ }\href
  {https://doi.org/10.1016/J.PHYSA.2008.12.033} {\bibfield  {journal} {\bibinfo
   {journal} {Physica A: Statistical Mechanics and its Applications}\ }\textbf
  {\bibinfo {volume} {388}},\ \bibinfo {pages} {1031} (\bibinfo {year}
  {2009}{\natexlab{a}})}\BibitemShut {NoStop}%
\bibitem [{\citenamefont {Zel'dovich}(1970)}]{Zeldovich1970}%
  \BibitemOpen
  \bibfield  {author} {\bibinfo {author} {\bibfnamefont {Y.~B.}\ \bibnamefont
  {Zel'dovich}},\ }\bibfield  {title} {\bibinfo {title} {{Gravitational
  instability: An approximate theory for large density perturbations.}},\
  }\href {https://ui.adsabs.harvard.edu/abs/1970A&A.....5...84Z/abstract}
  {\bibfield  {journal} {\bibinfo  {journal} {A\&A}\ }\textbf {\bibinfo
  {volume} {5}},\ \bibinfo {pages} {84} (\bibinfo {year} {1970})}\BibitemShut
  {NoStop}%
\bibitem [{\citenamefont {Shandarin}\ and\ \citenamefont
  {Zeldovich}(1989)}]{Shandarin1989}%
  \BibitemOpen
  \bibfield  {author} {\bibinfo {author} {\bibfnamefont {S.~F.}\ \bibnamefont
  {Shandarin}}\ and\ \bibinfo {author} {\bibfnamefont {Y.~B.}\ \bibnamefont
  {Zeldovich}},\ }\bibfield  {title} {\bibinfo {title} {{The large-scale
  structure of the universe: Turbulence, intermittency, structures in a
  self-gravitating medium}},\ }\href
  {https://doi.org/10.1103/RevModPhys.61.185} {\bibfield  {journal} {\bibinfo
  {journal} {Reviews of Modern Physics}\ }\textbf {\bibinfo {volume} {61}},\
  \bibinfo {pages} {185} (\bibinfo {year} {1989})}\BibitemShut {NoStop}%
\bibitem [{\citenamefont {Gurbatov}\ \emph {et~al.}(1989)\citenamefont
  {Gurbatov}, \citenamefont {Saichev}, \citenamefont {Shandarin}, \citenamefont
  {Gurbatov}, \citenamefont {Saichev},\ and\ \citenamefont
  {Shandarin}}]{Gurbatov1989}%
  \BibitemOpen
  \bibfield  {author} {\bibinfo {author} {\bibfnamefont {S.~N.}\ \bibnamefont
  {Gurbatov}}, \bibinfo {author} {\bibfnamefont {A.~I.}\ \bibnamefont
  {Saichev}}, \bibinfo {author} {\bibfnamefont {S.~F.}\ \bibnamefont
  {Shandarin}}, \bibinfo {author} {\bibfnamefont {S.~N.}\ \bibnamefont
  {Gurbatov}}, \bibinfo {author} {\bibfnamefont {A.~I.}\ \bibnamefont
  {Saichev}},\ and\ \bibinfo {author} {\bibfnamefont {S.~F.}\ \bibnamefont
  {Shandarin}},\ }\bibfield  {title} {\bibinfo {title} {{The large-scale
  structure of the universe in the frame of the model equation of non-linear
  diffusion}},\ }\href {https://doi.org/10.1093/mnras/236.2.385} {\bibfield
  {journal} {\bibinfo  {journal} {MNRAS}\ }\textbf {\bibinfo {volume} {236}},\
  \bibinfo {pages} {385} (\bibinfo {year} {1989})}\BibitemShut {NoStop}%
\bibitem [{\citenamefont {Gurbatov}\ \emph {et~al.}(1991)\citenamefont
  {Gurbatov}, \citenamefont {Malakhov},\ and\ \citenamefont
  {Saichev}}]{Gurbatov1991}%
  \BibitemOpen
  \bibfield  {author} {\bibinfo {author} {\bibfnamefont {S.}~\bibnamefont
  {Gurbatov}}, \bibinfo {author} {\bibfnamefont {A.}~\bibnamefont {Malakhov}},\
  and\ \bibinfo {author} {\bibfnamefont {A.~I.}\ \bibnamefont {Saichev}},\
  }\href {https://search.worldcat.org/title/23731964} {\emph {\bibinfo {title}
  {{Nonlinear random waves and turbulence in nondispersive media : waves, rays,
  particles}}}}\ (\bibinfo  {publisher} {Manchester University Press ;
  Distributed exclusively in the US and Canada by St. Martin's Press},\
  \bibinfo {year} {1991})\BibitemShut {NoStop}%
\bibitem [{\citenamefont {Vergassola}\ \emph {et~al.}(1994)\citenamefont
  {Vergassola}, \citenamefont {Dubrulle}, \citenamefont {Frisch}, \citenamefont
  {Noullez}, \citenamefont {Vergassola}, \citenamefont {Dubrulle},
  \citenamefont {Frisch},\ and\ \citenamefont {Noullez}}]{Vergassola1994}%
  \BibitemOpen
  \bibfield  {author} {\bibinfo {author} {\bibfnamefont {M.}~\bibnamefont
  {Vergassola}}, \bibinfo {author} {\bibfnamefont {B.}~\bibnamefont
  {Dubrulle}}, \bibinfo {author} {\bibfnamefont {U.}~\bibnamefont {Frisch}},
  \bibinfo {author} {\bibfnamefont {A.}~\bibnamefont {Noullez}}, \bibinfo
  {author} {\bibfnamefont {M.}~\bibnamefont {Vergassola}}, \bibinfo {author}
  {\bibfnamefont {B.}~\bibnamefont {Dubrulle}}, \bibinfo {author}
  {\bibfnamefont {U.}~\bibnamefont {Frisch}},\ and\ \bibinfo {author}
  {\bibfnamefont {A.}~\bibnamefont {Noullez}},\ }\bibfield  {title} {\bibinfo
  {title} {{Burgers' equation, Devil's staircases and the mass distribution for
  large-scale structures.}},\ }\href
  {https://ui.adsabs.harvard.edu/abs/1994A&A...289..325V/abstract} {\bibfield
  {journal} {\bibinfo  {journal} {A\&A}\ }\textbf {\bibinfo {volume} {289}},\
  \bibinfo {pages} {325} (\bibinfo {year} {1994})}\BibitemShut {NoStop}%
\bibitem [{\citenamefont {Valageas}\ and\ \citenamefont
  {Bernardeau}(2011)}]{Valageas2011}%
  \BibitemOpen
  \bibfield  {author} {\bibinfo {author} {\bibfnamefont {P.}~\bibnamefont
  {Valageas}}\ and\ \bibinfo {author} {\bibfnamefont {F.}~\bibnamefont
  {Bernardeau}},\ }\bibfield  {title} {\bibinfo {title} {{Density fields and
  halo mass functions in the geometrical adhesion toy model}},\ }\href
  {https://doi.org/10.1103/PhysRevD.83.043508} {\bibfield  {journal} {\bibinfo
  {journal} {Physical Review D}\ }\textbf {\bibinfo {volume} {83}},\ \bibinfo
  {pages} {043508} (\bibinfo {year} {2011})},\ \Eprint
  {https://arxiv.org/abs/1009.1974} {arXiv:1009.1974} \BibitemShut {NoStop}%
\bibitem [{\citenamefont {Melott}\ \emph {et~al.}(1994)\citenamefont {Melott},
  \citenamefont {Shandarin}, \citenamefont {Weinberg}, \citenamefont {Melott},
  \citenamefont {Shandarin},\ and\ \citenamefont {Weinberg}}]{Melott1994}%
  \BibitemOpen
  \bibfield  {author} {\bibinfo {author} {\bibfnamefont {A.~L.}\ \bibnamefont
  {Melott}}, \bibinfo {author} {\bibfnamefont {S.~F.}\ \bibnamefont
  {Shandarin}}, \bibinfo {author} {\bibfnamefont {D.~H.}\ \bibnamefont
  {Weinberg}}, \bibinfo {author} {\bibfnamefont {A.~L.}\ \bibnamefont
  {Melott}}, \bibinfo {author} {\bibfnamefont {S.~F.}\ \bibnamefont
  {Shandarin}},\ and\ \bibinfo {author} {\bibfnamefont {D.~H.}\ \bibnamefont
  {Weinberg}},\ }\bibfield  {title} {\bibinfo {title} {{A Test of the Adhesion
  Approximation for Gravitational Clustering}},\ }\href
  {https://doi.org/10.1086/174216} {\bibfield  {journal} {\bibinfo  {journal}
  {ApJ}\ }\textbf {\bibinfo {volume} {428}},\ \bibinfo {pages} {28} (\bibinfo
  {year} {1994})},\ \Eprint {https://arxiv.org/abs/9311075} {arXiv:9311075
  [arXiv:astro-ph]} \BibitemShut {NoStop}%
\bibitem [{\citenamefont {Sahni}\ \emph {et~al.}(1994)\citenamefont {Sahni},
  \citenamefont {Sathyaprakash},\ and\ \citenamefont {Shandarin}}]{Sahni1994}%
  \BibitemOpen
  \bibfield  {author} {\bibinfo {author} {\bibfnamefont {V.}~\bibnamefont
  {Sahni}}, \bibinfo {author} {\bibfnamefont {B.~S.}\ \bibnamefont
  {Sathyaprakash}},\ and\ \bibinfo {author} {\bibfnamefont {S.~F.}\
  \bibnamefont {Shandarin}},\ }\bibfield  {title} {\bibinfo {title} {{The
  Evolution of Voids in the Adhesion Approximation}},\ }\href
  {https://doi.org/10.1086/174464} {\bibfield  {journal} {\bibinfo  {journal}
  {ApJ}\ }\textbf {\bibinfo {volume} {431}},\ \bibinfo {pages} {20} (\bibinfo
  {year} {1994})},\ \Eprint {https://arxiv.org/abs/9403044} {arXiv:9403044
  [astro-ph]} \BibitemShut {NoStop}%
\bibitem [{\citenamefont {She}\ \emph {et~al.}(1992)\citenamefont {She},
  \citenamefont {Aurell},\ and\ \citenamefont {Frisch}}]{She1992}%
  \BibitemOpen
  \bibfield  {author} {\bibinfo {author} {\bibfnamefont {Z.~S.}\ \bibnamefont
  {She}}, \bibinfo {author} {\bibfnamefont {E.}~\bibnamefont {Aurell}},\ and\
  \bibinfo {author} {\bibfnamefont {U.}~\bibnamefont {Frisch}},\ }\bibfield
  {title} {\bibinfo {title} {{The inviscid Burgers equation with initial data
  of Brownian type}},\ }\href {https://doi.org/10.1007/BF02096551/METRICS}
  {\bibfield  {journal} {\bibinfo  {journal} {Communications in Mathematical
  Physics}\ }\textbf {\bibinfo {volume} {148}},\ \bibinfo {pages} {623}
  (\bibinfo {year} {1992})}\BibitemShut {NoStop}%
\bibitem [{\citenamefont {Gurbatov}\ \emph {et~al.}(1997)\citenamefont
  {Gurbatov}, \citenamefont {Simdyankin}, \citenamefont {Aurell}, \citenamefont
  {Frisch},\ and\ \citenamefont {T{\'{o}}th}}]{Gurbatov1997}%
  \BibitemOpen
  \bibfield  {author} {\bibinfo {author} {\bibfnamefont {S.~N.}\ \bibnamefont
  {Gurbatov}}, \bibinfo {author} {\bibfnamefont {S.~I.}\ \bibnamefont
  {Simdyankin}}, \bibinfo {author} {\bibfnamefont {E.}~\bibnamefont {Aurell}},
  \bibinfo {author} {\bibfnamefont {U.}~\bibnamefont {Frisch}},\ and\ \bibinfo
  {author} {\bibfnamefont {G.}~\bibnamefont {T{\'{o}}th}},\ }\bibfield  {title}
  {\bibinfo {title} {{On the decay of Burgers turbulence}},\ }\href
  {https://doi.org/10.1017/S0022112097006241} {\bibfield  {journal} {\bibinfo
  {journal} {Journal of Fluid Mechanics}\ }\textbf {\bibinfo {volume} {344}},\
  \bibinfo {pages} {339} (\bibinfo {year} {1997})},\ \Eprint
  {https://arxiv.org/abs/9709002} {arXiv:9709002 [physics]} \BibitemShut
  {NoStop}%
\bibitem [{\citenamefont {Bertoin}(1998)}]{Bertoin1998}%
  \BibitemOpen
  \bibfield  {author} {\bibinfo {author} {\bibfnamefont {J.}~\bibnamefont
  {Bertoin}},\ }\bibfield  {title} {\bibinfo {title} {{The inviscid burgers
  equation with Brownian initial velocity}},\ }\href
  {https://doi.org/10.1007/S002200050334/METRICS} {\bibfield  {journal}
  {\bibinfo  {journal} {Communications in Mathematical Physics}\ }\textbf
  {\bibinfo {volume} {193}},\ \bibinfo {pages} {397} (\bibinfo {year}
  {1998})}\BibitemShut {NoStop}%
\bibitem [{\citenamefont {Valageas}(2009{\natexlab{b}})}]{Valageas2009a}%
  \BibitemOpen
  \bibfield  {author} {\bibinfo {author} {\bibfnamefont {P.}~\bibnamefont
  {Valageas}},\ }\bibfield  {title} {\bibinfo {title} {{Statistical properties
  of the burgers equation with brownian initial velocity}},\ }\href
  {https://doi.org/10.1007/S10955-009-9685-5/METRICS} {\bibfield  {journal}
  {\bibinfo  {journal} {Journal of Statistical Physics}\ }\textbf {\bibinfo
  {volume} {134}},\ \bibinfo {pages} {589} (\bibinfo {year}
  {2009}{\natexlab{b}})}\BibitemShut {NoStop}%
\bibitem [{\citenamefont {Frachebourg}\ and\ \citenamefont
  {Martin}(2000)}]{Frachebourg2000a}%
  \BibitemOpen
  \bibfield  {author} {\bibinfo {author} {\bibfnamefont {L.}~\bibnamefont
  {Frachebourg}}\ and\ \bibinfo {author} {\bibfnamefont {P.~A.}\ \bibnamefont
  {Martin}},\ }\bibfield  {title} {\bibinfo {title} {{Exact statistical
  properties of the Burgers equation}},\ }\href
  {https://doi.org/10.1017/S0022112000001142} {\bibfield  {journal} {\bibinfo
  {journal} {Journal of Fluid Mechanics}\ }\textbf {\bibinfo {volume} {417}},\
  \bibinfo {pages} {323} (\bibinfo {year} {2000})},\ \Eprint
  {https://arxiv.org/abs/9905056} {arXiv:9905056 [cond-mat]} \BibitemShut
  {NoStop}%
\bibitem [{\citenamefont {Valageas}(2009{\natexlab{c}})}]{Valageas2009b}%
  \BibitemOpen
  \bibfield  {author} {\bibinfo {author} {\bibfnamefont {P.}~\bibnamefont
  {Valageas}},\ }\bibfield  {title} {\bibinfo {title} {{Some statistical
  properties of the Burgers equation with white-noise initial velocity}},\
  }\href {https://doi.org/10.1007/S10955-009-9809-Y/METRICS} {\bibfield
  {journal} {\bibinfo  {journal} {Journal of Statistical Physics}\ }\textbf
  {\bibinfo {volume} {137}},\ \bibinfo {pages} {729} (\bibinfo {year}
  {2009}{\natexlab{c}})},\ \Eprint {https://arxiv.org/abs/0903.0956}
  {arXiv:0903.0956} \BibitemShut {NoStop}%
\bibitem [{\citenamefont {Avellaneda}\ and\ \citenamefont
  {Weinan}(1995)}]{Avellaneda1995}%
  \BibitemOpen
  \bibfield  {author} {\bibinfo {author} {\bibfnamefont {M.}~\bibnamefont
  {Avellaneda}}\ and\ \bibinfo {author} {\bibfnamefont {E.}~\bibnamefont
  {Weinan}},\ }\bibfield  {title} {\bibinfo {title} {{Statistical properties of
  shocks in Burgers turbulence}},\ }\href
  {https://doi.org/10.1007/BF02104509/METRICS} {\bibfield  {journal} {\bibinfo
  {journal} {Communications in Mathematical Physics}\ }\textbf {\bibinfo
  {volume} {172}},\ \bibinfo {pages} {13} (\bibinfo {year} {1995})}\BibitemShut
  {NoStop}%
\bibitem [{\citenamefont {Molchan}(1997)}]{Molchan1997}%
  \BibitemOpen
  \bibfield  {author} {\bibinfo {author} {\bibfnamefont {G.~M.}\ \bibnamefont
  {Molchan}},\ }\bibfield  {title} {\bibinfo {title} {{Burgers equation with
  self-similar Gaussian initial data: Tail probabilities}},\ }\href
  {https://doi.org/10.1007/BF02732428/METRICS} {\bibfield  {journal} {\bibinfo
  {journal} {Journal of Statistical Physics}\ }\textbf {\bibinfo {volume}
  {88}},\ \bibinfo {pages} {1139} (\bibinfo {year} {1997})}\BibitemShut
  {NoStop}%
\bibitem [{\citenamefont {Valageas}(2009{\natexlab{d}})}]{Valageas2009c}%
  \BibitemOpen
  \bibfield  {author} {\bibinfo {author} {\bibfnamefont {P.}~\bibnamefont
  {Valageas}},\ }\bibfield  {title} {\bibinfo {title} {{Quasilinear regime and
  rare-event tails of decaying Burgers turbulence}},\ }\href
  {https://doi.org/10.1103/PhysRevE.80.016305} {\bibfield  {journal} {\bibinfo
  {journal} {Physical Review E}\ }\textbf {\bibinfo {volume} {80}},\ \bibinfo
  {pages} {016305} (\bibinfo {year} {2009}{\natexlab{d}})}\BibitemShut
  {NoStop}%
\bibitem [{\citenamefont {Carraro}\ and\ \citenamefont
  {Duchon}(1998)}]{Carraro1998}%
  \BibitemOpen
  \bibfield  {author} {\bibinfo {author} {\bibfnamefont {L.}~\bibnamefont
  {Carraro}}\ and\ \bibinfo {author} {\bibfnamefont {J.}~\bibnamefont
  {Duchon}},\ }\bibfield  {title} {\bibinfo {title} {{{\'E}quation de Burgers
  avec conditions initiales {\`a} accroissements ind{\'e}pendants et
  homog{\`e}nes}},\ }\href {https://doi.org/10.1016/S0294-1449(98)80030-9}
  {\bibfield  {journal} {\bibinfo  {journal} {Annales de l'Institut Henri
  Poincar{\'e} C}\ }\textbf {\bibinfo {volume} {15}},\ \bibinfo {pages} {431}
  (\bibinfo {year} {1998})}\BibitemShut {NoStop}%
\bibitem [{\citenamefont {Bertoin}(2001)}]{Bertoin2001}%
  \BibitemOpen
  \bibfield  {author} {\bibinfo {author} {\bibfnamefont {J.}~\bibnamefont
  {Bertoin}},\ }\bibfield  {title} {\bibinfo {title} {{Some Properties of
  Burgers Turbulence with White or Stable Noise Initial Data}},\ }in\ \href
  {https://doi.org/10.1007/978-1-4612-0197-7_12} {\emph {\bibinfo {booktitle}
  {{L{\'e}vy Processes}}}}\ (\bibinfo  {publisher} {Birkh{\"a}user, Boston,
  MA},\ \bibinfo {year} {2001})\ pp.\ \bibinfo {pages} {267--279}\BibitemShut
  {NoStop}%
\bibitem [{\citenamefont {Bernard}\ and\ \citenamefont
  {Gawe¸dzki}(1998)}]{Bernard1998}%
  \BibitemOpen
  \bibfield  {author} {\bibinfo {author} {\bibfnamefont {D.}~\bibnamefont
  {Bernard}}\ and\ \bibinfo {author} {\bibfnamefont {K.~G.}\ \bibnamefont
  {Gawe¸dzki}},\ }\bibfield  {title} {\bibinfo {title} {{Scaling and exotic
  regimes in decaying Burgers turbulence}},\ }\href@noop {} {\bibfield
  {journal} {\bibinfo  {journal} {J. Phys. A: Math. Gen}\ }\textbf {\bibinfo
  {volume} {31}},\ \bibinfo {pages} {8735} (\bibinfo {year}
  {1998})}\BibitemShut {NoStop}%
\bibitem [{\citenamefont {Gurbatov}(1999)}]{Gurbatov1999}%
  \BibitemOpen
  \bibfield  {author} {\bibinfo {author} {\bibfnamefont {S.~N.}\ \bibnamefont
  {Gurbatov}},\ }\bibfield  {title} {\bibinfo {title} {{Universality classes
  for self-similarity of noiseless multi-dimensional Burgers turbulence and
  interface growth}},\ }\href {https://doi.org/10.1103/PhysRevE.61.2595}
  {\bibfield  {journal} {\bibinfo  {journal} {Physical Review E - Statistical
  Physics, Plasmas, Fluids, and Related Interdisciplinary Topics}\ }\textbf
  {\bibinfo {volume} {61}},\ \bibinfo {pages} {2595} (\bibinfo {year}
  {1999})},\ \Eprint {https://arxiv.org/abs/9912011v1} {arXiv:9912011v1
  [chao-dyn]} \BibitemShut {NoStop}%
\bibitem [{\citenamefont {Gueudr{\'e}}\ and\ \citenamefont {{Le
  Doussal}}(2014)}]{Gueudre2014}%
  \BibitemOpen
  \bibfield  {author} {\bibinfo {author} {\bibfnamefont {T.}~\bibnamefont
  {Gueudr{\'e}}}\ and\ \bibinfo {author} {\bibfnamefont {P.}~\bibnamefont {{Le
  Doussal}}},\ }\bibfield  {title} {\bibinfo {title} {{Statistics of shocks in
  a toy model with heavy tails}},\ }\href
  {https://doi.org/10.1103/PhysRevE.89.042111} {\bibfield  {journal} {\bibinfo
  {journal} {Physical Review E - Statistical, Nonlinear, and Soft Matter
  Physics}\ }\textbf {\bibinfo {volume} {89}},\ \bibinfo {pages} {42111}
  (\bibinfo {year} {2014})}\BibitemShut {NoStop}%
\bibitem [{\citenamefont {Molchanov}\ \emph {et~al.}(1997)\citenamefont
  {Molchanov}, \citenamefont {Surgailis},\ and\ \citenamefont
  {Woyczynski}}]{Molchanov1997}%
  \BibitemOpen
  \bibfield  {author} {\bibinfo {author} {\bibfnamefont {S.~A.}\ \bibnamefont
  {Molchanov}}, \bibinfo {author} {\bibfnamefont {D.}~\bibnamefont
  {Surgailis}},\ and\ \bibinfo {author} {\bibfnamefont {W.~A.}\ \bibnamefont
  {Woyczynski}},\ }\bibfield  {title} {\bibinfo {title} {{The large-scale
  structure of the universe and quasi-Voronoi tessellation of shock fronts in
  forced Burgers turbulence in Rd}},\ }\href
  {https://doi.org/10.1214/AOAP/1034625260} {\bibfield  {journal} {\bibinfo
  {journal} {Annals of Applied Probability}\ }\textbf {\bibinfo {volume} {7}},\
  \bibinfo {pages} {200} (\bibinfo {year} {1997})}\BibitemShut {NoStop}%
\bibitem [{\citenamefont {Albeverio}\ \emph {et~al.}(1994)\citenamefont
  {Albeverio}, \citenamefont {Molchanov},\ and\ \citenamefont
  {Surgailis}}]{Albeverio1994}%
  \BibitemOpen
  \bibfield  {author} {\bibinfo {author} {\bibfnamefont {S.}~\bibnamefont
  {Albeverio}}, \bibinfo {author} {\bibfnamefont {S.~A.}\ \bibnamefont
  {Molchanov}},\ and\ \bibinfo {author} {\bibfnamefont {D.}~\bibnamefont
  {Surgailis}},\ }\bibfield  {title} {\bibinfo {title} {{Stratified structure
  of the Universe and Burgers' equation-a probabilistic approach}},\ }\href
  {https://doi.org/10.1007/BF01268990/METRICS} {\bibfield  {journal} {\bibinfo
  {journal} {Probability Theory and Related Fields}\ }\textbf {\bibinfo
  {volume} {100}},\ \bibinfo {pages} {457} (\bibinfo {year}
  {1994})}\BibitemShut {NoStop}%
\bibitem [{\citenamefont {Valageas}(2007)}]{Valageas2007}%
  \BibitemOpen
  \bibfield  {author} {\bibinfo {author} {\bibfnamefont {P.}~\bibnamefont
  {Valageas}},\ }\bibfield  {title} {\bibinfo {title} {{Using the Zeldovich
  dynamics to test expansion schemes}},\ }\href
  {https://doi.org/10.1051/0004-6361:20078065} {\bibfield  {journal} {\bibinfo
  {journal} {Astronomy \& Astrophysics}\ }\textbf {\bibinfo {volume} {476}},\
  \bibinfo {pages} {31} (\bibinfo {year} {2007})}\BibitemShut {NoStop}%
\bibitem [{\citenamefont {Molchanov}\ \emph {et~al.}(1995)\citenamefont
  {Molchanov}, \citenamefont {Surgailis},\ and\ \citenamefont
  {Woyczynski}}]{Molchanov1995}%
  \BibitemOpen
  \bibfield  {author} {\bibinfo {author} {\bibfnamefont {S.~A.}\ \bibnamefont
  {Molchanov}}, \bibinfo {author} {\bibfnamefont {D.}~\bibnamefont
  {Surgailis}},\ and\ \bibinfo {author} {\bibfnamefont {W.~A.}\ \bibnamefont
  {Woyczynski}},\ }\bibfield  {title} {\bibinfo {title} {{Hyperbolic
  asymptotics in Burgers' turbulence and extremal processes}},\ }\href
  {https://doi.org/10.1007/BF02099589/METRICS} {\bibfield  {journal} {\bibinfo
  {journal} {Communications in Mathematical Physics}\ }\textbf {\bibinfo
  {volume} {168}},\ \bibinfo {pages} {209} (\bibinfo {year}
  {1995})}\BibitemShut {NoStop}%
\end{thebibliography}%

\end{document}